\documentclass[prd,superscriptaddress,showpacs,amssymb,amsmath,amsfonts,aps,altaffilletter,nofootinbib,letterpaper,twocolumn]{revtex4-1}

\usepackage{color}
\usepackage{amsfonts}
\usepackage{graphicx}
\usepackage{rotating}
\usepackage{fancyhdr}
\usepackage{float}
\usepackage{ulem}
\usepackage[raggedright]{subfigure}
\usepackage{acronym}
\usepackage{multirow}
\usepackage{array}
\usepackage{hyperref}
\usepackage{appendix}
\usepackage{float}
\restylefloat{table}

\makeatletter
\makeatletter
\def\@fnsymbol#1{\ifcase#1\or * \or  $+$ \or  \$ \or \#  \or \dag \or \ddag \or
$\mathsection$ \or $ \mathparagraph$ \or $\|$  \or \textordfeminine \or \textbul
let   
\or ** \or $++$ \or  \$\$ \or \#\#  \or \dag\dag \or \ddag\ddag \or
$\mathsection\mathsection$ \or $ \mathparagraph\mathparagraph$ \or $\|\|$  \or 
\textordfeminine\textordfeminine \or \textbullet \textbullet \or *** \or $+++$ 
\or  \$\$\$ \or \#\#  \or \dag\dag \or \ddag\ddag \or
$\mathsection \mathsection\mathsection$ \or $ \mathparagraph 
\mathparagraph\mathparagraph$ \or $\|\|\|$  \or 
\textordfeminine\textordfeminine\textordfeminine \or 
\textbullet\textbullet\textbullet \or \else \@ctrerr\fi}
\makeatother

\renewcommand{\today}{\number\day\space\ifcase\month\or
  January\or February\or March\or April\or May\or June\or
  July\or August\or September\or October\or November\or December\fi
  \space\number\year}

\begin{document}

\title{Multivariate Classification with Random Forests for Gravitational Wave Searches of Black Hole Binary Coalescence}

\author{%
Paul T. Baker$^{1}$,
Sarah Caudill$^{2}$,
Kari A. Hodge$^{3}$,
Dipongkar Talukder$^{4}$,
Collin Capano$^{5}$,
and Neil J. Cornish$^{1}$%
}\noaffiliation

\affiliation {Montana State University, Bozeman, MT 59717, USA }
\affiliation {Leonard E. Parker Center for Gravitation, Cosmology, \& Astrophysics, University of Wisconsin--Milwaukee, Milwaukee, WI 53201, USA }
\affiliation {California Institute of Technology, Pasadena, CA 91125, USA }
\affiliation {University of Oregon, Eugene, OR 97403, USA }
\affiliation {Maryland Center for Fundamental Physics \& Joint Space Science Institute, Department of Physics, University of Maryland, College Park, 20742, USA }

\date[\relax]{Date: \today }

\begin{abstract}
Searches for gravitational waves produced by coalescing black hole binaries with total masses $\gtrsim25\,$M$_\odot$ use matched filtering with templates of short duration.
Non-Gaussian noise bursts in gravitational wave detector data can mimic short signals and limit the sensitivity of these searches.
Previous searches have relied on empirically designed statistics incorporating signal-to-noise ratio and signal-based vetoes to separate gravitational wave candidates from noise candidates.
We report on sensitivity improvements achieved using a multivariate candidate ranking statistic derived from a supervised machine learning algorithm.
We apply the random forest of bagged decision trees technique to two separate searches in the high mass $\left( \gtrsim25\,\mathrm{M}_\odot \right)$ parameter space.
For a search which is sensitive to gravitational waves from the inspiral, merger, and ringdown (IMR) of binary black holes with total mass between $25\,$M$_\odot$ and $100\,$M$_\odot$, we find sensitive volume improvements as high as $70_{\pm 13}-109_{\pm 11}$\% when compared to the previously used ranking statistic.
For a ringdown-only search which is sensitive to gravitational waves from the resultant perturbed intermediate mass black hole with mass roughly between $10\,$M$_\odot$ and $600\,$M$_\odot$, we find sensitive volume improvements as high as $61_{\pm 4}-241_{\pm 12}$\% when compared to the previously used ranking statistic.
We also report how sensitivity improvements can differ depending on mass regime, mass ratio, and available data quality information.
Finally, we describe the techniques used to tune and train the random forest classifier that can be generalized to its use in other searches for gravitational waves.
\end{abstract}

\maketitle

\section{Introduction}\label{sec:overview}
We are rapidly approaching the era of advanced gravitational-wave detectors.
Advanced LIGO~\cite{2010aligo} and Advanced Virgo~\cite{2014avirgo} are expected to begin operation in 2015.
Within the next decade, these will be joined by the KAGRA~\cite{2014kagra} and LIGO-India~\cite{2014indigo} detectors.
The coalescence of compact binaries containing neutron stars and/or stellar mass black holes are expected to be a strong and promising source for the first detection of gravitational waves~\cite{2010rates}.
Higher mass sources with total masses $\gtrsim25\,$M$_\odot$ including binary black holes (BBHs) and intermediate mass black holes (IMBHs) are less certain but still potentially strong sources~\cite{2010rates,2013s6highmass,2014rdsearch}.
Discovery and new science will be possible with detection of gravitational waves from these objects~\cite{1993cbcscience, 2009gwscience}.
\par

Measurement of gravitational waves requires exquisitely sensitive detectors as well as advanced data analysis techniques~\cite{2009ligo}.
By digging into detector noise for weak signals rather than waiting for a rare, loud event, we increase detection rates.
Unfortunately, detector noise can be non-stationary and non-Gaussian, leading to loud, short duration noise transients.
Such behavior is particularly troublesome for higher mass searches where the expected in-band signal is of similar duration as noise transients.
Traditional searches for compact binary coalescence have utilized multi-detector coincidence, carefully designed ranking statistics, and other data quality methods~\cite{20091styrS5lowmass, 2009s5lowmass186, 2012S6lowmass, 2011s5highmass, 2013s6highmass, 2014rdsearch}.
However, in many searches performed to-date over initial LIGO and Virgo data, the sensitivity was limited by an accidental coincidence involving a non-Gaussian transient noise burst~\cite{2011s5highmass, 2013s6highmass,2014rdsearch}.
\par

Only recently have gravitational-wave searches begun to utilize methods that work with the full multidimensional parameter space of classification statistics for candidate events.
Previous studies have shown multivariate methods give detection probability improvement over techniques based on single parameter thresholds~\cite{2008cannonbayes}, \cite{2014kari}.
\par

Machine learning has a wealth of tools available for the purpose of multivariate statistical classification~\cite{2012amlmla, 2014narskybook}.
These include but are not limited to artificial neural networks~\cite{1989nn,2009nn}, support vector machines~\cite{1995svm,2000svm}, and random forests of decision trees~\cite{2001breimanrf}.
Such methods have already proven useful in a number of other fields with large data sets and background contamination including optical and radio astronomy ~\cite{2014mlapulsars, 2012ptfbloom, 2013mlabrink} and high energy physics~\cite{2009mlaheph,2005narskypresent}.
Within the field of gravitational wave physics, a search for gravitational-wave bursts associated with gamma ray bursts found a factor of $\sim$3 increase in sensitive volume when using a multivariate analysis with boosted decision trees~\cite{2013mvscburst}.
Applications of artificial neural networks to a search for compact binary coalescence signals associated with gamma ray bursts found smaller improvements~\cite{2014ANN}.
Machine learning algorithms have successfully been applied to the problem of detector noise artifact classification~\cite{2013mvscglitch}.
Additionally, a search for bursts of gravitational waves from cosmic string cusps~\cite{2013cosmicstring} used the multivariate technique described in~\cite{2008cannonbayes}.
\par

In this paper, we focus on the development and sensitivity improvements of a multivariate analysis applied to matched filter searches for gravitational waves produced by coalescing black hole binaries with total masses $\gtrsim25\,$M$_\odot$.
In particular, we focus on the application to two separate searches in this parameter space.
The first, designated the IMR search, looks for gravitational waves from the inspiral, merger, and ringdown of BBHs with total mass between $25\,$M$_\odot$ and $100\,$M$_\odot$.
The second, designated the ringdown-only search, looks for gravitational waves from the resultant perturbed IMBH with mass roughly between $10\,$M$_\odot$ and $600\,$M$_\odot$.
These investigations are performed over data collected by LIGO and Virgo between 2009 and 2010 so that comparisons can be made with previous IMR and ringdown-only search results~\cite{2013s6highmass,2014rdsearch}.
Using a random forest of bagged decision trees (RFBDT) supervised machine learning algorithm (MLA), we explore sensitivity improvements over each search's previous classification statistic.
Additionally, we describe techniques used to tune and train the RFBDT classifier that can be generalized to its use in other searches for gravitational waves.
\par

In Sec.~\ref{sec:detection}, we frame the general detection problem in gravitational-wave data analysis and motivate the need for multivariate classification.
In Sec.~\ref{sec:data}, we describe our data set.
In Sec.~\ref{sec:detstat}, we explain the method used to classify gravitational-wave candidates in matched-filter searches.
In Sec.~\ref{sec:mla}, we review RFBDTs as used in these investigations.
In Sec.~\ref{sec:tuning}, we discuss the training set, the multidimensional space used to characterize candidates, and the tunable parameters of the classifier.
In Sec.~\ref{sec:results}, we describe the improvement in sensitive volume obtained by the IMR and ringdown-only searches over LIGO and Virgo data from 2009 to 2010 when using RFBDTs.
Finally, in Sec.~\ref{sec:summary} we summarize our results.
\par

\section{The Detection Problem}\label{sec:detection}
Searches for gravitational waves are generally divided based on astrophysical source.
The gravitational waveform from compact binary coalescence has a well-defined model~\cite{2006cbcreview, 2012cbcreview}.
Thus searches for these types of signals use the method of matched-filtering with a template bank of model waveforms.
This is the optimal method for finding modeled signals with known parameters buried in Gaussian noise~\cite{1960matchedfilter, 1962wainstein}.
However, if the parameters are not known, matched filtering is not optimal~\cite{2009prix}, and additional techniques must be employed to address the extraction of weak and/or rare signals from non-Gaussian, non-stationary detector noise, the elimination or identification of false alarms, and the ranking of gravitational-wave candidates by significance. This paper presents the construction of an {\it ad hoc} statistic, automated through machine learning, that can tackle these issues.
\par

\subsection{Searches for compact binary coalescence}\label{sec:cbc}
The coalescence of compact binaries generates a gravitational-wave signal composed of inspiral, merger and ringdown phases~\cite{2006cbcreview, 2012cbcreview}.
The low frequency inspiral phase marks the period during which the compact objects orbit each other, radiating energy and angular momentum as gravitational waves~\cite{2002inspiralreview}.
The signal for low mass systems in the LIGO and Virgo frequency sensitivity bands (i.e., above the steeply rising seismic noise at 40\,Hz for initial detectors or 10\,Hz for advanced detectors~\cite{iligoaligonoise}) is dominated by the inspiral phase.
Several searches have looked for the inspiral from low mass systems with component masses $>1\,$M$_\odot$ and total mass $< 25\,$M$_\odot$~\cite{2009s5lowmass186, 20091styrS5lowmass, 2012S6lowmass}.
The higher frequency merger phase marks the coalescence of the compact objects and the peak gravitational-wave emission~\cite{2010Rnrreview1, 2009nrreview, 2011nrreview}.
Since the merger frequency is inversely proportional to the mass of the binary, the signal for high mass systems in the LIGO and Virgo sensitivity bands could include inspiral, merger and ringdown phases.
Searches for high mass signals including all three phases have been performed for systems with total mass between 25$\,$M$_\odot$ and 100$\,$M$_\odot$~\cite{2011s5highmass, 2013s6highmass}.
We designate this as the IMR search.
Systems accessible to LIGO and Virgo with even higher total masses will only have a ringdown phase in-band, during which the compact objects have already formed a single perturbed black hole~\cite{1999rdreview, 2007rdwaveform}.
Searches for ringdown signals have looked for perturbed black holes with total masses roughly in the range 10$\,$M$_\odot$ to 600$\,$M$_\odot$ and dimensionless spins in the range 0 to 0.99~\cite{2009s4ringdown, 2014rdsearch}.
The dimensionless spin is defined as $\hat{a}=cS/GM^2$ for black hole mass $M$ and spin angular momentum $S$.
We designate this as the ringdown-only search.
\par

Each of these searches use a matched-filter algorithm with template banks of model waveforms to search data from multiple gravitational-wave detectors.
The output is a SNR time series for each detector.
We record local maxima, called triggers, in the SNR time series that fall above a predetermined threshold.
Low mass searches use template banks of inspiral model waveforms generated at 3.5 post-Newtonian order in the frequency domain~\cite{1995inspiralwaveform,2004inspiralwaveform}.
These waveforms typically remain in the initial LIGO/Virgo frequency sensitivity band for tens of seconds providing a natural defense against triggers arising from short bursts of non-Gaussian noise.
\par

The templates for IMR searches include the full inspiral-merger-ringdown waveform, computed analytically and tuned against numerical relativity results. For these investigations, the non-spinning EOBNRv1 family of IMR waveforms was used ~\cite{2007eobnrv1}.
The templates, like those for the low mass search, are described by the chirp mass $\mathcal{M}=\eta^{3/5}M$ and symmetric mass ratio $\eta=m_1m_2/M^2$ of the component objects (where $M=m_1+m_2$)~\cite{findchirp}.
The duration of high mass waveforms in-band for the initial detectors is much shorter than the duration for low mass waveforms, making the IMR search susceptible to triggers associated with short bursts of non-Gaussian noise.
\par

The templates for the search for perturbed black holes, with even higher total mass, is based on black hole perturbation theory and numerical relativity.
A perturbed Kerr black hole will emit gravitational waves in a superposition of quasinormal modes of oscillation characterized by a frequency $f_{\ell mn}$ and damping time $\tau_{\ell mn}$~\cite{1973teukolsky, 1999rdreview}.
Numerical simulations have demonstrated that the $\left(\ell, m, n\right)=\left(2,2,0\right)$ dominates the gravitational-wave emission~\cite{2007dominantlmn, 2007rdwaveform}.
From here on, we will designate $f_{220}$ as $f_{0}$ and write the damping time $\tau_{220}$ in terms of the quality factor $Q=Q_{220}=\pi f_{220}\tau_{220}$.
Ringdown model waveforms decay on the timescale $0.0001\lesssim \tau/\mathrm{s} \lesssim 0.1$, again making this search susceptible to contamination from short noise bursts.
\par

The matched-filter algorithms are described in \cite{findchirp, 2009s4ringdown}.
Further details on the templates and template bank construction in the IMR and ringdown-only searches can be found in \cite{2013s6highmass, 2014rdsearch}.
\par

Matched filtering alone cannot completely distinguish triggers caused by gravitational waves from those caused by noise.
Thus tools such as data quality vetoes, multi-detector coincidence, and SNR consistency checks are needed~\cite{2010vetoes, 2013ihopepaper}.
Additionally, a $\chi^2$ time-frequency signal consistency test augments searches with broadband signal including the IMR search but is less useful for short, quasi-monochromatic ringdown signals~\cite{2005allen}.
Finally, each search uses a detection statistic to summarize the separation of signal from background.
Details on the construction of a detection statistic are provided in Section~\ref{sec:detstat}.
\par

In general, coincidence tests are applied to single detector triggers to check for multi-detector consistency.
The low and high mass searches use an ellipsoidal coincidence test ({\it ethinca}~\cite{ethinca}) that requires consistent values of template masses and time of arrival.
The ringdown-only search coincidence test similarly calculates the distance $ds^2$ between two triggers by checking simultaneously for time and template coincidence (d$f_0$ and d$Q$)~\cite{2014rdsearch}.
When three detectors are operating, if each pair of triggers passes the coincidence test, we store a triple coincidence.
We also store double coincidences for particular network configurations as outlined in Section~\ref{sec:data}.
\par

\subsection{Signal and background}\label{sec:sigts}
Evaluating the performance of a detection statistic and training the machine learning classifier require the calculation of detection efficiency at an allowed level of background contamination.
In the absence of actual gravitational-wave events, we determine detection efficiency through the use of simulated signals (``injections") added to the detectors' data streams.
To estimate the search background, we generate a set of accidental coincidences using the method of time-shifted data.
\par

The simulated signal set is added to the data and a separate search is run.
Triggers are recorded corresponding to times when injections were made.
The simulated signals are representative of the expected gravitational waveforms detectable by a search.
For the IMR and ringdown-only searches, the simulated signals include waveforms from the EOBNRv2 family~\cite{2011eobnr} for systems whose component objects are not spinning and from the IMRPhenomB family~\cite{2011imrphenomb} for systems whose component objects have aligned, anti-aligned, or no spins.
Additionally, for the ringdown-only search, we inject ringdown-only waveforms.
For a discussion of injection waveform parameters, see Section~\ref{sec:sigtraining}.
Considerations for the injection sets used in training the classifier are discussed in Section~\ref{sec:training} and in computing search sensitivity are discussed in Section~\ref{sec:results}.
\par

The background rate of accidental trigger coincidence between detectors is evaluated using the method of time-shifted data.
We shift the data by intervals of time longer than the light travel time between detectors and then perform a separate search.
Any multi-detector coincidence found in the time-shifted search is very likely due to non-Gaussian glitches.
We perform searches over 100 sets of time-shifted data and recorded the accidental coincidences found by the algorithm.
Details of the method are provided in Section~IIIB of~\cite{2013s6highmass} and IIIC of~\cite{2014rdsearch}.
For a discussion of the set of accidental coincidences used in training the classifier, see Section~\ref{sec:training}.
\par

\section{Data Set}\label{sec:data}
We performed investigations using data collected by the LIGO and Virgo detectors between July 2009 and October 2010~\cite{2012s6}.
In~\cite{2014rdsearch}, we designate this time as Period~2 to distinguish it from the analysis of Period~1 data collected between November 2005 and September 2007.
All results reported here consider only Period~2 data.
For continuity, we will continue to designate our data analysis time as Period~2.
\par

Period~2 covers LIGO's sixth science run~\cite{2009eLIGO}.
During this time, the $4\,$km detectors in Hanford, Washington (H1) and in Livingston, Louisiana (L1) were operating.
The $3\,$km Virgo detector (V1) in Cascina, Italy conducted its second and third science run during this time~\cite{2011vsr2}.
The investigations were performed using data from the coincident search networks of the H1L1V1, H1L1, H1V1, and L1V1 detectors.
Coincidences were stored for all triple and double detector combinations.
\par

Data was analyzed separately using the IMR and the ringdown-only search pipelines from the analyses reported in~\cite{2013s6highmass} and~\cite{2014rdsearch}.
In order to combat noise transients, three levels of data quality vetoes are applied to remove noise from LIGO-Virgo data when searches are performed.
Details of vetoes are provided in~\cite{2010vetoes} and specific descriptions of the use of the vetoes for these Period~2 analyses can be found in \cite{2013s6highmass, 2014kari, 2014rdsearch}.\footnote{The naming convention for veto categorization can vary across searches.
We use the convention for Veto Levels 1, 2, and 3 as defined in Section~V of~\cite{2010vetoes}}
We analyze the performance of RFBDT classification after the first and second veto levels have been applied and compare this to the performance after the first, second, and third veto levels have been applied.
After removal of the first and second veto levels, we were left with 0.48 years of analysis time and after additional removal of the third veto level, we were left with 0.42 years of analysis time.
We designate the search over Period~2 after the removal of the first and second veto levels as Veto Level~2 and after the removal of the first, second, and third veto levels as Veto Level~3.
\par

In order to capture the variability of detector noise and sensitivity, we divided Period~2 into separate analysis periods of $\sim$1 to 2 months.
We then estimated the volume to which each search was sensitive by injecting simulated waveforms into the data and testing our ability to recover them.
Details of the method are provided in Section~V of~\cite{2013s6highmass, 2014rdsearch}.
We repeat this procedure in Section~\ref{sec:results} in order to quantify the improvement in sensitive volume obtained by each search over LIGO and Virgo data from 2009 to 2010 when using RFBDTs.
\par

\section{Detection statistics}\label{sec:detstat}
We must rank all coincidences based on their likelihood of being a signal.
Gravitational-wave data analysis has no dearth of statistics to classify gravitational-wave candidates as signal or background, and often, the ranking statistic will be empirically designed as a composite of other statistics.
If the noise in the detector data was Gaussian, a matched-filter SNR would be a sufficient ranking statistic.
However, since detector noise is non-Gaussian and non-stationary, we often re-weight the SNR by additional statistics that improve our ability to distinguish signal from background.
The exact form will depend on the nature of the signal for which we are searching.
A good statistic for differentiating long inspirals may not work well for short ringdowns.
\par

Searches for low mass binaries have ranked candidates using matched-filter SNR weighted by the $\chi^2$ signal consistency test value (e.g., {\it effective} SNR~\cite{20091styrS5lowmass} and {\it new} SNR~\cite{2012S6lowmass}).
IMR searches have used similar statistics~\cite{inspiralonlyS4, S5hm, 2013s6highmass}.
Previous ringdown-only searches and studies have used SNR-based statistics to address the non-Gaussianity of the data without the use of additional signal-based waveform consistency tests (e.g., the chopped-L statistic for double~\cite{2009s4ringdown} and triple~\cite{2013talukder, 2014imr} coincident triggers).
However, ranking statistics can be constructed using multivariate techniques to incorporate the full discriminatory power of the multidimensional parameter space of gravitational-wave candidates.
Several searches have utilized this including~\cite{2013mvscburst, 2013cosmicstring}.
In Section~\ref{sec:mla}, we detail the implementation of a multivariate statistical classifier using RFBDTs for the most recent IMR and ringdown-only searches.
\par

The final statistic used to rank candidates in order of significance is known as a detection statistic.
We combine the ranking statistics for each trigger into a coincident statistic (i.e., a statistic that incorporates information from all detectors' triggers found in coincidence).
This coincident statistic is then used to calculate a combined false alarm rate, the final detection statistic of the search.
We determine the coincident statistic $R$ for three different types of coincidences: gravitational-wave search candidates, simulated waveform injection coincidences, and time-shifted coincidences.
We then determine a false alarm rate (FAR) for each type of coincidence by counting the number of time-shifted coincidences found in the analysis time $T$ for each of the coincident search networks given in Section~\ref{sec:data}.
For each of the different types of coincidences in each of the search networks, we determine the FAR with the expression:
\begin{equation}\label{eq:far}
\mathrm{FAR}=\frac{\sum\limits_{k=1}^{100} N_k(R \ge R^*)}{T}
\end{equation}
where $N_k$ is the measured number of coincidences with $R \ge R^*$ in the $k^{th}$ shifted analysis for a total of 100 time-shifted analyses.
We then rank coincidences by their FARs across all search networks into a combined ranking, known as combined FAR~\cite{2009keppelthesis}.
This is the final detection statistic used in these investigations.
\par

\section{Machine Learning Algorithm}\label{sec:mla}
In order to compute the FAR for any candidate, we use the RFBDT algorithm to assign a probability that any candidate is a gravitational-wave signal.
This random forest technology is a well-developed improvement over the classical decision tree.
Each event is characterized by a feature vector, containing parameters thought to be useful for distinguishing signal from background.
A decision tree consists of a series of binary splits on these feature vector parameters.
A single decision tree is a weak classifier, as it is susceptible to false minima and over-training~\cite{2005statpattern} and generally performs worse than neural networks~\cite{2005narskypresent}.
\par

Random forest technology combats these issues and is considered more stable in the presence of outliers and in very high dimensional parameter spaces than other machine learning algorithms~\cite{2005statpattern, 2014narskybook}.
We use the \textsf{StatPatternRecognition} (\textsf{SPR}) software package\footnote{\tt http://statpatrec.sourceforge.net} developed for high energy physics data analysis.
This analysis uses the Random Forest of Bootstrap AGGregatED (bagged) Decision Trees algorithm.
The method of bootstrap aggregation or ``bagging'' as described below, tends to perform better in high noise situations than other random forest methods such as boosting~\cite{1999BauerKohaviVoting}.
\par

\subsection{Random forests}\label{sec:rf}
\begin{figure}
    \centering
    \includegraphics[width=.45\textwidth]{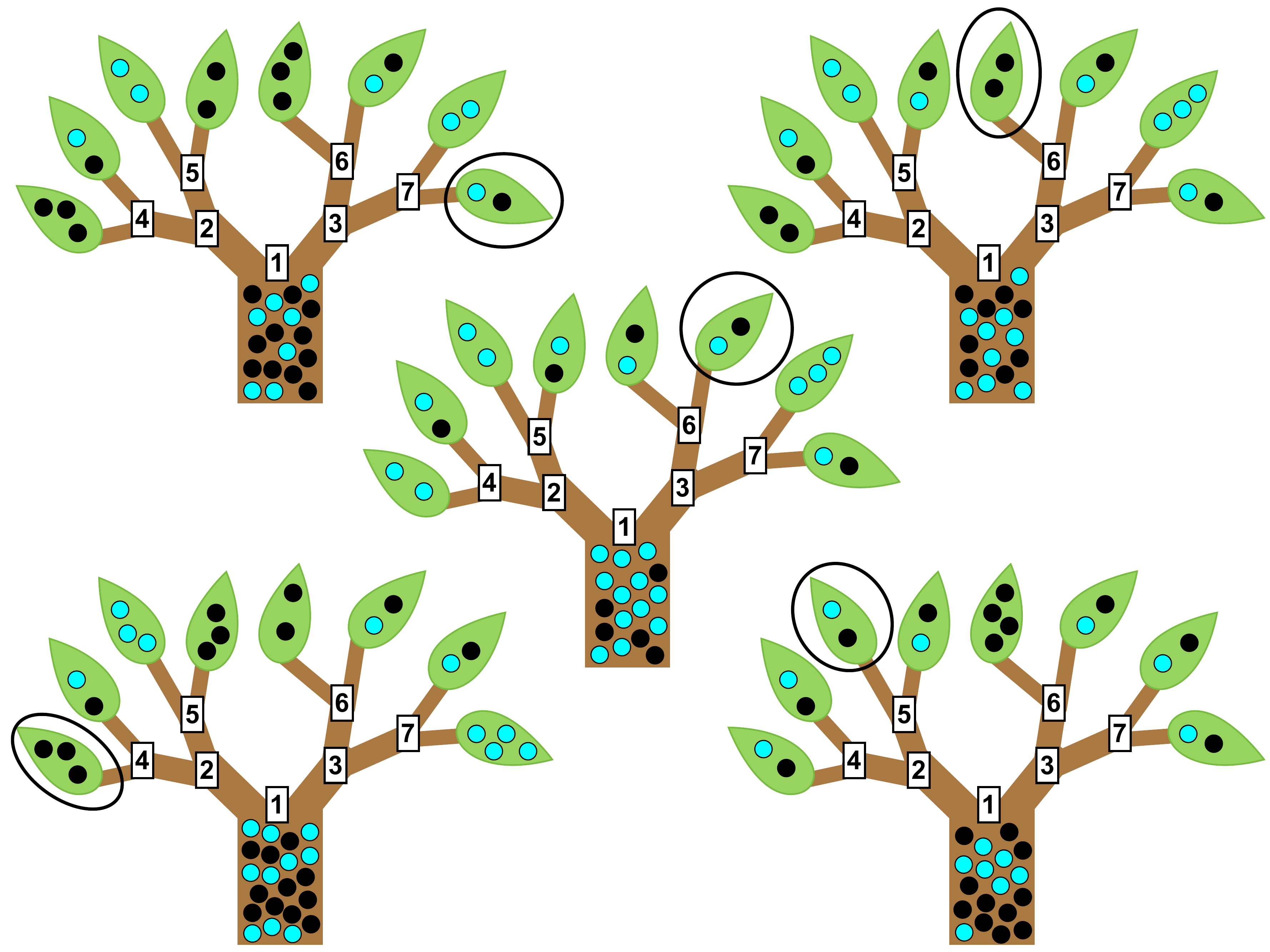}
    \caption{A cartoon example of a random forest. There are five decision trees in this random forest. Each was trained on a training set of objects belonging to either the black set or the cyan set. Note that the training set of each decision tree is different from the others. At each numbered node, or split in the tree, a binary decision based on a threshold of a feature vector parameter value is imposed. The decisions imposed at each node will differ for the different trees. When no split on a node can reduce the entropy or it contains fewer events than a preset limit, it is no longer divided and becomes a ``leaf''. Consider an object that we wish to classify as black or cyan. Suppose the object ends up in each circled leaf. Then the probability that the object is black is the fraction of black objects in all leaves, $p_\mathrm{forest}=73\%$.
      }
    \label{fig:forest}
  \end{figure}

A random forest of bagged decision trees uses a collection of many decision trees that are built from a set of training data.
The training data is composed of the feature vectors of events that we know \emph{a priori} to belong to either the set of signals or background (i.e., coincidences associated with simulated injections and coincidences associated with time-shifted data).
Then the decision trees are used to assign a probability that an event that we wish to classify belongs to the class of signal or background. A cartoon diagram is presented in Fig.~\ref{fig:forest}.
\par

To construct a decision tree, we make a series of one-dimensional splits on a random subset of parameters from the feature vector.
Each split determines a branching point, or node.
Individual splits are chosen to best separate signal and background given the available parameters.
There are several methods to determine the optimal parameter threshold at each node.
We measure the goodness of a threshold split based on receiver operating characteristic curves (as described in Section~\ref{sec:roc}).
The \textsf{SPR} software package~\cite{2005statpattern} provides several options for optimization criterion.
We found that the Gini index~\cite{1912gini} and the negative cross-entropy provided comparable, suitable performance for both searches. Thus we arbitrarily chose the Gini index for the IMR search and the negative cross-entropy for the ringdown-only search. Additional discussion is given in Sect.~\ref{sec:criterion}.
The Gini index is defined by
\begin{equation}
G(p)=-2p\bar{p},
\end{equation}
where $p$ is the fraction of events in a node that are signals and $\bar{p}=1-p$ is the fraction of events in the node that are background. Splits are made only if they will minimize the Gini index.
The negative cross-entropy function is defined by
\begin{equation}
H(p) = -p \log_2 p - \bar{p} \log_2 \bar{p}.
\end{equation}
As $H$ is symmetric against the exchange of $p$ and $\bar{p}$, if a node contains as many signal events as background, then $H\left(\frac{1}{2}\right)=1$.
A perfectly sorted node has $H(1)=H(0)=0$.
By minimizing the negative cross-entropy of our nodes, we find the optimal sort.
\par

When no split on a node can reduce the entropy or it contains fewer events than a preset limit, it is no longer divided.
At this point, the node becomes a ``leaf''.
The number of events in the preset limit is known as the ``minimum leaf size''.
When all nodes have become leaves, we have created a decision tree.
This process is repeated to create each decision tree in the forest.
Each individual tree is trained on a bootstrap replica, or a resampled set of the original training set, so that each tree will have a different set of training events.
Furthermore, a different randomly-chosen subset of feature vector parameters is chosen at each node to attempt the next split.
Thus each decision tree in the forest is unique.
Results from each tree can be averaged to reduce the variance in the statistical classification.
This is the method of bootstrap aggregation or ``bagging''.
\par

The forest can then be used to classify an event of unknown class.
The event is placed in the initial node of each tree and is passed along the various trees' branches until it arrives at a leaf on each tree.
To compute the ranking statistic for an event from a forest of decision trees, we find its final leaf in all trees.
The probability that an event is a signal is given by the fraction of signal events in all leaves,
\begin{equation}\label{eq:forest_prob}
p_\mathrm{forest}=\frac{\sum s_i}{\sum s_i + b_i} = \frac{1}{N} \sum s_i
\end{equation}
where $s_i$ and $b_i$ are the number of signal and background events in the $i$th leaf and $N$ is the total number of signal and background events in all final leaves.
The final ranking statistic, $M_\mathrm{forest}$, for a forest of decision trees is given by the ratio of the probability that the event is a signal to the probability that the event is background,
\begin{equation}\label{eq:forest_rank}
M_\mathrm{forest}=\frac{p_\mathrm{forest}}{1-p_\mathrm{forest}}.
\end{equation}
This ranking statistic is a probability ratio, indicating how much the signal model is favored over the background model.
\par

In order to test the performance of the random forest, we determine how well it sorts events that we know \emph{a priori} are signal or background.
Rather than generate a new training set of simulated injections and time-shifted data, we may sort the training set used in construction of the forest.
To prevent over-estimation of classifier performance, decision trees cannot be used to classify the same events on which they were trained.
Thus we use a round-robin approach to iteratively classify events using a random forest trained on a set excluding those events.
We construct ten random forests each using 90\% of the training events such that the remaining 10\% of events may be safely classified.
In this way we can efficiently verify the performance of the random forest using only the original training events.
\par

Each forest is trained on and used to evaluate a particular type of double coincidence from the detector network (i.e., H1L1, H1V1, L1V1), as each pair of detectors produces unique statistics.
Triple coincidences are split into their respective doubles, as there is not sufficient triple-coincident background to train a separate forest.
For a triple coincidence with triggers from detectors $a$, $b$, and $c$, the ranking statistic $M_\mathrm{forest, triple}$ will be the product of $M_\mathrm{forest, double}$ for each pair of triggers,
\begin{equation}\label{eq:triple_forest}
M_\mathrm{forest, triple}=\prod_i M_\mathrm{forest, i}
\end{equation}
where $i \in$ \{$ab$, $ac$, $bc$\} denotes the possible pairs of double coincident triggers in a triple coincidence.
\par

\section{Tuning}\label{sec:tuning}
There are several issues to consider when optimizing the performance of RFBDT classifiers.
Performance of the algorithm is dependent on the quality of the training set (i.e., how well the training data represent the actual population we wish to detect).
Additionally, we must select appropriate statistics to include in the feature vector of each coincidence.
Finally, RFBDT classifiers have several parameters that must be tuned to optimize the sorting performance.
These include number of trees, number of sampled parameters from the feature vector at each node, and minimum leaf size.
Improperly choosing these meta parameters will lead to a poorly trained classifier.
\par

In general there are two types of mistrained classifiers.
An over-trained classifier separates the training data well, but the sort is specific to those data.
An over-trained classifier may provide very high or very low $M_\mathrm{forest}$, but these values contain a large systematic bias.
Any events that were not well represented by this training set will be misclassified.
An under-trained classifier has not gleaned enough information from the training data to sort properly.
In this case the classifier is unsure of which set any event belongs, assigning intermediate values of $M_\mathrm{forest}$ to all.
\par

\begin{table}[H]
\caption{Summary of random forest parameters.}
\label{tab:summaryrf}
\begin{ruledtabular}
\begin{tabular}{lcc}
Search & IMR & Ringdown-only \\
\hline
Number of trees	& 100 &	2000	\\
Minimum leaf size & 5	& 65 \\
Total number of parameters & 15	& 24	\\
Number of randomly sampled \\ \-\ \-\ \-\ parameters per node	& 6 &	14	\\
Criterion for optimization	& Gini- & cross- \\
	&  index & entropy \\
\end{tabular}
\end{ruledtabular}
\end{table}

\subsection{Figure of merit}\label{sec:roc}
We evaluate the performance of different tunings using receiver operating characteristic (ROC) curves, separately for each search.
In general, these curves show detection rate as a function of contamination rate as a discriminant function is varied.
For our purposes, thresholds on the combined FAR serve as the varying discriminant function.
Thus, both the detection and contamination rates are functions of the combined FAR.

Since we seek to improve the sensitivity of our searches, we reject the traditional definition of detection rate and instead define a quantity that depends on sensitive volume.
We use a fractional volume computed at each combined FAR threshold,
\begin{equation}\label{eq:fracvol}
\frac{V}{V_\mathrm{tot}}=\frac{\sum_i \epsilon_i r_i^3}{\sum_i r_i^3}
\end{equation}
where $i$ sums over all injections recovered as coincidences by the analysis pipeline, and $r_i$ is the physical distance of the injection.
For each combined FAR threshold $\lambda^*$, $\epsilon_i$ counts whether injection $i$ was  found with a combined FAR $\lambda_i$ less than or equal to $\lambda^*$,
\begin{displaymath}
   \epsilon_i = \left\{
     \begin{array}{lr}
       1 & : \lambda_i \le \lambda^*\\
       0 & : \lambda_i > \lambda^*
     \end{array}.
   \right.
\end{displaymath}
\par

In the following sections, we explore tunings and performance for the RFBDT algorithm for different total masses and mass ratios as well as at Veto Level~2 and~3 in order to understand how the application of vetoes affects the RFBDTs.
\par


\subsection{Training set}\label{sec:training}
The training of the classifier utilizes the signal and background data sets as described in Sect.~\ref{sec:sigts}.
In the following discussions, we consider several issues that arise in the construction of training sets for gravitational-wave classification when using RFBDTs.
\par

\subsubsection{Signal training}\label{sec:sigtraining}
In order to train the classifier on the appearance of signal, we injected sets of simulated waveforms into the data and recorded those found in coincidence by the searches.
\par

Both searches injected sets of waveforms from the EOBNRv2 and IMRPhenomB families.
The total mass $M$, mass ratio $q$, and component spin $\hat{a}_{1,2}$ distributions of these waveforms are given in Table~\ref{tab:summaryimrts}.
We define the mass ratio as $q=m_>/m_<$ where $m_>=\mathrm{max}(m_1, m_2)$ and $m_<=\mathrm{min}(m_1, m_2)$.
The component spins are $\hat{a}_{1,2}=cS_{1,2}/Gm_{1,2}^2$ for the spin angular momenta $S_{1,2}$ and masses $m_{1,2}$ of the two binary components.
From this, we define the mass-weighted spin parameter
\begin{equation}
\chi_s=\frac{m_1\hat{a}_1+m_2\hat{a}_2}{m_1+m_2}.
\end{equation}
Additionally, for the ringdown-only search, we injected two sets of ringdown-only waveforms as described in Table~\ref{tab:summaryrdts}.
The two sets gave coverage of the ringdown template bank in $\left(f_0, Q\right)$-space and of the potential $\left(M_f, \hat{a}\right)$-space accessible to the ringdown-only search where $M_f$ is the final black hole mass and $\hat{a}$ is the dimensionless spin parameter.
All injections were given isotropically-distributed sky location and source orientation parameters.
As described below, only injections that are cleanly found by the search algorithm are used in training the classifiers.
\par

For performance investigations in Section~\ref{sec:results}, we determine search sensitivities using all injections found by the searches' matched filtering pipelines (i.e., not just those that are cleanly found).
The IMR search considers full coalescence waveforms from the EOBNRv2 and IMRPhenomB families.
The ringdown-only search considers only EOBNRv2 waveforms.
These injection sets and their parameters are given in Section~\ref{sec:results}.
\par

\begin{table*}[t]
\caption{Summary of full coalescence signal training set. These injections are parameterized by total binary mass $M$, mass ratio $q$, and mass-weighted spin parameter $\chi_s$.}\label{tab:summaryimrts}
\begin{ruledtabular}
\begin{tabular}{lcccc}
& \multicolumn{2}{c}{EOBNRv2} & \multicolumn{2}{c}{IMRPhenomB} \\
Search & IMR & Ringdown-only & IMR & Ringdown-only \\
\hline
Mass distribution:	& uniform in $\left(m_1,m_2\right)$ & uniform in $\left(M, q\right)$ & uniform in $\left(M, q\right)$ & uniform in $\left(M, q\right)$ \\
Total mass range: & $M/\mathrm{M}_\odot \in [25, 100]$ & $M/\mathrm{M}_\odot \in [50, 450]$ & $M/\mathrm{M}_\odot \in [25, 100]$ & $M/\mathrm{M}_\odot \in [50, 450]$\\
Mass ratio range: & $q \in [1, 10]$ & $q \in [1, 10]$ & $q \in [1, 10]$ & $q \in [1, 10]$\\
Spin parameter distribution:	& non-spinning & non-spinning & uniform in $\chi_s$ & uniform in $\chi_s$\\
Spin parameter range: & $\chi_s=0$ & $\chi_s=0$ & $\chi_s \in [-0.85, 0.85]$ & $\chi_s \in [0, 0.85]$ \\
\end{tabular}
\end{ruledtabular}
\end{table*}

\begin{table}[t]
\caption{Summary of ringdown-only signal training set. Injection set~1 corresponds to coverage of the ringdown template bank in $\left(f_0, Q\right)$-space. Injection set~2 corresponds roughly to the potential $\left(M_f, \hat{a}\right)$-space accessible to the ringdown-only search.}
\label{tab:summaryrdts}
\begin{ruledtabular}
\begin{tabular}{lcc}
& Injection set 1& Injection set 2 \\
\hline
Distribution: & uniform in $\left(f_0, Q\right)$ & uniform in $\left(M_f, \hat{a}\right)$\\
Parameter 1: & $f_0/\mathrm{Hz} \in [50, 2000]$ & $M_f/\mathrm{M}_\odot \in [50, 900]$\\
Parameter 2: & $Q \in [2, 20]$ & $\hat{a} \in [0, 0.99]$\\
\end{tabular}
\end{ruledtabular}
\end{table}

To identify triggers associated with simulated waveform injections made into the data, we use a small time window of width $\pm$1.0 second around the injection time.
We record the parameters of the trigger with the highest SNR within this time window and associate it with the injection.
Unfortunately, when injections are made into real data containing non-Gaussian noise, the injection may occur near a non-Gaussian feature or glitch in the data.
In the case where the SNR of the injection trigger is smaller than that of the glitch trigger, the recorded trigger will correspond to the glitch trigger and will not accurately represent the simulated waveform.
When using injections to train the classifier on the appearance of gravitational-wave signals, we must be careful to exclude any injections in a window contaminated by a glitch.
\par

\begin{figure*}[t]
    \centering
\subfigure[]{\label{fig:ifara}\includegraphics[width=.45\textwidth]{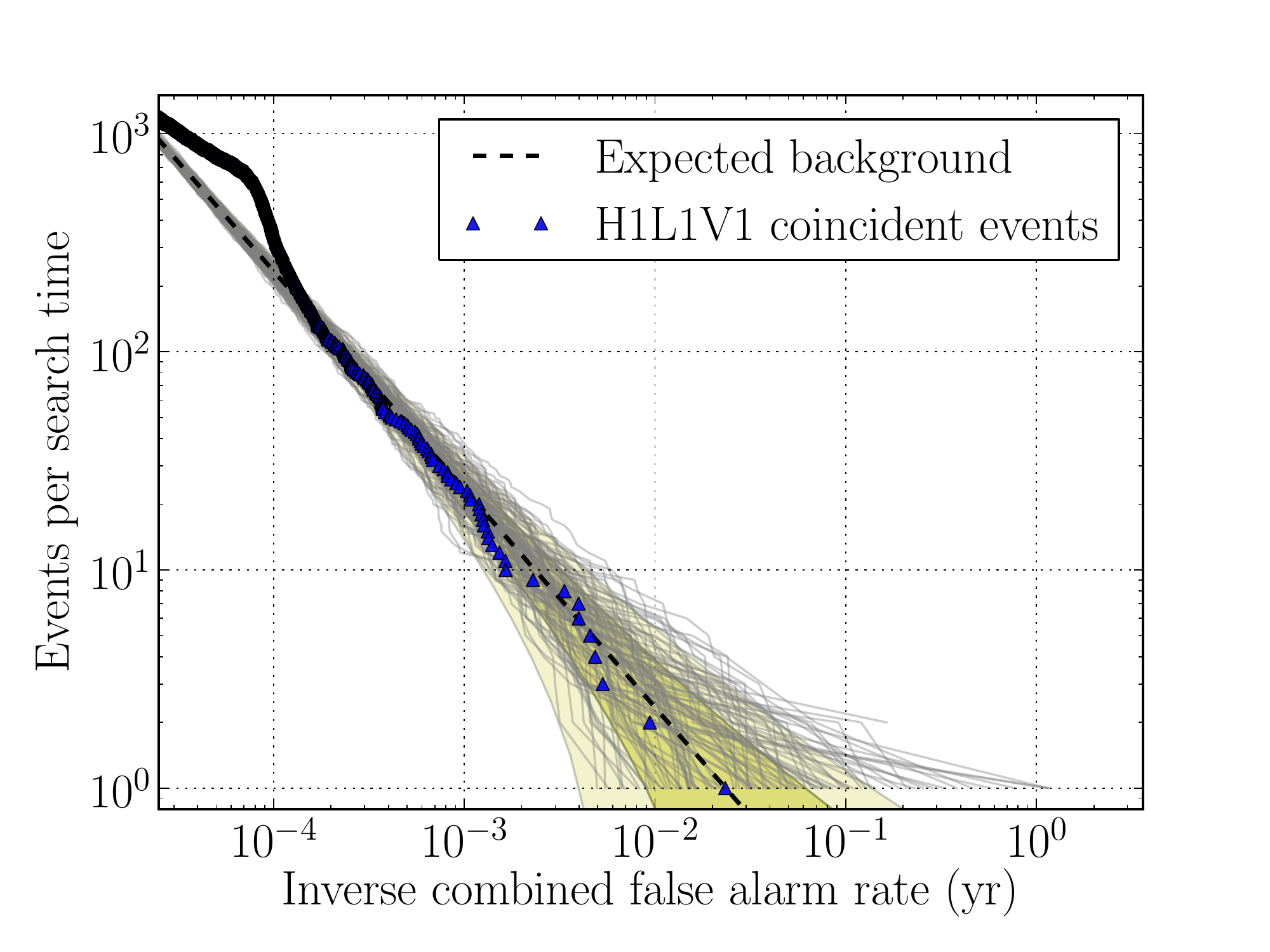}}
    ~
    \subfigure[]{\label{fig:ifarb}\includegraphics[width=.45\textwidth]{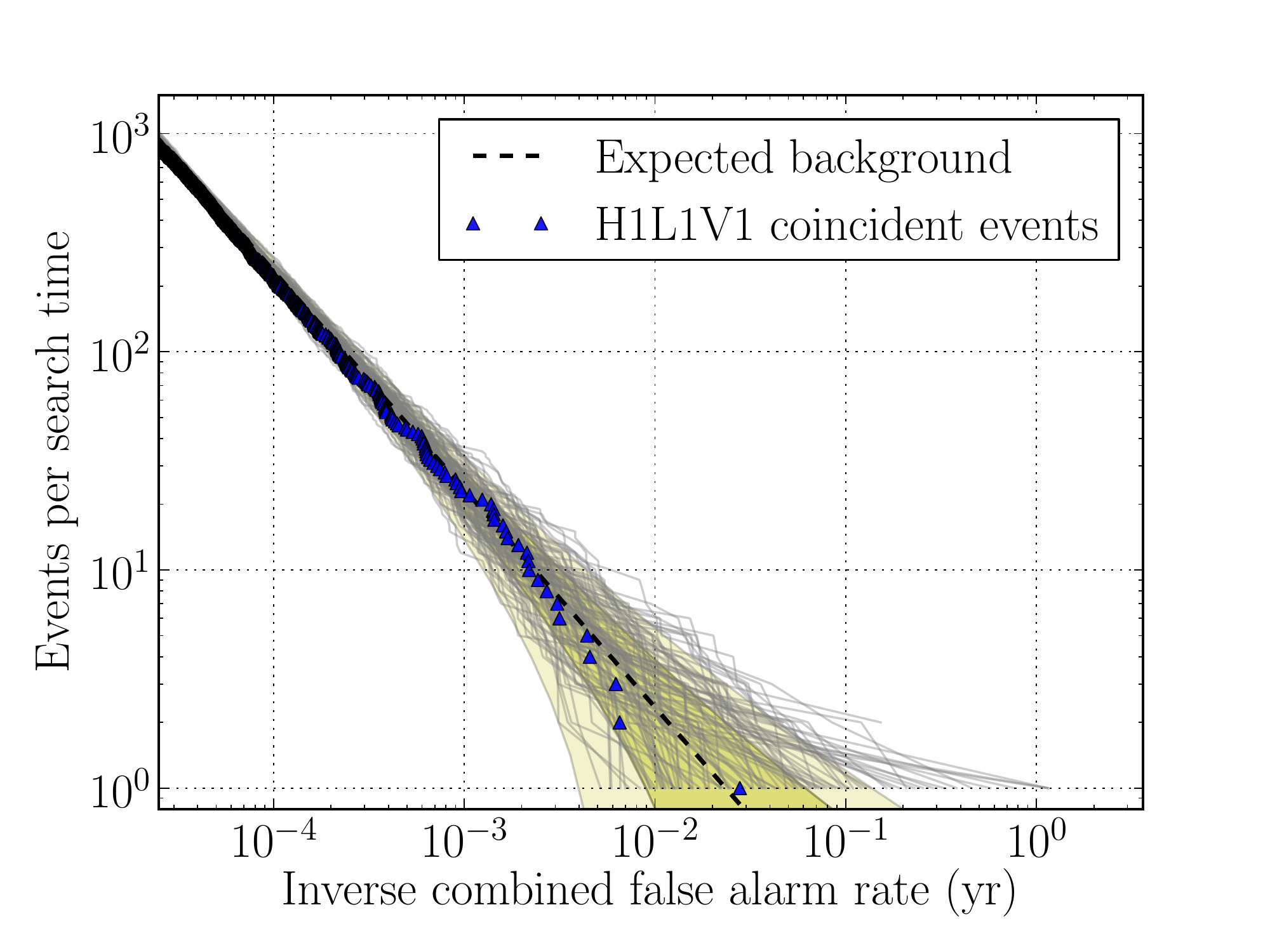}}
    \caption{Cumulative distributions of coincident events found as a function of inverse combined false alarm rate.
The data plotted here show results for a ringdown-only search over $\sim9$ days of H1L1V1 network data at Veto Level 3.
Blue triangles represent the coincident events found by the ringdown-only search~\cite{2014rdsearch}.
Grey lines plot the coincident events in each of the 100 time-shift experiments.
Yellow contours mark the 1$\sigma$ and 2$\sigma$ region of the expected background from accidental coincidences.
Figure~\ref{fig:ifara} shows results of the search obtained with a RFBDT classifier trained on a contaminated injection set.
Figure~\ref{fig:ifarb} shows results when the classifier is trained on a clean injection set.
A RFBDT classifier trained on a clean injection set properly ranks H1L1V1 coincidences with low significance so that there is not a ``bump" in the distribution at low combined FAR.}
    \label{fig:ifarbump}
  \end{figure*}

Figure~\ref{fig:ifarbump} demonstrates the issue that can arise when using a contaminated signal training set.
These plots show the cumulative distributions of coincident events found as a function of inverse combined false alarm rate for a small chunk of the H1L1V1 network search at Veto Level 3.
The ranking statistic used in both Fig.~\ref{fig:ifara} and Fig.~\ref{fig:ifarb} is $M_\mathrm{forest}$.
However, results for Fig.~\ref{fig:ifara} were obtained with a RFBDT classifier trained on injections identified using an injection-finding window of width $\pm$1.0 second (i.e., a contaminated injection set).
Results for Fig.~\ref{fig:ifarb} were obtained with a RFBDT classifier trained on injections identified using a narrower injection-finding window of width $\pm$0.01 seconds and after removing any injections made within $\pm$0.5 seconds of a glitch (i.e., a clean injection set).
In Fig.~\ref{fig:ifara}, we see that there is an excursion of H1L1V1 gravitational-wave candidate coincidences from the 2$\sigma$ region of the expected background at low values of inverse combined FAR.
This excursion for coincidences with low significance is caused by a population of injections that were misidentified because of a nearby glitch in the data.
The RFBDT classifier was taught that these glitches designated as injections should be classified as signal.
Thus, when similar glitches were found as coincidences in the H1L1V1 network search at Veto Level 3, they were given a boost in their $M_\mathrm{forest}$ rank.
However, in Fig.~\ref{fig:ifarb}, we see that by excluding these misidentified injections from the training set, the low significance H1L1V1 coincidences now fall within the 2$\sigma$ region of the expected background.
\par

We developed a software tool\footnote{{\tt https://ligo-vcs.phys.uwm.edu/cgit/lalsuite/tree/ pylal/bin/ligolw\_dbinjfind}} for use with the LALSuite gravitational-wave data analysis routines\footnote{{\tt https://www.lsc-group.phys.uwm.edu/daswg/projects/ lalsuite.html}} to construct clean injection sets.
 Using this tool, we investigated two time window parameters that can be tuned: the width of the injection-finding time window and the width of the injection-removal time window.

The injection-finding time window is motivated by the fact that a trigger due to an injection should be found in the data within a few milliseconds of the injection time given the light travel time between detectors.
Thus, in Gaussian detector noise, a few millisecond-wide injection-finding window should be sufficient.
However, due to non-Gaussian, non-stationary detector noise, the coincidence of triggers associated with an injection could be overshadowed if a loud glitch trigger is nearby. Thus, we allow a much larger window.
When conducting searches for gravitational waves, this window is typically set to $\pm$1.0 second from the injection time.
However, such a large window results in a contaminated signal training set as we see in Fig.~\ref{fig:ifarbump}.
\par

The injection-removal time window is motivated by the fact that a significant trigger found by the search before injections are performed is a potential contaminating trigger for any injection made similarly in time.
A simple time window is used to cross-check whether an injection trigger found by the search could be attributed to a trigger found in detector data before injections were performed.
\par

We investigated the performance of the RFBDT classifier for the ringdown-only search separately for detection of $q=1$ and $q=4$ EOBNRv2 simulated waveforms at Veto Level 3.
We performed several tuning runs, adjusting the size of the injection-finding and injection-removal windows.
We found that an injection-finding window of $\pm$0.01 seconds around an injection and an injection-removal window of  $\pm$0.5 seconds around an injection were the most effective at combating the excess of foreground triggers at low significance.
These settings were used in designing both the IMR and ringdown-only signal training sets.
\par

\subsubsection{Background training set}\label{sec:bkgdtraining}
The background training set composed of accidental coincidences is not noticeably contaminated by signal.
Since this background is constructed by time-shifting the data, it is possible that a real gravitational-wave signal could result in time-shifted triggers contaminating the background training set.
However, given the rare detection rates for gravitational waves in the detector data analyzed here, it is unlikely that such a contamination has occurred.
However, in the advanced detector era, when gravitational-wave detection is expected to be relatively common, this issue will need to be revisited.
\par

An additional issue to consider for the background training set concerns the size of the set.
In references~\cite{2013s6highmass, 2014rdsearch} and briefly in Section~\ref{sec:data}, we describe the procedure used to compute the upper limit on BBH and IMBH coalescence rates.
The typical procedure involves analyzing data in periods of $\sim$1 to 2 months.
For these investigations, we ran the RFBDT classifier for each $\sim$1 to 2 month chunk.
However, the size of the background training set for these $\sim$1 to 2 months analyses can be as small as 1\% of the total background training set available for the entire Period~2 analysis.
Thus, for the ringdown-only search, we took the additional step of examining the performance of the RFBDT classifier in the case where the monthly analyses used background training sets from their respective months and in the case where the monthly analyses used background training sets from the entire Period~2 analysis.
As we report in Section~\ref{sec:results}, using a background training set composed of time-shifted coincidences from the entire Period~2 analysis does not result in a clear sensitivity improvement.
\par

\subsection{Feature vector}\label{sec:featurevect}
A multivariate statistical classifier gives us the ability to use all available gravitational-wave data analysis statistics to calculate a combined FAR.
These may include single trigger statistics such as SNR and the $\chi^2$ signal consistency test~\cite{2005allen} as well as empirically-designed composite statistics that were previously used by each search as a classification statistic.
The classifier will inherit the distinguishing power of the composite statistics as well as any other information we provide from statistics that may not have been directly folded into the composite statistics.
These could include information that highlights inconsistencies in the single triggers' template parameters or alerts us to the presence of bad data quality.
The set of all statistics characterize the feature space and each coincidence identified by the search is described by a feature vector.
\par

As explained in Sect.~\ref{sec:rf}, a different subset of feature vector parameters are chosen at each node.
Selecting the optimal size of the subset can increase the randomness of the forest and reduce concerns of overfitting.
We discuss the tuning of this number in Sect.~\ref{sec:sampled}.
\par

Also, note that the RFBDT algorithm can only make plane cuts through the feature space.
It cannot reproduce a statistic that is composed of a non-linear combination of other statistics.
As we describe in more detail in Appendix~\ref{sec:appendix1} and~\ref{sec:appendix2}, if we know {\it a priori} a useful functional form for a non-linear composite statistic, we should include that statistic in the feature vector.
Such a statistic can only ever be approximated by the plane cuts.
Nevertheless, we design feature vectors with a large selection of statistics in the hope that some combination may be useful.
\par

Details of the parameters chosen to characterize coincidences with the RFBDT classifier for the IMR and ringdown-only searches are given in Appendices~\ref{sec:appendix1} and~\ref{sec:appendix2} and in~\cite{2014kari, 2013bakerthesis}.
The parameters are summarized here in Tables~\ref{tab:paramshm} and~\ref{tab:rdparams}.
These include information from the template parameters and SNR of each single detector trigger as well as composite statistics that combine information from these.
Additionally, a few parameters attempt to quantify the quality of the data.
\par

One of the data quality parameters for the ringdown-only search is a binary value used to indicate whether a trigger in a coincidence occurred during a time interval flagged for noise transients.
The flagged intervals were defined using the hierarchical method for vetoing noise transients known as {\it hveto} as described in~\cite{2011hveto}.
The LIGO and Virgo gravitational-wave detectors have hundreds of auxiliary channels monitoring local environment and detector subsystems.
The {\it hveto} algorithm identifies auxiliary channels that exhibit a significant correlation with transient noise present in the gravitational-wave channel and that have a negligible sensitivity to gravitational-waves.
If a trigger in the gravitational-wave channel is found to have a statistical relationship with auxiliary channel glitches, a flagged time interval is defined.
In Sect.~\ref{sec:rdresults}, we explore the performance of the RFBDT classifier before and after the addition of the {\it hveto} parameter to the feature vector.
This investigation was done to explore the ability of the classifier to incorporate data quality information.
\par

Significant work has been done to identify glitches in the data using multivariate statistical classifiers~\cite{2013mvscglitch} and Bayesian inference~\cite{2014bayeswave}.
With more development, this work could be used to provide information to a multivariate classifier used to identify gravitational waves, allowing for powerful background identification and potentially significant improvement to the sensitivity of the search.
\par

\begin{table*}[t]
\caption{Feature vector parameters for the IMR search's RFBDT classifier. The quantities indexed by $i$ are included for both detectors $a$ and $b$.}\label{tab:paramshm}
\begin{ruledtabular}
\renewcommand{\arraystretch}{1.8}
\begin{tabular}{lcl}
 Quantity & Definition & Description \\
\colrule
$\rho_i$	&	$\frac{|\left< x_i, h_i\right>|}{\sqrt{\left< h_i, h_i\right>}}$	& \multicolumn{1}{m{10.5cm}}{Signal-to-noise ratio of trigger found in detector $i$; found by filtering data $x_i(t)$ with template $h_i\left(t,\,\mathcal{M},\,\eta\right)$; described in Ref.~\cite{findchirp} }\\
$\chi^2_i$ & 	$10\left[\sum\limits_{j=0}^{10} (\rho_{j}-\rho_i/10)^2\right]$ & \multicolumn{1}{m{10.5cm}}{Quantity measuring how well the data matches the template in $j$ frequency bins for detector $i$; derived in Ref.~\cite{2005allen} }\\
$\rho_{\text{eff},i}$	&	$\frac{\rho_i}{\left[\frac{\chi^2_i}{10}\left(1+\rho_i^2/50\right)\right]^{1/4}}$	& \multicolumn{1}{m{10.5cm}}{Effective signal-to-noise ratio of the trigger found in detector $i$; used as a ranking statistic in Ref.~\cite{S5hm} }\\
$r^2_i$ veto duration & $\left\{t:\frac{\chi^2_i(t)}{10+\rho_i(t)^2}>10\right\}$ &   \multicolumn{1}{m{10.5cm}}{Duration $t$ that a weighted-$\chi^2$ time-series in detector $i$ goes above a threshold of 10; described in Ref.~\cite{2007andythesis}} \\
$\chi^2_{\text{continuous},i}$ & $\sum \left| \left<x_i(t)x_i(t+\tau)\right>-\rho_i\right| ^2$ & \multicolumn{1}{m{10.5cm}}{Sum of squares of the residual of the SNR time series and the autocorrelation time series of a single detector trigger from detector $i$; described in Ref.~\cite{2008chadthesis}} \\
$\rho_{\text{\text{high,combined}}}$ & $\sqrt{\sum\limits_i^N \rho^2_{\text{high},i}}$ &  \multicolumn{1}{m{10.5cm}}{Combined IMR search's re-weighed signal-to-noise ratio used as a ranking statistic in Ref.~\cite{2013s6highmass} where $i$ sums over $N$ triggers found in coincidence} \\
d$t$		&	$\left|t_a-t_b\right|$						&	 \multicolumn{1}{m{10.5cm}}{Absolute value of time difference between triggers in detectors $a$ and $b$}	\\
d$\mathcal{M}_\mathrm{rel}$		&	 $\frac{\left|\mathcal{M}_a-\mathcal{M}_b\right|}{\mathcal{M}_{\text{average}}}$						&	 \multicolumn{1}{m{10.5cm}}{Absolute value of the relative difference in the chirp mass of the templates matched to the data in detectors $a$ and $b$}	\\
d$\mathcal{\eta}_\mathrm{rel}$		&	 $\frac{\left|\eta_a-\eta_b\right|}{\eta_\text{average}}$						&	 \multicolumn{1}{m{10.5cm}}{Absolute value of the symmetric mass ratio of the templates matched to the data in detectors $a$ and $b$}	\\
{\it e-thinca} & $E$ & \multicolumn{1}{m{10.5cm}}{Value of the ellipsoidal coincidence test, which measures the distance of the two matched templates in time-mass parameter space; derived in Ref.~\cite{ethinca}} \\
\end{tabular}
\end{ruledtabular}
\end{table*}

\begin{table*}[htbp]
\caption{Feature vector parameters for the ringdown-only search's RFBDT classifier. The quantities indexed by $i$ are included for both detectors $a$ and $b$.}\label{tab:rdparams}
\begin{ruledtabular}
\renewcommand{\arraystretch}{1.8}
\begin{tabular}{lcl}
 Quantity & Definition & Description \\
\colrule
$\rho_i$	&	$\frac{|\left< x_i, h_i\right>|}{\sqrt{\left< h_i, h_i\right>}}$	& \multicolumn{1}{m{10.5cm}}{Signal-to-noise ratio of trigger found in detector $i$; found by filtering data $x_i(t)$ with template $h\left(t,\,f_i,\,Q_i\right)$}\\
d$t$		&	$\left|t_a-t_b\right|$						&	 \multicolumn{1}{m{10.5cm}}{Absolute value of time difference between triggers in detectors $a$ and $b$}	\\
d$f$		&	 $\left|f_a-f_b\right|$						&	 \multicolumn{1}{m{10.5cm}}{Absolute value of template frequency difference between triggers in detectors $a$ and $b$}	\\
d$Q$	&	$\left|Q_a-Q_b\right|$					&	  \multicolumn{1}{m{10.5cm}}{Absolute value of template quality factor difference between triggers in detectors $a$ and $b$}\\
d$s^2$	&	 $g_{ij}$d$p^i$d$p^j$			&	 \multicolumn{1}{m{10.5cm}}{Three-dimensional metric distance between two triggers in $\left( f_0,\,Q,\,t\right)$-space for $p\in\left( f_0,\,Q,\,t\right)$; outlined in Ref.~\cite{2014rdsearch}}\\
$g_{tt}$	&	$\pi^2 f_0^2\frac{1+4Q^2}{Q^2}$				& Metric coefficient  in $(t,\,t)$-space\\
$g_{f_0f_0}$	&	$\frac{1+6Q^2+16Q^4}{4f_0^2\left( 1+2Q^2\right)}$	&  Metric coefficient  in $(f_0,\,f_0)$-space\\
$g_{QQ}$	&	$\frac{1+28Q^4+128Q^6+64Q^8}{4Q^2\left( 1+6Q^2+8Q^4\right)}$	&  Metric coefficient  in $(Q,\,Q)$-space\\
$g_{tf_0}$	&	$2\pi Q \frac{1+4Q^2}{1+2Q^2}$	&  Metric coefficient  in $(t,\,f_0)$-space\\
$g_{tQ}$	&	$2\pi f_0 \frac{1-2Q^2}{\left(1+2Q^2\right)^2}$	&  Metric coefficient  in $(t,\,Q)$-space\\
$g_{f_0Q}$	&	$\frac{1+2Q^2+8Q^4}{2f_0Q\left( 1+2Q^2\right)^2}$	&  Metric coefficient  in $(f_0,\,Q)$-space\\
$\xi$	&	max$\left( \frac{\rho_a}{\rho_b},\, \frac{\rho_b}{\rho_a} \right)$	& Maximum of the ratio of signal-to-ratios for triggers $a$ to $b$ or $b$ to $a$\\
${\rho_N}^2$	&	$\sum_{i}^N {\rho_i}^2$	& Combined network signal-to-noise ratio for $N$ triggers found in coincidence \\
$\rho_\mathrm{S4}$	&	$\rho_\mathrm{S4,triple},\, \rho_\mathrm{S4,double}$	& Detection statistic used in Ref.~\cite{2009s4ringdown}; outlined in Eq.~\eqref{eq:ntwksnr} and~(\ref{eq:s4doub})\\
$\rho_\mathrm{S5/S6}$	&	$\rho_\mathrm{S5/S6,triple},\, \rho_\mathrm{S4,double}$	& Detection statistic described in Ref.~\cite{2013talukder}; outlined in Eq.~\eqref{eq:s5s6trip} and~(\ref{eq:s4doub})\\
$D_i$	&	$\frac{\sigma_i}{\rho_i}\left( 1\,\mathrm{Mpc}\right)$	& \multicolumn{1}{m{10.5cm}}{Effective distance of trigger found with signal-to-noise ratio $\rho_i$ in detector $i$ that has a sensitivity $\sigma_i$ to a signal at $1\,$Mpc}\\
d$D$	&	$\left|D_a-D_b\right|$	&  \multicolumn{1}{m{10.5cm}}{Absolute value of effective distance difference between triggers in detectors $a$ and $b$}\\
$\kappa$	&	max$\left( \frac{D_a}{D_b},\, \frac{D_b}{D_a} \right)$	&Maximum of the ratio of effective distances for triggers $a$ to $b$ or $b$ to $a$ \\
$n_i$	& $n_i\left(\left| t \right| < 0.5\,\mathrm{ms}\right)$ & \multicolumn{1}{m{10.5cm}}{Count of the number of triggers in detector $i$ clustered over a time interval of $0.5\,$ms using the SNR peak-finding algorithm in Ref.~\cite{2008gogginthesis}}\\
hveto$_i$	& $\left\{ \begin{array}{lr}
       1 & : \mathrm{{\it hveto}\,\,flag\,\,on} \\
       0 & : \mathrm{{\it hveto}\,\,flag\,\,off}
     \end{array}
   \right.$& \multicolumn{1}{m{10.5cm}}{Binary value used to indicate whether a trigger in detector $i$ occurred during a {\it hveto} time interval~\cite{2011hveto} flagged for noise transients}\\
\end{tabular}
\end{ruledtabular}
\end{table*}

\subsection{Random forest parameters}

A summary of the tunable parameters selected for the RFBDT algorithm for each search is given in Table~\ref{tab:summaryrf}.

\subsubsection{Number of trees}\label{sec:trees}
We can adjust the number of trees in our forest to provide a more stable $M_\mathrm{forest}$ statistic.
Increasing the number of trees results in an increased number of training events folded into the $M_\mathrm{forest}$ statistic calculation.
However, the training data contains a finite amount of information and adding a large number of additional trees will ultimately reproduce results found in earlier trees.
Furthermore, adding more trees will increase the computational cost of training linearly.
\par

\begin{figure}
    \centering
    \label{fig:b}\includegraphics[width=.45\textwidth]{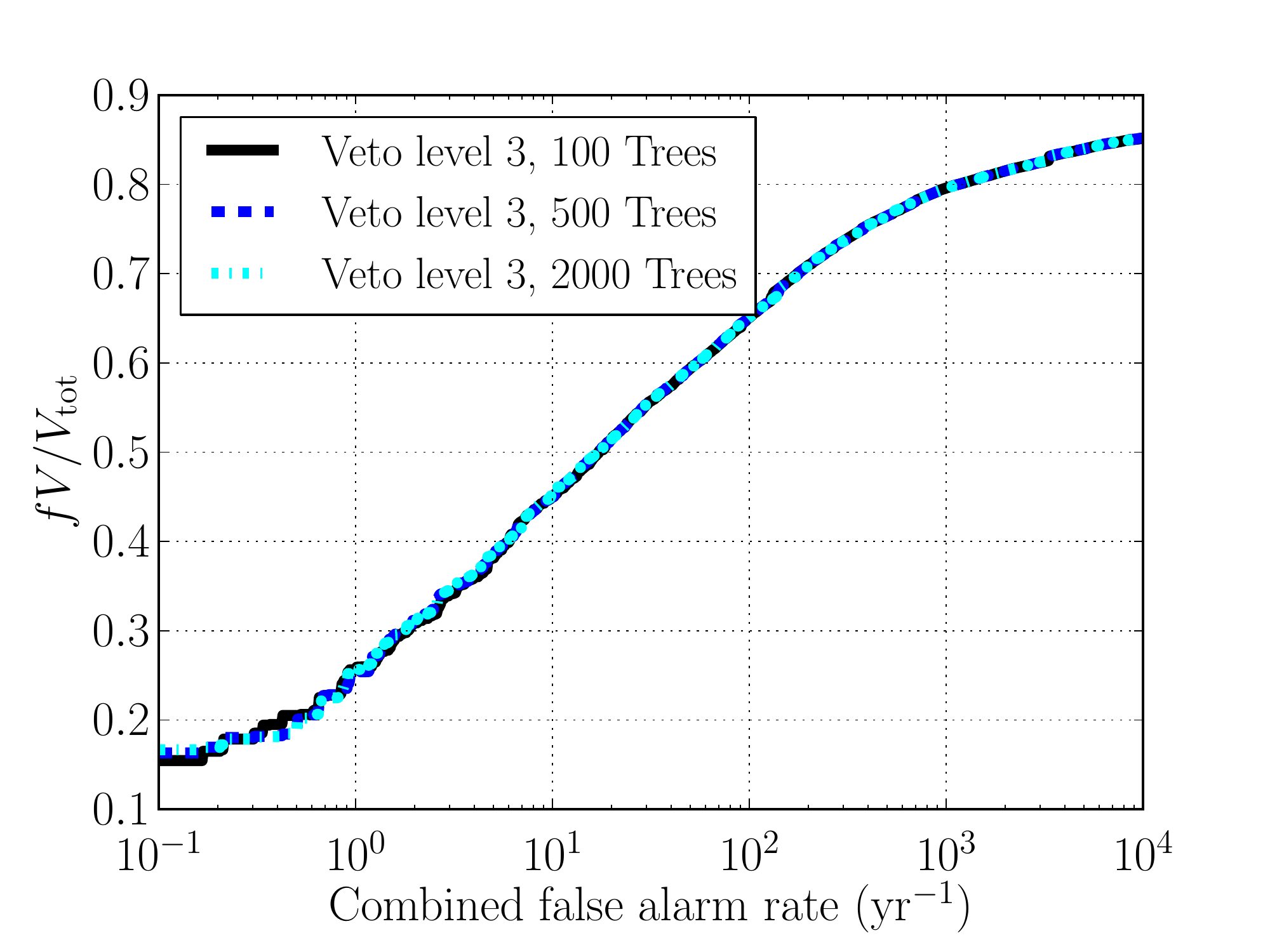}
    \caption{Investigation of the effect of using a different number of trees on the recovery of $q=1$ EOBNRv2 simulated waveforms at Veto Level~3.
        In general, we find that the use of more than 100 trees gives roughly the same sensitivity regardless of mass ratio or veto level.
        In this ROC, to adjust for the loss in analysis time in moving from Veto Level~2 to Veto Level~3, we scale the volume fraction in Eq.~\eqref{eq:fracvol} by the ratio of analysis times $f=t_\mathrm{VL3}/t_\mathrm{VL2}$.
        From the analysis times reported in Section~\ref{sec:data}, we find $f=0.88$.
      }
    \label{fig:tree1}
  \end{figure}

In Fig.~\ref{fig:tree1}, we investigate the effect of using a different number of trees for the ringdown-only search on the recovery of $q=1$ EOBNRv2 waveforms at Veto Level 3.
We find no significant improvement for using more than 100 trees.
Similar results were obtained at Veto Level~2 and for the recovery of $q=4$ EOBNRv2 waveforms.
The IMR search trained classifiers with 100 trees in each forest.
Initially, for the ringdown-only search, we selected to use 2000 trees in order to offset possible loss in sensitivity due to needing a larger leaf size as described in Sect.~\ref{sec:leaf}.
However, we ultimately found that this did not change the sensitivity.
Since computational costs were not high, we left the forest size as 2000 trees for the ringdown-only search.

\subsubsection{Minimum leaf size}\label{sec:leaf}
The minimum leaf size defines the stopping point for the splitting of nodes.
We define the minimum number of events allowed in a node before it becomes a leaf.
When all nodes become leaves, the recursive splitting of the tree stops.
\par

The choice of leaf size is affected by how representative the training data are of actual data and by how many coincident events are in the training data.
If the leaf size is too small, the forest will be over-trained.
In this case the sort is specific to the training data and may be systematically wrong for anything else.
If the leaf size is too large, the forest will be under-trained.
The individual decision trees did not have enough chances to make good cuts on the training data.
We will be left unsure if any coincident event is signal or background.
\par

We are limited by the size of the background training set of time-shifted data.
In each monthly analysis, the size of the background training set varied between thousands to hundreds of thousands of coincident events depending on veto level and analyzed networks.
A leaf size of 5 worked very well for the IMR search's trees, but investigations on the ringdown-only search with a leaf size of 5 showed that such a small choice led to an over-trained forest.
Some signal \emph{and} background coincidences were given an infinite $M_\mathrm{forest}$ rank (i.e., the classifier was 100\% sure that the coincidence was signal).
By exploring leaf sizes around 0.1-1\% the size of the varied background training sets, we found that a leaf size of 65 eliminated the over-training and also gave good performance for the ringdown-only search.
\par

\subsubsection{Number of sampled parameters}\label{sec:sampled}
At each node, we choose a random subset of parameters to use for splitting.
Out of $N_v$ total feature vector parameters, we select $m$ randomly and evaluate the split criteria for each.
Thus, a different set of $m$ parameters is available for picking the optimal parameter and its threshold for each branching point.
\par

If $m$ is too large, each node will have the same parameters available to make the splits.
This can lead to the same few parameters being used over and over again, and the forest will not fully explore the space of possible cuts.
Furthermore, because individual trees will be making cuts based on the same parameters, all of the trees in the forest will be very similar.
This is an example of over-training.
\par

If $m$ is too small, each node would have very few options to make the splits.
The classifier would be forced to use poor parameters at some splits, resulting in inefficient cuts.
The tree can run up against the leaf size limit before the training events were well-sorted.
This is an example of under-training.
The classification in this case would be highly dependent on the presence (or lack thereof) of poor parameters.
\par

A general rule-of-thumb for a good number of random sampled parameters is $\sim\sqrt{N_v}$~\cite{2014pyastro}.
For the IMR search, of the 15 parameters that make up the feature vector, we empirically found good performance for a selection of 6 randomly chosen parameters at each node.
For the ringdown-only search, 14 out of the total 24 feature vector parameters gave good performance.
\par

\subsubsection{Criterion for optimization}\label{sec:criterion}
The optimization criterion is used to select the best thresholds on parameters and proceeds the selection of random sampled parameters for each node.
The RFBDT algorithm provides several methods to determine the optimal parameter thresholds.
These are grouped by whether the output is composed of a discrete set or a continuous set of $M_\mathrm{forest}$ rankings.
While some of the discrete statistics performed well, we preferred to draw rankings from a continuous set.
Of the optimization criteria that gave continuous statistics, Gini index~\cite{1912gini} and negative cross-entropy (defined in Sect.~\ref{sec:rf}) gave good performance and were comparable to each other for both searches. Additionally, in order to obtain a good average separation between signal and background, the suggested optimization criteria are either Gini index or negative cross-entropy~\cite{2005statpattern}.
Thus, these two statistics were chosen for the IMR and ringdown-only searches, respectively. The choices were arbitrary in the sense that either optimization criteria would have been suitable for either search. Splits were only made if they minimized the Gini index or the negative cross-entropy.
\par

\section{Results}\label{sec:results}

\subsection{IMR search}\label{sec:hmresults}
In order to assess the sensitivity improvements of the IMR search to waveforms from BBH coalescing systems with non-spinning components, we use the same set of EOBNRv2 injections used to compute the upper limits on BBH coalescence rates in Sect.~VB of~\cite{2013s6highmass}.
These injections were distributed approximately uniformly over the component masses $m_1$ and $m_2$ within the ranges $1\le m_i/\mathrm{M}_\odot \le 99$ and $20\le M/\mathrm{M}_\odot \le 109$.
Additionally, we use the same set of IMRPhenomB injections used to make statements on sensitivity to spinning and non-spinning BBH coalescences in Sect.~VC of~\cite{2013s6highmass}.
We use a non-spinning set and a spinning set of IMRPhenomB injections, both uniformly distributed in total mass $25\le M/\mathrm{M}_\odot \le 100$ and uniformly distributed in $q/(q + 1) = m_1/M$ for a given $M$, between the limits $1\le q < 4$.
In addition, the spinning injections were assigned (anti-)aligned spin parameter $\chi_s$ uniformly distributed between -0.85 and 0.85.
\par

The previous IMR search over Period~2 data~\cite{2013s6highmass} used the combined signal-to-noise and $\chi^2$-based ranking statistic $\rho_{\text{\text{high,combined}}}$ for FAR calculations.
For more details on $\rho_{\text{\text{high,combined}}}$, see Table~\ref{tab:paramshm} and Appendix~\ref{sec:appendix1}.
Here, we report on a re-analysis that replaces $\rho_{\text{\text{high,combined}}}$ with the ranking statistic calculated by the RFBDT, $M_\mathrm{forest}$, as described in Sect.~\ref{sec:rf}.
Additionally, we have chosen a different FAR threshold for calculating sensitivity, rather than the loudest event statistic typically used in calculating upper limits in~\cite{2013s6highmass}.
The threshold that we use is the expected loudest FAR,
\begin{equation}
\breve{\mathrm{FAR}} = 1/T
\end{equation}
where $T$ is the total time of the analysis chunk being considered.
For a listing of $\breve{\mathrm{FAR}}$ for each analysis chunk and a comparison with the loudest event statistic, see Table~8.1 of~\cite{2014kari}.
\par
  
  Improvements in the following section are reported with uncertainties determined using the statistical uncertainty originating from the finite number of injections that we have performed in these investigations.
  
  \subsection{IMR search sensitive $VT$ improvements}\label{sec:hmsensvt}
  
\begin{figure*}
  \centering
   \subfigure[]{\label{fig:mtotvola}\includegraphics[width=.47\textwidth]{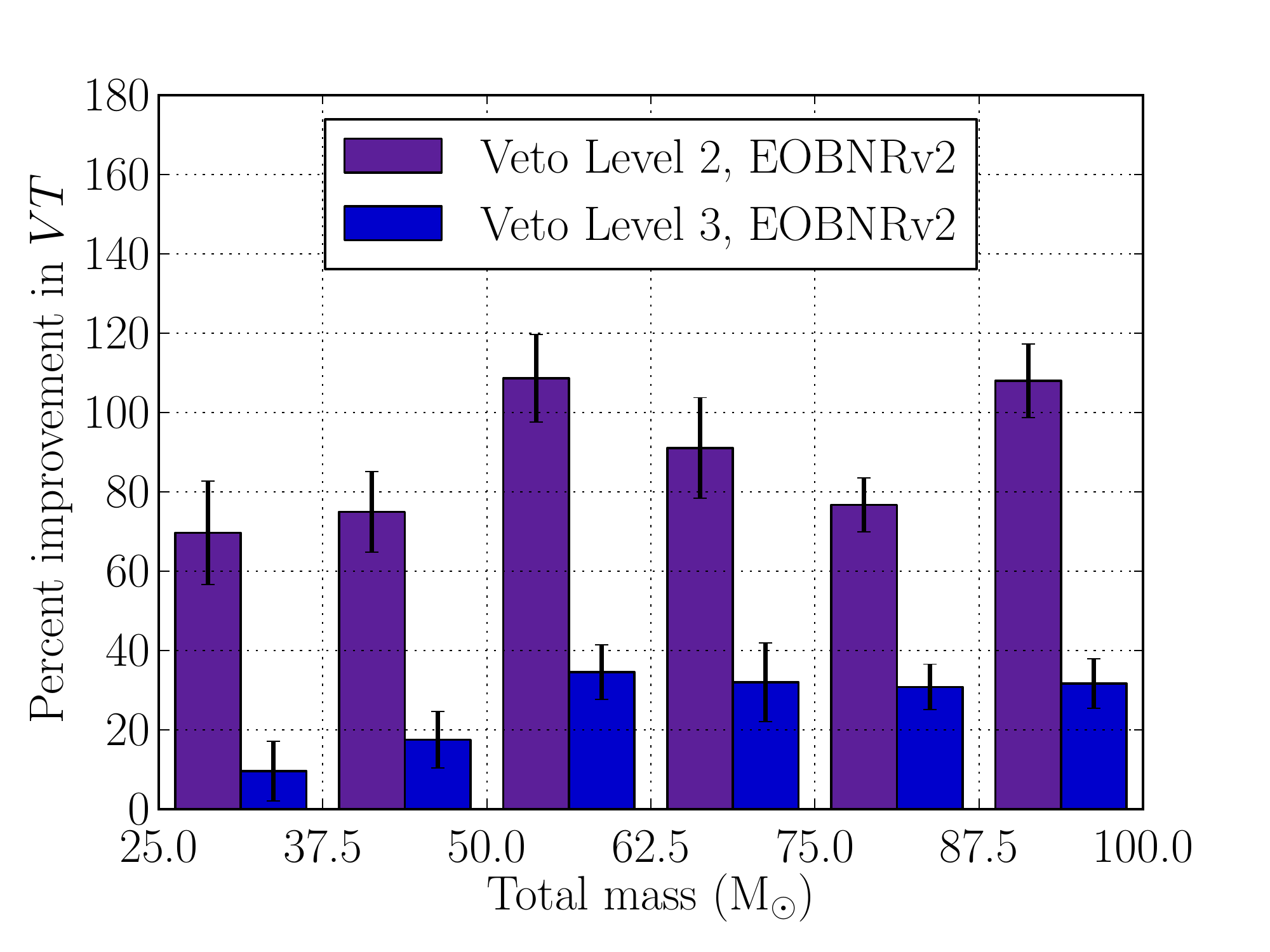}}
   \subfigure[]{\label{fig:mtotvolb}\includegraphics[width=.47\textwidth]{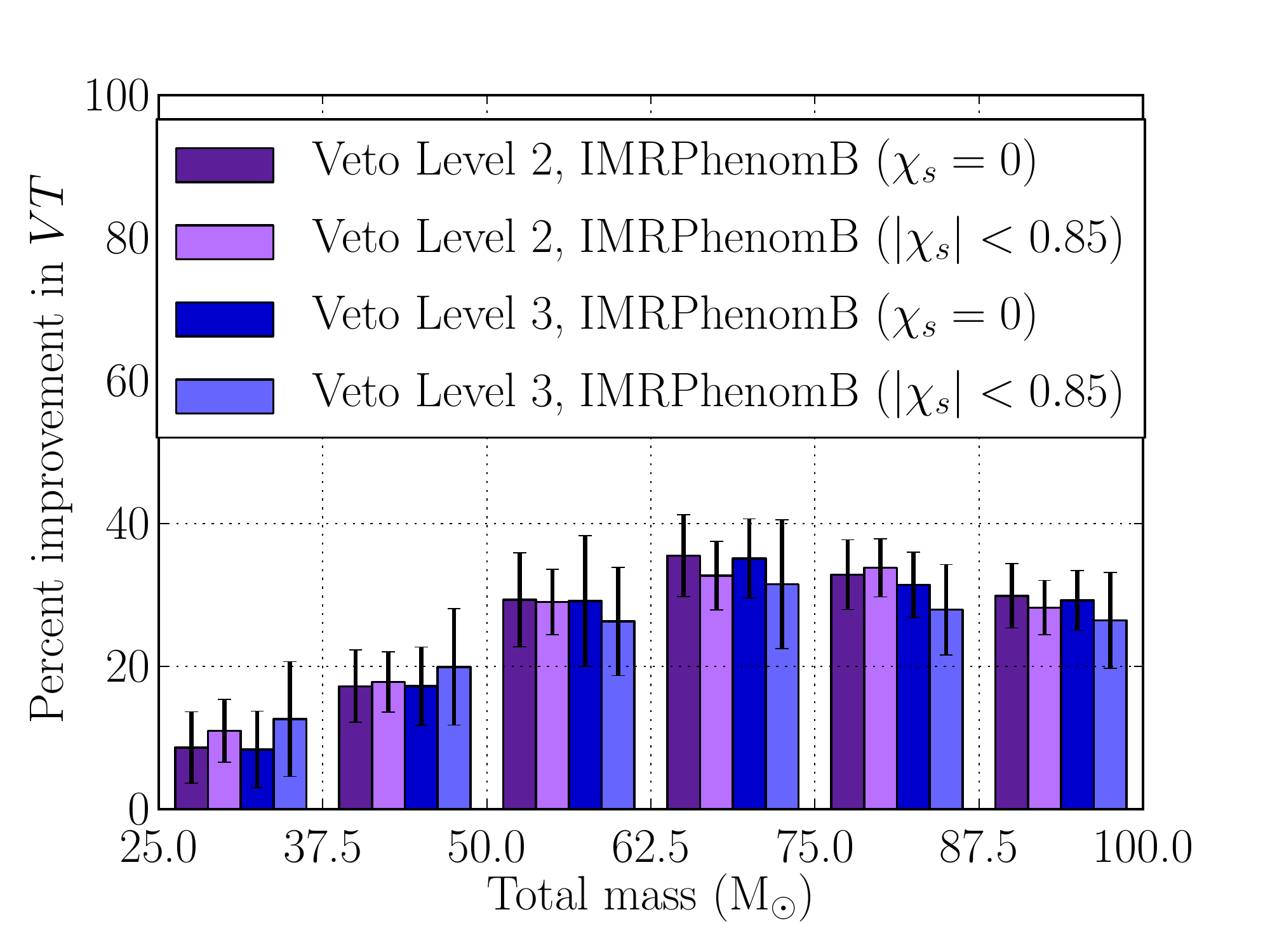}}
   \caption{Percent improvements (over the use of the $\rho_{\text{\text{high,combined}}}$~\cite{2013s6highmass} ranking statistic at Veto Levels~2 and~3) in sensitive volume multiplied by analysis time $(VT)$ for the recovery of EOBNRv2 ({\it left}) simulated waveforms at Veto Levels~2 and~3 and the recovery of IMRPhenomB ({\it right}) waveforms at Veto Levels~2 and~3 by the IMR search.
     Results for IMRPhenomB are shown for signals with spin parameter $\left( \left| \chi_s \right|<0.85 \right)$ and no spin $\left( \chi_s=0\right)$.
     The quantity $VT$ gives us a measure of the true sensitivity of the search and allows us to compare performances across veto levels.
     Results are shown for total binary masses from $25\le M/\mathrm{M}_\odot \le 100$ in mass bins of width $12.5\,\mathrm{M}_\odot$.
   }
   \label{fig:hmmtotalvol}
\end{figure*}

  Figure~\ref{fig:hmmtotalvol} demonstrates the percent improvements in sensitive volume multiplied by analysis time $(VT)$ when using the $M_\mathrm{forest}$ ranking statistic, rather than the $\rho_{\text{\text{high,combined}}}$ ranking statistic.
Results are shown at both Veto Levels~2 and~3 for total binary masses from $25\le M/\mathrm{M}_\odot \le 100$ in mass bins of width $12.5\,\mathrm{M}_\odot$.
Improvements for EOBNRv2 waveforms are shown in Fig.~\ref{fig:mtotvola} and for IMRPhenomB are shown in Fig.~\ref{fig:mtotvolb}.
The use of the $M_\mathrm{forest}$ ranking statistic gives improvements in $VT$ over the use of $\rho_{\text{\text{high,combined}}}$ at both Veto Levels~2 and~3.
The largest improvements are seen for total masses larger than $50\mathrm{M}_\odot$.
The IMR search is more sensitive in these higher mass regions.
Thus, larger improvement is found where the search is more sensitive.
\par

For EOBNRv2 waveforms, larger improvements are seen at  Veto Level~2 than at Veto Level~3.
At Veto Level~2, $VT$ improvements ranged from $70_{\pm 13}-109_{\pm 11}$\% for EOBNRv2 waveforms and from roughly $9_{\pm 5}-36_{\pm 6}$\% for IMRPhenomB waveforms.
At Veto Level~3, $VT$ improvements ranged from $10_{\pm 8}-35_{\pm 7}$\% for EOBNRv2 waveforms and remained roughly the same for IMRPhenomB waveforms.
More investigation is needed to understand why IMRPhenomB improvements are not as strong as EOBNRv2 improvements. One contributing factor could be component spin, which introduces several competing effects on the search including increased horizon distance with positive $\chi_s$, decreased sensitivity due to reduced overlap with EOBNRv1 templates, and higher signal-based $\chi^2$ test values~\cite{2013s6highmass}. It is currently unclear if any of these effects reduce the potential percent improvement seen with the $M_\mathrm{forest}$ ranking statistic.
\par

\begin{figure*}
    \centering
    \subfigure[]{\label{fig:m1m2vola}\includegraphics[width=.49\textwidth]{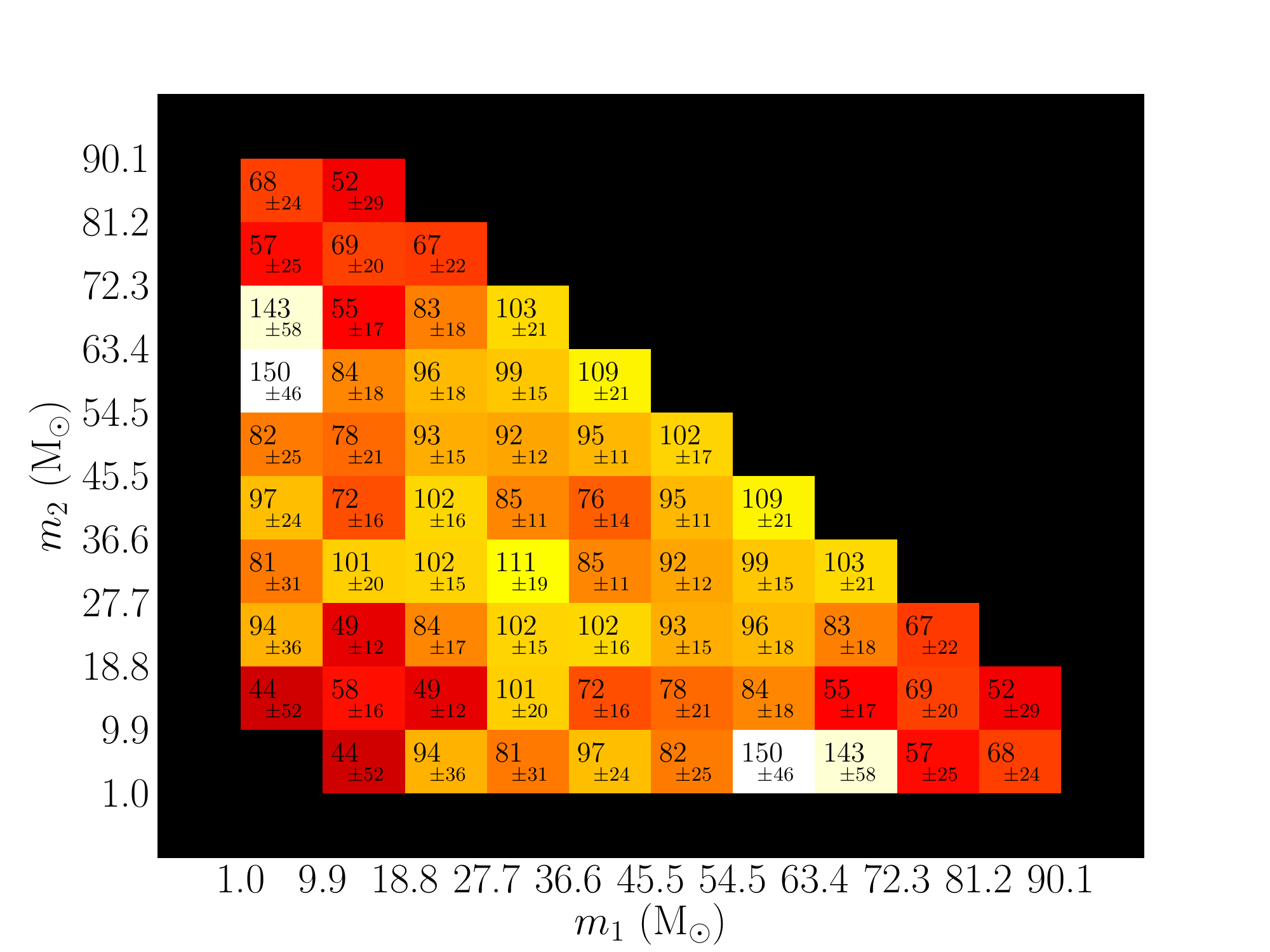}}
    \subfigure[]{\label{fig:m1m2volb}\includegraphics[width=.49\textwidth]{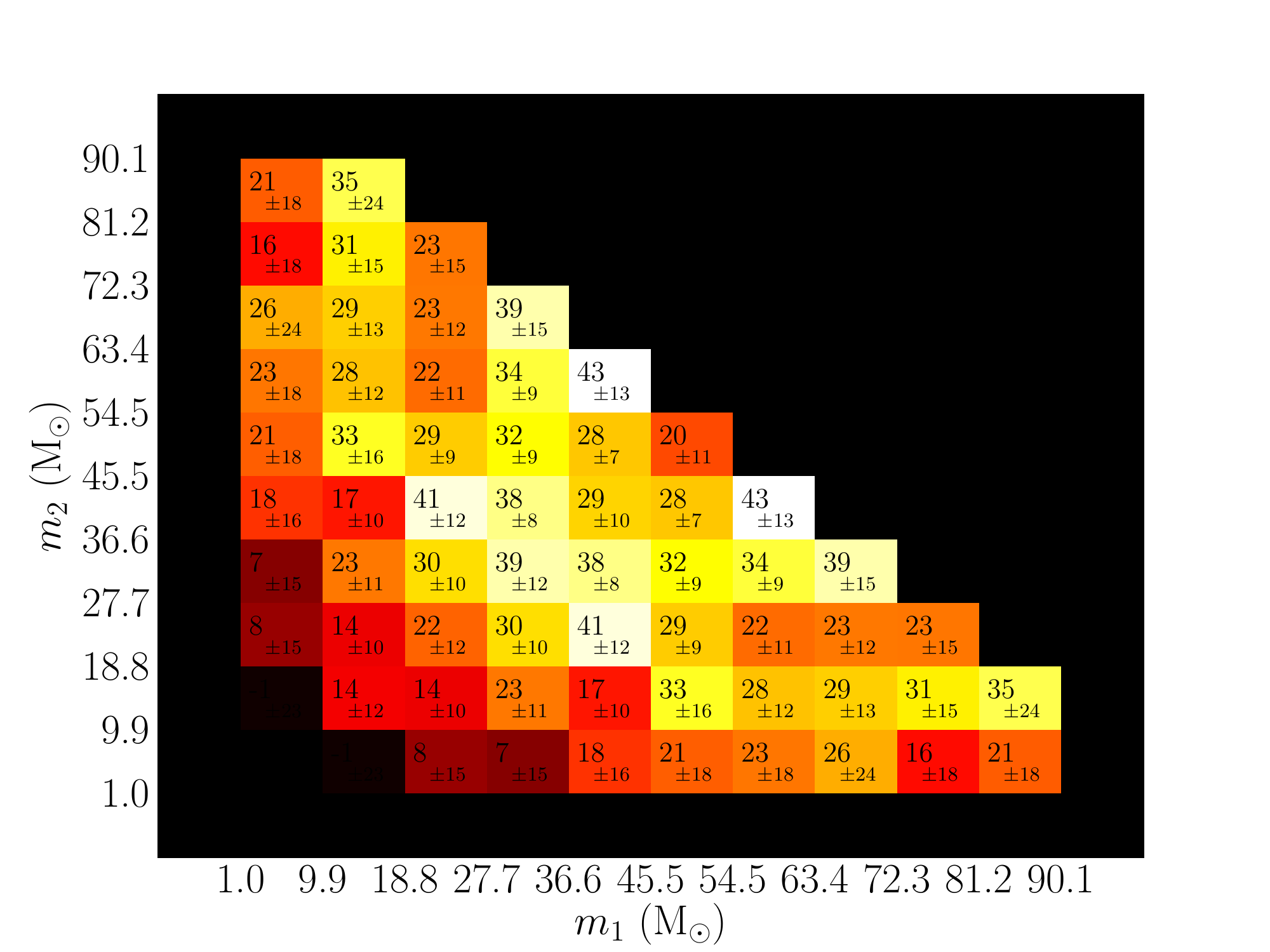}}
    \caption{Percent improvements  (over the use of the $\rho_{\text{\text{high,combined}}}$~\cite{2013s6highmass} ranking statistic at Veto Levels~2 and~3) in sensitive volume multiplied by analysis time $(VT)$ for the recovery of EOBNRv2 simulated waveforms at Veto Levels~2 ({\it left}) and~3 ({\it right}) by the IMR search.
      Percent improvement results are shown as a function of binary component masses.
      Note that the color scales of Figs.~\ref{fig:m1m2vola} and~\ref{fig:m1m2volb} are not equivalent.
    }
    \label{fig:hmm1m2vol}
  \end{figure*}

For more detail, Fig.~\ref{fig:hmm1m2vol} shows the percent improvements in $VT$ for EOBNRv2 waveforms as a function of component masses.
At Veto Level~2 in Fig.~\ref{fig:mtotvola}, we see that every mass bin sees a percent improvement in $VT$.
At Veto Level~3 in Fig.~\ref{fig:mtotvolb}, again we see that the improvements are smaller than at Veto Level~2.
In fact, no improvement is found for the lowest mass bin centered on $\left(5.5, 14.4\right)\,\mathrm{M}_\odot$.
\par
  
In Table~\ref{tab:hmvetoresults}, we explore the percent $VT$ improvements obtained with $M_\mathrm{forest}$ at different veto levels.
The improvements reported are made with respect to the sensitive volumes achieved with the $\rho_{\text{\text{high,combined}}}$ ranking statistic at Veto Level~2.
These values are presented as a means of comparing sensitivity between Veto Level~2 and Veto Level~3.
We see that $M_\mathrm{forest}$ at Veto Level~2 shows greater improvement and hence a more stringent upper limit than $M_\mathrm{forest}$ at Veto Level~3.
This is in contrast to the better performance of $\rho_{\text{\text{high,combined}}}$ at Veto Level~3 than at Veto Level~2.
For the standard IMR search with the $\rho_{\text{\text{high,combined}}}$ ranking statistic, the additional vetoing of poor quality data at Veto Level~3 was performed with the goal of preventing high SNR noise events from contaminating the list of gravitational-wave candidate events and reducing the sensitivity of the search.
However, for the random forest technique, those high SNR noise events are down-weighted in significance due to information contained in other parameters in the feature vector.
As a search at Veto Level~2 has more analysis time, it has the potential to have better sensitivity than a search at Veto Level~3.
In Table~\ref{tab:hmvetoresults}, we see that the use of the  $M_\mathrm{forest}$ ranking statistic for the IMR search has resulted in a better search sensitivity at Veto Level~2.
As we discuss in Sect.~\ref{sec:rdresults}, the ringdown-only search did not see the same behavior at Veto Level~2.
The information contained in the ringdown-only search's feature vector may not have had sufficient signal and background separation information to overcome the level of background contamination present at Veto Level~2 as compared to Veto Level~3.
\par  

\begin{table*}[t]
\caption{Percent $VT$ improvements over the use of the $\rho_{\text{\text{high,combined}}}$ ranking statistic in the IMR search at Veto Level~2 for EOBNRv2 waveforms. Note that these percent improvements should not be compared with values reported in Fig.~\ref{fig:hmmtotalvol} and~\ref{fig:hmm1m2vol} but are rather presented as a means of comparing sensitivity between Veto Level~2 and Veto Level~3. We see that $M_\mathrm{forest}$ at Veto Level~2 shows greater improvement and hence a more stringent upper limit than $M_\mathrm{forest}$ at Veto Level~3.}\label{tab:hmvetoresults}
\begin{ruledtabular}
\begin{tabular}{lccc}
 & $\rho_{\text{\text{high,combined}}}$ & $M_\mathrm{forest}$ & $M_\mathrm{forest}$ \\
Mass bin $\left( \mathrm{M}_\odot \right)$ & Veto Level 3 & Veto Level 2 & Veto Level 3 \\
\hline
 25.0 - 37.5 & 28 $\pm$ 8\% & 70 $\pm$ 13\% & 40 $\pm$ 9\%\\
37.5 - 50.0 & 36 $\pm$ 8\% & 75 $\pm$ 10\% & 60 $\pm$ 10\%\\
50.0 - 62.5 & 35 $\pm$ 7\% & 109 $\pm$ 11\% & 82 $\pm$ 10\%\\
62.5 - 75.0 & 33 $\pm$ 10\% & 91 $\pm$ 13\% & 76 $\pm$ 13\%\\
75.0 - 87.5 & 15 $\pm$ 5\% & 77 $\pm$ 7\% & 50 $\pm$ 6\%\\
87.5 - 100.0 & 41 $\pm$ 7\% & 108 $\pm$ 9\% & 86 $\pm$ 9\%\\
\end{tabular}
\end{ruledtabular}
\end{table*}
  
\subsection{Ringdown-only search}\label{sec:rdresults}
In order to assess the sensitivity improvements of the ringdown-only search to waveforms from binary IMBH coalescing systems with non-spinning components, we use the same set of EOBNRv2 injections used to compute the upper limits on IMBH coalescence rates in Section~V of~\cite{2014rdsearch}.
Due to the variation in ringdown-only search sensitivity over different mass ratios, we chose to explore sensitivity improvements separately for $q=1$ and $q=4$. This variation occurs because the total ringdown efficiency depends on symmetric mass ratio so that extreme mass ratio systems will not be detectable unless the system is sufficiently close~\cite{2014rdsearch}.
The injection sets were distributed uniformly over a total binary mass range from $50 \le M/\mathrm{M}_\odot \le 450$ and upper limits were computed in mass bins of width $50\,$M$_\odot$.
The final black hole spins of these injections can be determined from the mass ratios and zero initial component spins~\cite{2009finalspin}.
For $q=1$, we find $\hat{a}=0.69$, and for $q=4$, we find $\hat{a}=0.47$.
\par

Previous investigations of ranking statistics for the ringdown-only search~\cite{2012talukderthesis, 2012caudillthesis, 2013talukder} found that $\rho_\mathrm{S5/S6}$ provided better sensitivity than the $\rho_\mathrm{S4}$ ranking statistic used as a detection statistic in~\cite{2009s4ringdown}.
Thus, here we report on sensitivities based on combined FARs computed using $\rho_\mathrm{S5/S6}$ as a ranking statistic and using $M_\mathrm{forest}$ as a ranking statistic.
We follow the same loudest event statistic procedure used in~\cite{2014rdsearch} for calculating upper limits. Improvements in the following section are reported with uncertainties determined using the statistical uncertainty originating from the finite number of injections that we have performed in these investigations.
\par

Our complete investigations involve evaluating the performance of the RFBDT classifier for ringdown-only searches over Period~2 data using five separate ranking statistics, described below.
Additionally, we explore the improvement separately for recovery of $q=1$ and $q=4$ EOBNRv2 simulated waveforms as well as for Veto Level~2 and Veto Level~3 searches.
\par

The first ringdown-only search, to which we will compare performance, utilized the SNR-based statistic $\rho_\mathrm{S5/S6}$ to rank both double and triple coincident events.
Details of this ranking statistic are given in Appendix~\ref{sec:appendix2} and in~\cite{2012talukderthesis, 2012caudillthesis, 2013talukder}.
In each of the investigative runs that follow, this statistic becomes a parameter that is added to the feature vector of each coincident event.
A summary of the runs is given in Table~\ref{tab:summaryrdruns}.
\par

Run~1 uses a RFBDTs with 2000 trees, a leaf size of 65, and a random selection of 14 parameters out of the 24 total parameters listed in Sect.~\ref{sec:featurevect} except the {\it hveto} parameter.
The training set was composed of a clean signal set as outlined in Sect.~\ref{sec:sigtraining} and background set trained separately for each $\sim$1-2 month chunk of Period~2 as outlined in Sect.~\ref{sec:bkgdtraining}.
\par

Run~2 is identical to Run~1 except that the background training set of the RFBDTs is composed of all Period~2 background coincident events rather than each corresponding $\sim$1-2 month set of background coincident events.
We say that the RFBDTs is trained on the ``full background set."
\par

Run~3 is identical to Run~1 except that the {\it hveto} parameter is included in the feature vector of each coincident event.
This investigation was done to explore the ability of the RFBDT to incorporate data quality information.
\par

Run~4 combines the exceptions of Run~2 and Run~3.
Thus, this investigation includes a RFBDT classifier trained on ``full background set" and feature vectors that include the {\it hveto} parameter.
\par

\begin{table}[t]
\caption{Summary of ringdown-only search investigations}\label{tab:summaryrdruns}
\begin{ruledtabular}
\begin{tabular}{lcc}
Run & Full background training set & {\it hveto} parameter \\
\hline
1 & No & No \\
2 & Yes & No \\
3 & No & Yes \\
4 & Yes & Yes \\
\end{tabular}
\end{ruledtabular}
\end{table}

\subsection{Ringdown-only sensitive $VT$ improvements}\label{sec:rdsensvt}

\begin{figure*}
    \centering
    \subfigure[$\,\,\,q=1$]{\label{fig:cat3vola}\includegraphics[width=.45\textwidth]{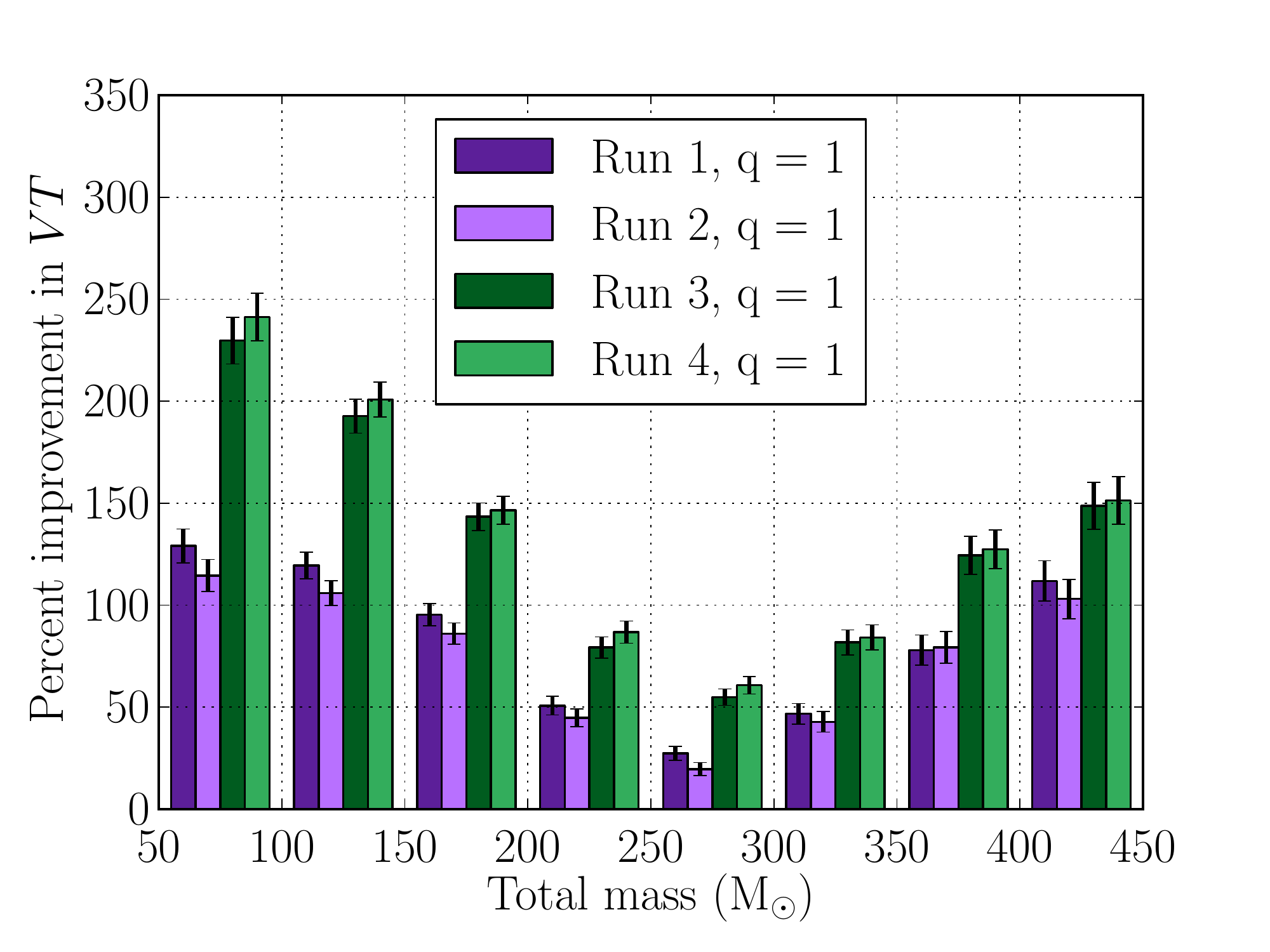}}
    \subfigure[$\,\,\,q=4$]{\label{fig:cat3volb}\includegraphics[width=.45\textwidth]{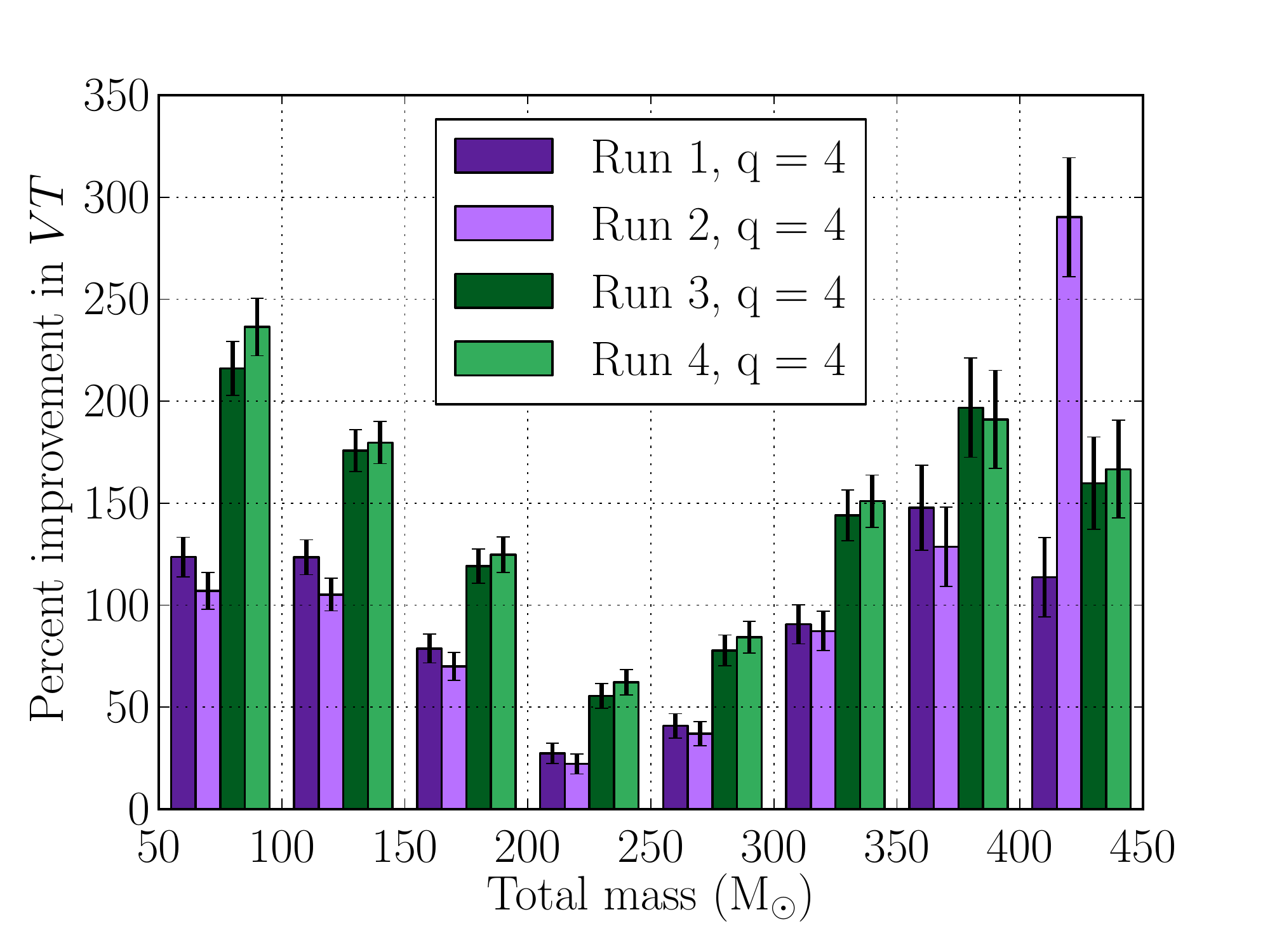}}
    \caption{Percent improvements (over the use of the $\rho_\mathrm{S5/S6}$ ranking statistic~\cite{2013talukder} at Veto Level~2) in sensitive volume multiplied by analysis time $(VT)$  for the recovery of $q=1$ ({\it left}) and $q=4$ ({\it right}) EOBNRv2 simulated waveforms at Veto Level~2.
      The quantity $VT$ gives us a measure of the true sensitivity of the search and allows us to compare performances across veto levels.
      Results are shown for total binary masses from $50\le M/\mathrm{M}_\odot \le 450$ in mass bins of width $50\,\mathrm{M}_\odot$.
    }
    \label{fig:cat3volrd}
  \end{figure*}

\begin{figure*}
    \centering
    \subfigure[$\,\,\,q=1$]{\label{fig:cat4vola}\includegraphics[width=.45\textwidth]{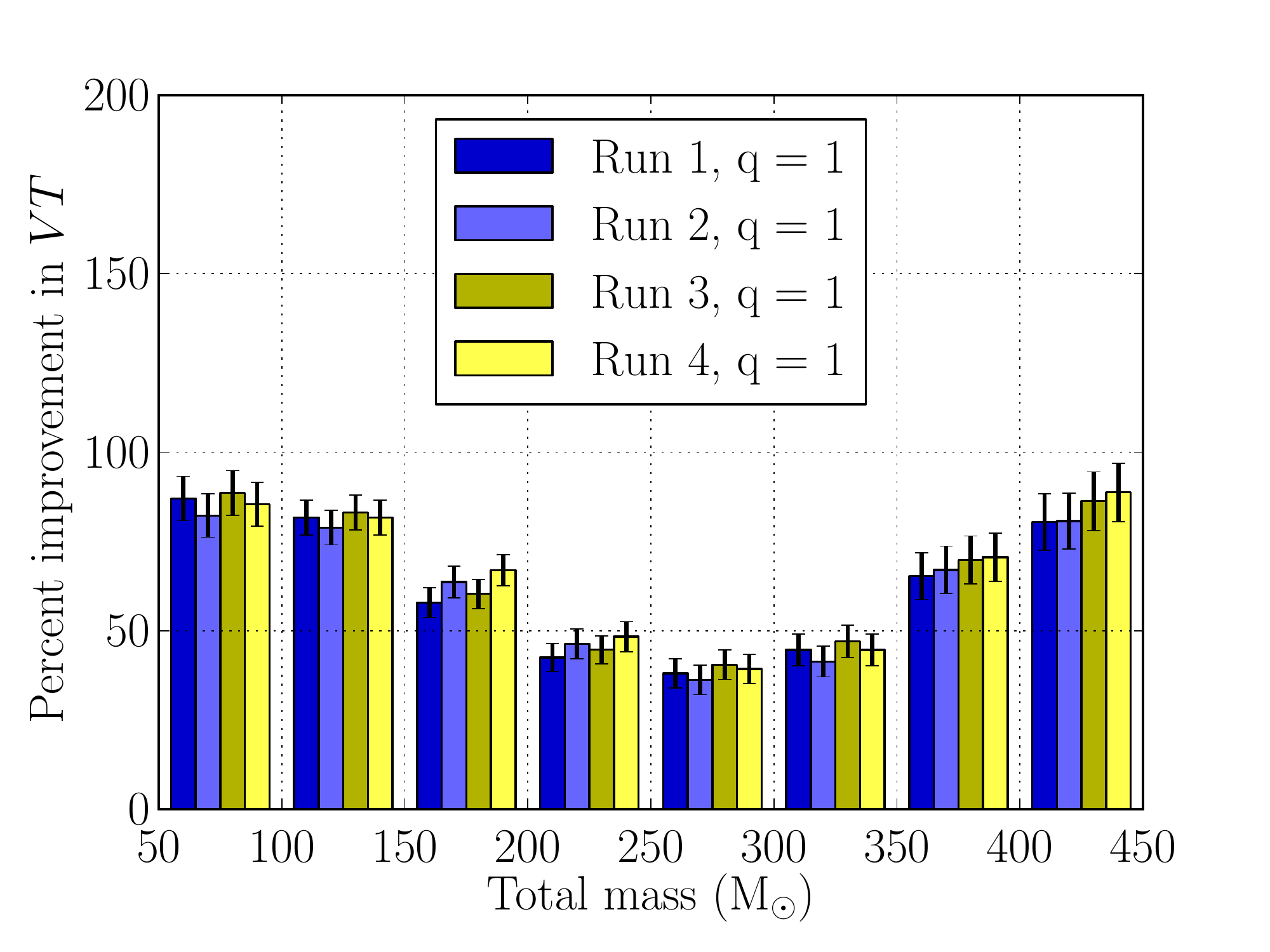}}
    \subfigure[$\,\,\,q=4$]{\label{fig:cat4volb}\includegraphics[width=.45\textwidth]{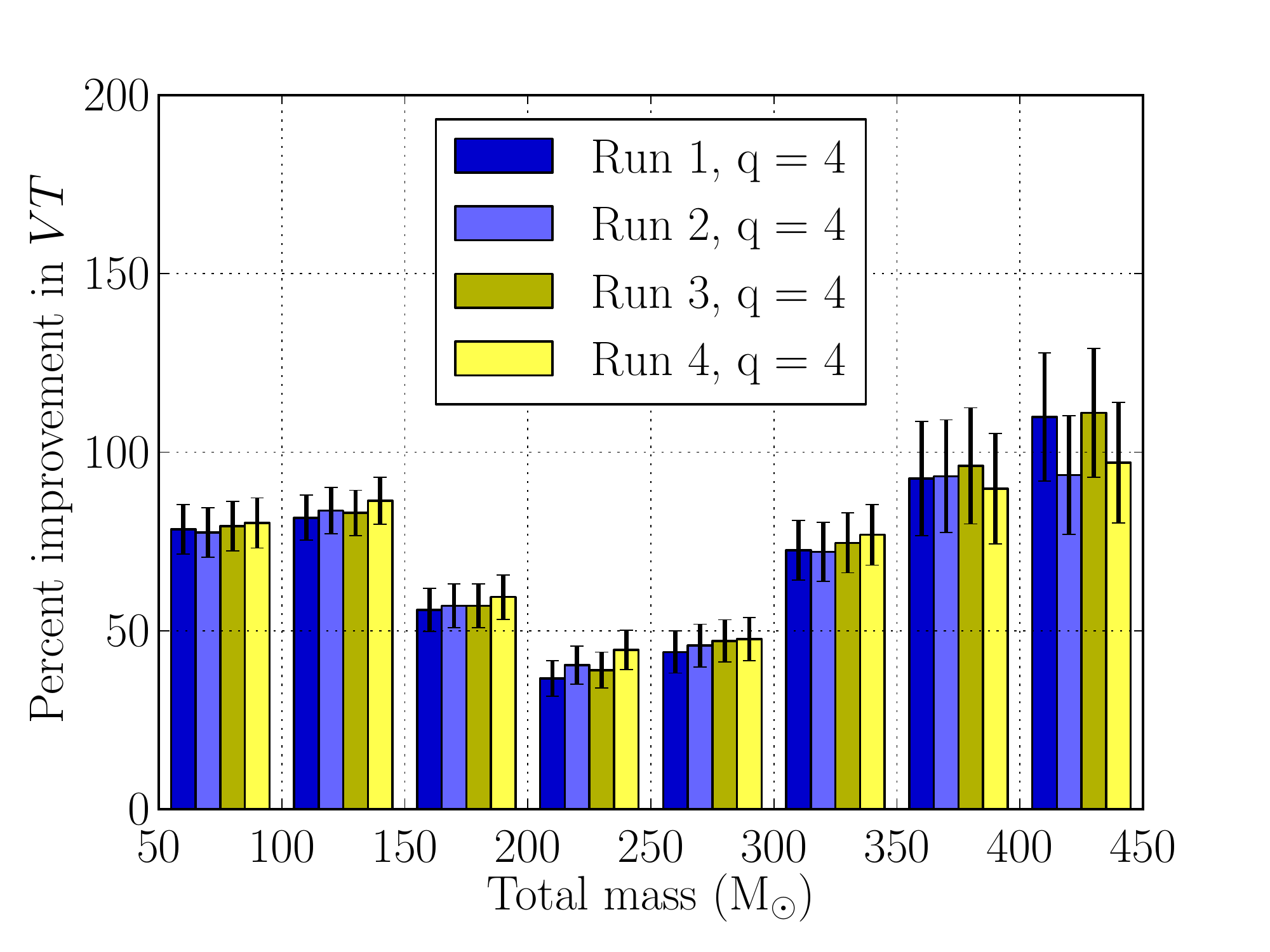}}
    \caption{Percent improvements (over the use of the $\rho_\mathrm{S5/S6}$ ranking statistic~\cite{2013talukder} at Veto Level~3) in sensitive volume multiplied by analysis time $(VT)$  for the recovery of $q=1$ ({\it left}) and $q=4$ ({\it right}) EOBNRv2 simulated waveforms at Veto Level~3.
      The quantity $VT$ gives us a measure of the true sensitivity of the search and allows us to compare performances across veto levels.
      Results are shown for total binary masses from $50\le M/\mathrm{M}_\odot \le 450$ in mass bins of width $50\,\mathrm{M}_\odot$.
    }
    \label{fig:cat4volrd}
  \end{figure*}

Figures~\ref{fig:cat3volrd} and~\ref{fig:cat4volrd} demonstrate the percent improvements in sensitive volume multiplied by analysis time $(VT)$ when using the $M_\mathrm{forest}$ ranking statistic, rather than the $\rho_\mathrm{S5/S6}$ ranking statistic at Veto Levels~2 and~3, respectively.
\par

Figure~\ref{fig:cat3volrd} focuses on the comparison of Runs 1-4 over $\rho_\mathrm{S5/S6}$ at Veto Level~2 for each mass ratio.
Here we see that all runs perform better than $\rho_\mathrm{S5/S6}$ at Veto Level~2.
The largest percent improvements are seen in the lowest and highest mass bins.
These are the mass regions where the ringdown-only search is least sensitive.
Thus, in these regimes, small changes in $VT$ lead to large percent improvements.
This is the reason for the seemingly large percent improvement in Fig.~\ref{fig:cat3volb} for Run~2.
In general, Run~3 and~4 that include the {\it hveto} parameter in the feature vector outperform Run~1 and~2 that do not include the {\it hveto} parameter.
Run~4 most consistently shows the largest $VT$ improvements although the differences are not large at Veto Level~3.
At Veto Level~2, $VT$ improvements ranged from $61_{\pm 4}-241_{\pm 12}$\% for $q=1$ and from $62_{\pm 6}-236_{\pm 14}$\% for $q=4$.
\par

We also note in Fig.~\ref{fig:cat3volrd} that Run~2 is slightly worse than Run~1.
This is due to the fact that, generally, it is advantageous to break large analyses up into several smaller chunks to account for sensitivity changes over the run.
By training the RFBDTs on the ``full background set," we subjected the entire training set to background triggers from the least sensitive times (i.e., times when the background triggers most resembled signal) which resulted in an overall decrease in sensitive volume.
In Run~1, these troublesome background triggers would be isolated in the separate training sets for each $\sim$1-2 month chunk of Period~2.
However, note that training the RFBDTs on the ``full background set" with an {\it hveto} data quality parameter in the feature vectors results in Run~4 being more sensitive than Run~3.
\par

Figure~\ref{fig:cat4volrd} focuses on the comparison of Runs 1-4 over $\rho_\mathrm{S5/S6}$ at Veto Level~3 for each mass ratio.
Again we see that all runs perform better than $\rho_\mathrm{S5/S6}$ at Veto Level~3 (although percent improvements are not as large as those seen at Veto Level~2), and the largest percent improvements are seen in the lowest and highest mass bins.
However, at Veto Level~3, we find that the addition of the {\it hveto} data quality parameter in the feature vectors of Run~3 and~4 do not give significant improvements over Run~1 and~2.
This fact indicates that the {\it hveto} parameter provides no additional information on the most significant glitches for the ringdown-only search that is not already included in the vetoes at Veto Level~3.
Although the difference is not large, in general, Run~3 and~4 still outperform Run~1 and~2.
At Veto Level~3, $VT$ improvements ranged from $39_{\pm 4}-89_{\pm 8}$\% for $q=1$ and from $39_{\pm 5}-111_{\pm 18}$\% for $q=4$.
\par
  
In Table~\ref{tab:rdvetoresults}, we explore the percent $VT$ improvements obtained with $M_\mathrm{forest}$ at different veto levels with and without the {\it hveto} parameter.
The improvements reported are made with respect to the $\rho_\mathrm{S5/S6}$ ranking statistic at Veto Level~2.
These values are presented as a means of comparing sensitivity between Veto Level~2 and Veto Level~3.
Here we make several observations.
First, unlike the behavior we observed for the IMR search, we see that Run~1 at Veto Level~3 shows greater improvement and hence a more stringent upper limit than Run~1 at Veto Level~2.
Thus, the removal of poor data quality at Veto Level~3 is an important step for improving the sensitivity of the ringdown-only search.
Second, we can compare Run~1 at Veto Level~3 with Run~3 at Veto Level~2.
This comparison allows us to gauge whether the gain in analysis time we get by including {\it hveto} data quality information in the feature vector at Veto Level~2 outweighs the boost in sensitive volume we gain by removing data flagged by Veto Level~3.
We see that Run~1 at Veto Level~3 gives consistently larger percent $VT$ improvements than Run~3 at Veto Level~2.
Thus, adding the {\it hveto} data quality information in the feature vector does not match the sensitivity improvements from the application of data quality vetoes.
However, we note that the {\it hveto} data quality information was not specifically tuned for the ringdown-only search nor is it meant to be an exhaustive data quality investigation.
Further exploration with more sophisticated data quality information is needed in order to determine whether the classifier can incorporate data quality information and approach the sensitivity achieved by the use of data quality vetoes.
\par

 \begin{table*}[t]
\caption{Percent $VT$ improvements over the use of the $\rho_\mathrm{S5/S6}$ ranking statistic at Veto Level~2 for $q=1$ EOBNRv2 waveforms. Note that these percent improvements should not be compared with values reported in Fig.~\ref{fig:cat3volrd} and~\ref{fig:cat4volrd} but are rather presented as a means of comparing sensitivity between Veto Level~2 and Veto Level~3. We see that Run~1 at Veto Level~3 shows greater improvement and hence a more stringent upper limit than Run~1 at Veto Level~2, unlike the IMR search as shown in Table~\ref{tab:hmvetoresults}. Additionally, we see that Run~1 at Veto Level~3 gives consistently larger percent $VT$ improvements than Run~3 at Veto Level~2, indicating that the {\it hveto} parameter in the feature vector does not give the same sensitivity improvements as the application of traditional vetoes.}\label{tab:rdvetoresults}
\begin{ruledtabular}
\begin{tabular}{lcccccc}
 &    $\rho_\mathrm{S5/S6}$     &     Run 1    &    Run 1       &      Run 3     &    Run 3 \\
Mass bin $\left( \mathrm{M}_\odot \right)$ & Veto Level 3 & Veto Level 2 & Veto Level 3 & Veto Level 2 & Veto Level 3 \\ 
\hline
 50.0 - 100.0 & 113 $\pm$ 7\%  & 129 $\pm$ 8\%& 299 $\pm$ 14\% & 230 $\pm$ 11\% & 302 $\pm$ 14\%\\
100.0 - 150.0 & 92 $\pm$ 6\% & 120 $\pm$ 7\%& 250 $\pm$ 10\% & 193 $\pm$ 8\% & 252 $\pm$ 10\% \\
150.0 - 200.0 & 88 $\pm$ 5\% & 95 $\pm$ 6\%& 196 $\pm$ 8\% & 143 $\pm$ 7\% & 201 $\pm$ 8\% \\
200.0 - 250.0 & 63 $\pm$ 5\% & 51 $\pm$ 5\%& 133 $\pm$ 7\% & 79 $\pm$ 5\%   & 136 $\pm$ 7\% \\
250.0 - 300.0 & 60 $\pm$ 4\% & 27 $\pm$ 3\%& 121 $\pm$ 7\% & 55 $\pm$ 4\%   & 125 $\pm$ 7\% \\
300.0 - 350.0 & 66 $\pm$ 6\% & 47 $\pm$ 5\%& 139 $\pm$ 8\% & 82 $\pm$ 6\%   & 143 $\pm$ 8\% \\
350.0 - 400.0 & 73 $\pm$ 8\% & 78 $\pm$ 7\%& 186 $\pm$ 12\% & 124 $\pm$ 9\% & 194 $\pm$ 12\% \\
400.0 - 450.0 & 70 $\pm$ 8\% & 112 $\pm$ 10\%& 207 $\pm$ 14\% & 149 $\pm$ 11\% & 218 $\pm$ 14\% \\
\end{tabular}
\end{ruledtabular}
\end{table*}

\section{Summary}\label{sec:summary}
This paper presents the development and sensitivity improvements of a multivariate analysis applied to matched filter searches for gravitational waves produced by coalescing black hole binaries with total masses $\gtrsim25\,$M$_\odot$.
We focus on the applications to the IMR search which looks for gravitational waves from the inspiral, merger, and ringdown of BBHs with total mass between $25\,$M$_\odot$ and $100\,$M$_\odot$ and to the ringdown-only search which looks for gravitational waves from the resultant perturbed IMBH with mass roughly between $10\,$M$_\odot$ and $600\,$M$_\odot$.
These investigations were performed over data collected by LIGO and Virgo between 2009 and 2010 so that comparisons can be made with previous IMR and ringdown-only search results~\cite{2013s6highmass,2014rdsearch}.
We discuss several issues related to tuning RFBDT multivariate classifiers in matched-filter IMR and ringdown-only searches.
We determine the sensitivity improvements achieved through the use of a RFBDT-derived ranking statistic over empirical SNR-based ranking statistics while considering the application of data quality vetoes.
Additionally, we present results for several modifications on the basic RFBDT implementation including the use of an expansive training set and data quality information.
\par

When optimizing the performance of RFBDT classifiers, we found that a RFBDT classifier with 100 trees, a leaf size of 5, and 6 randomly sampled parameters from the feature vector gave good performance for the IMR search while a RFBDT classifier with 2000 trees, a leaf size of 65, and 14 randomly sampled parameters from the feature vector gave good performance for the ringdown-only search.
In both cases, we used a training set of ``clean" signal designed to carefully remove contamination from glitches within the software injection-finding time window.
This technique eliminated the excursion of gravitational-wave candidate coincidences from the $2\sigma$ region of the expected background at low values of inverse combined FAR as demonstrated in Fig.~\ref{fig:ifarbump}.
Additionally, we examined the performance of the RFBDT classifier in the case where the monthly analyses used background training sets from their respective months and in the case where the monthly analyses used background training sets from the entire Period 2 analysis (i.e., the "full background set").
We found that using the full background training set does not result in a clear sensitivity improvement unless a data quality {\it hveto} parameter is introduced in the feature vector.
\par

For the IMR search, we performed a re-analysis replacing $\rho_\mathrm{high,combined}$ with the ranking statistic calculated by the RFBDT, $M_\mathrm{forest}$.
Comparisons with $\rho_\mathrm{high,combined}$ were made separately at each veto level.
For EOBNRv2 waveforms, the percent improvements in $VT$ were largest at Veto Level~2.
Depending on mass bin, the $VT$ improvements ranged from $70_{\pm 13}-109_{\pm 11}$\% at Veto Level~2 and from $10_{\pm 8}-35_{\pm 7}$\% at Veto Level~3.
For IMRPhenomB waveforms, $VT$ improvements ranged from $9_{\pm 5}-36_{\pm 6}$\% regardless of veto level.
Additionally, we made comparisons across veto levels, using the performance of $\rho_\mathrm{high,combined}$ at Veto Level~2 as the standard.
We found that $M_\mathrm{forest}$ at Veto Level 2 shows greater improvement and hence a more stringent upper limit than $M_\mathrm{forest}$ at Veto Level~3.
This is in contrast to the better performance of $\rho_\mathrm{high,combined}$ at Veto Level~3 than at Veto Level~2.
\par

For the ringdown-only search, we evaluated the performance of the RFBDT classifier using five separate ranking statistics.
Comparisons were made with respect to a ringdown-only search that used the $\rho_\mathrm{S5/S6}$ ranking statistic~\cite{2012talukderthesis, 2012caudillthesis, 2013talukder}.
The additional four searches used the $M_\mathrm{forest}$ ranking statistic for various instantiations of the RFBDT classifier.
Comparisons with $\rho_\mathrm{S5/S6}$ were made separately at each veto level.
At Veto Level~2, we found that a RFBDT classifier trained on the full background set and including the data quality {\it hveto} parameter in the feature vector resulted in $VT$ improvements in the range $61_{\pm 4}-241_{\pm 12}$\% for $q=1$ EOBNRv2 waveforms and in the range $62_{\pm 6}-236_{\pm 14}$\% $q=4$ EOBNRv2 waveforms.
At Veto Level~3, this same configuration resulted in $VT$ improvements in the range $39_{\pm 4}-89_{\pm 8}$\% for $q=1$ EOBNRv2 waveforms and in the range $39_{\pm 5}-111_{\pm 18}$\% $q=4$ EOBNRv2 waveforms.
Again, we made comparisons across veto levels, using the performance of $\rho_\mathrm{S5/S6}$ at Veto Level~2 as the standard.
Unlike the IMR search, we found that $M_\mathrm{forest}$ at Veto Level 3 shows greater improvement and hence a more stringent upper limit than $M_\mathrm{forest}$ at Veto Level~2.
Additionally, we found that adding an {\it hveto} parameter at Veto Level~2 does not result in the same increase in sensitivity obtained by applying level 3 vetoes to a search using the basic implementation of the RFBDT classifier.
With more sophisticated methods for adding data quality information to the feature vector, we may see additional improvements or different behavior.
Further exploration is needed.
\par

In general, for each search, we found that the RFBDT multivariate classifier results in a considerably more sensitive search than the empirical SNR-based statistic at both veto levels.
The software for constructing clean injection sets and the RFBDTs is now implemented in the LALSuite gravitational-wave data analysis routines for use with other matched-filter searches.
More investigations will be needed to understand whether lower mass searches for gravitational waves from binary coalescence would benefit from the use of multivariate classification with supervised MLAs.
For higher mass searches, particularly those susceptible to contamination from noise transients, RFBDT multivariate classifiers have proven to be a valuable tool for improving search sensitivity.
\par

\begin{center}
{\bf Acknowledgments}
\end{center}

We gratefully acknowledge the National Science Foundation for funding LIGO, and LIGO Scientific Collaboration and the Virgo Collaboration for access to this data.
PTB and NJC were supported by NSF award PHY-1306702.
SC was supported by NSF awards PHY-0970074 and PHY-1307429.
DT was supported by NSF awards PHY-1205952 and PHY-1307401.
CC was partially supported by NSF grants PHY-0903631 and PHY-1208881.
This document has been assigned LIGO laboratory document number P1400231.
The authors would like to acknowledge Thomas Dent, Chad Hanna, and Kipp Cannon for work during the initial phase of this analysis.
The authors would also like to thank Alan Weinstein, Gregory Mendell, and Marco Drago for useful discussion and guidance.


%
\appendix

\section{IMR search feature vector}
\label{sec:appendix1}

\begin{figure*}
    \centering
     \subfigure[]{\label{fig:bvol}\includegraphics[width=.45\textwidth]{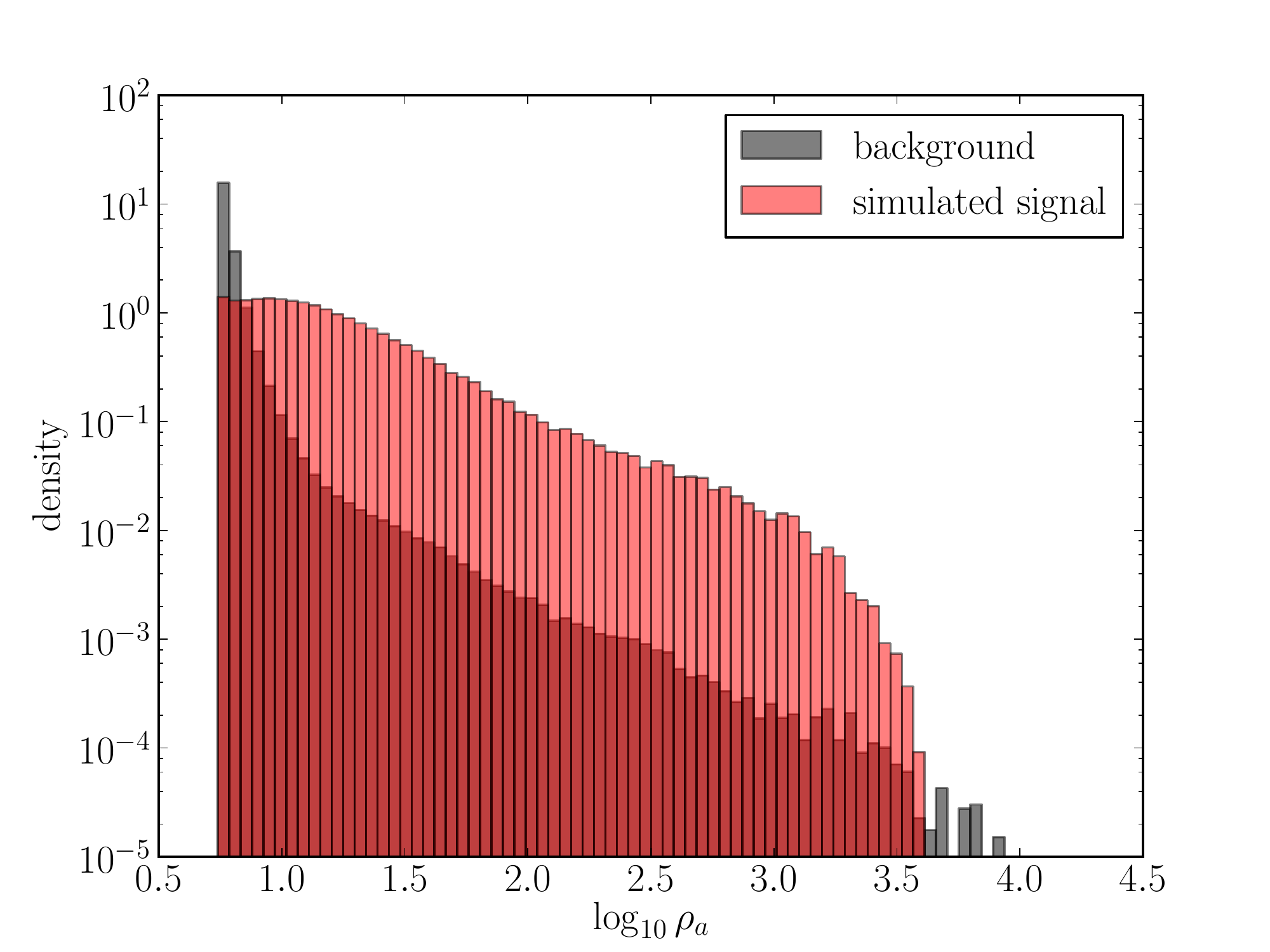}}
     \subfigure[]{\label{fig:bvol}\includegraphics[width=.45\textwidth]{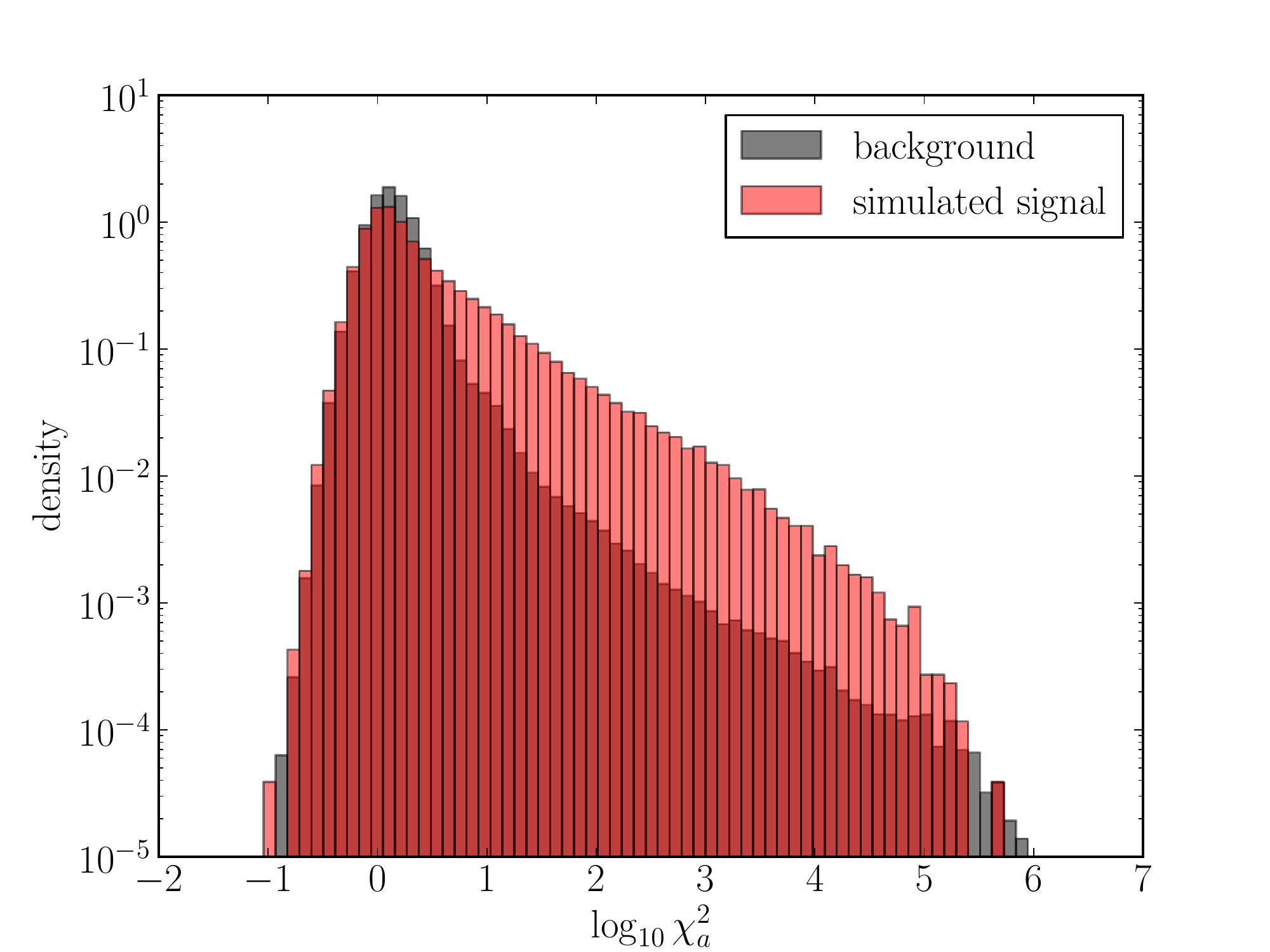}}
     \subfigure[]{\label{fig:avol}\includegraphics[width=.45\textwidth]{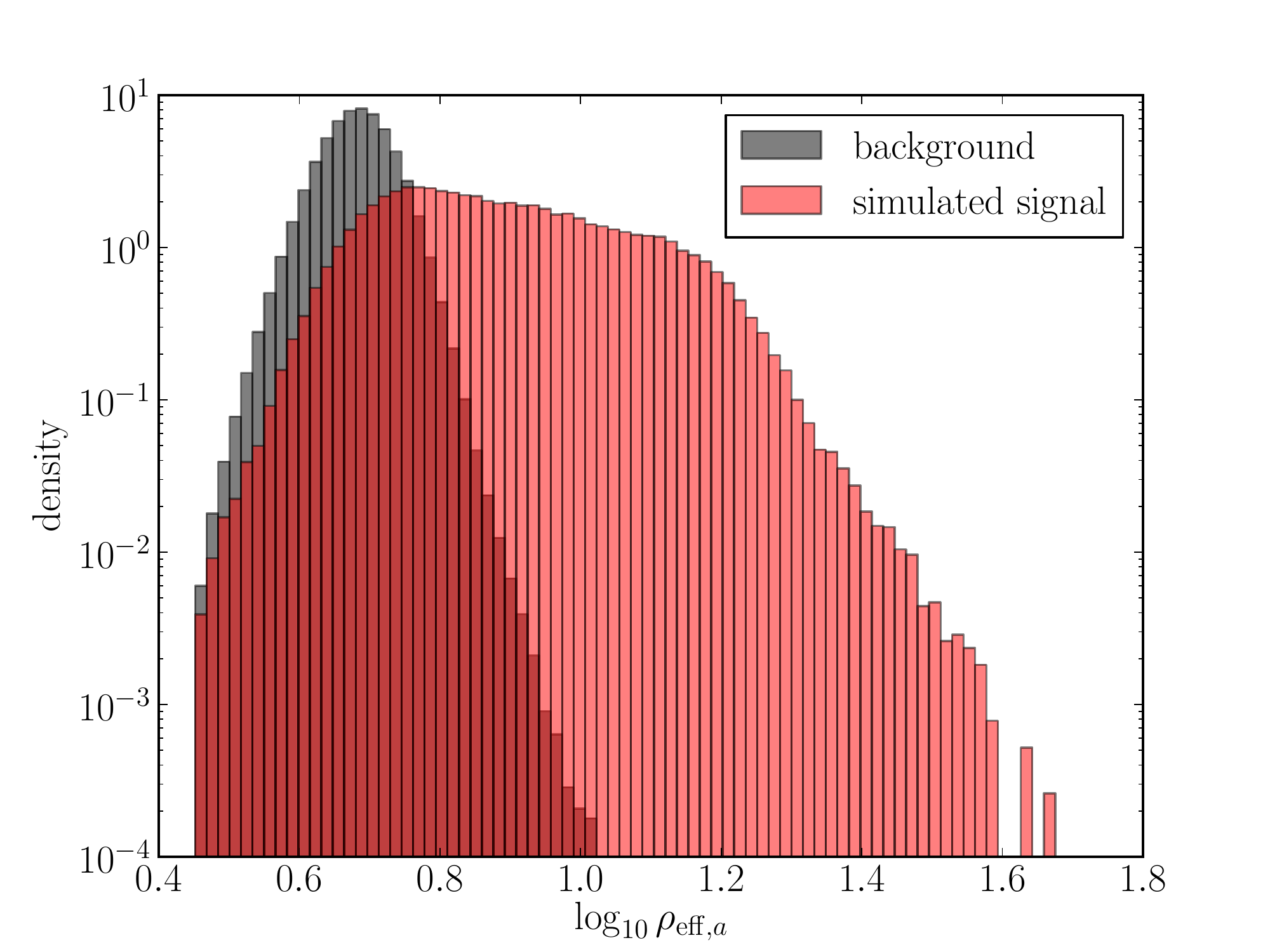}}
      \subfigure[]{\label{fig:bvol}\includegraphics[width=.45\textwidth]{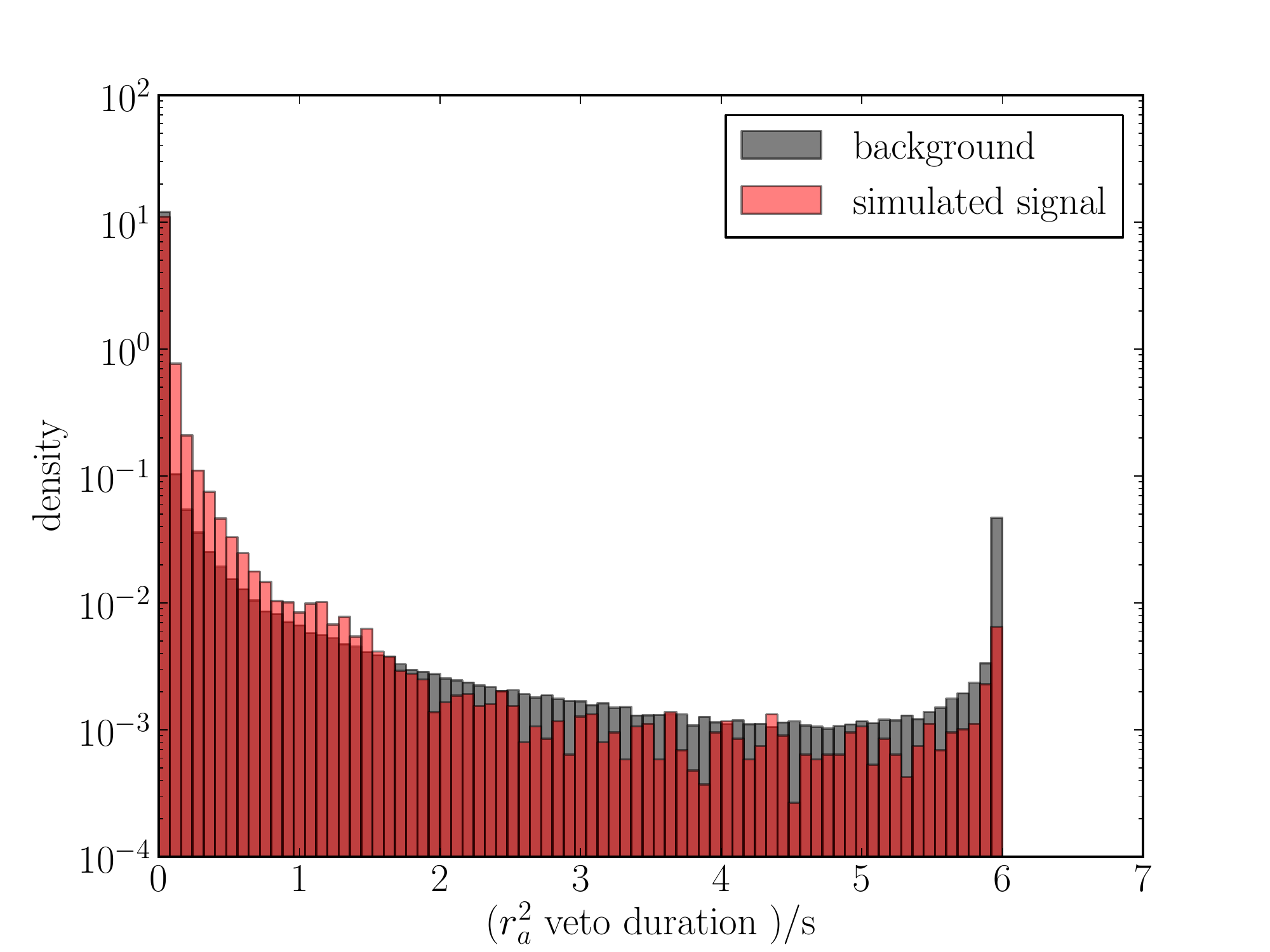}}
    \subfigure[]{\label{fig:bvol}\includegraphics[width=.45\textwidth]{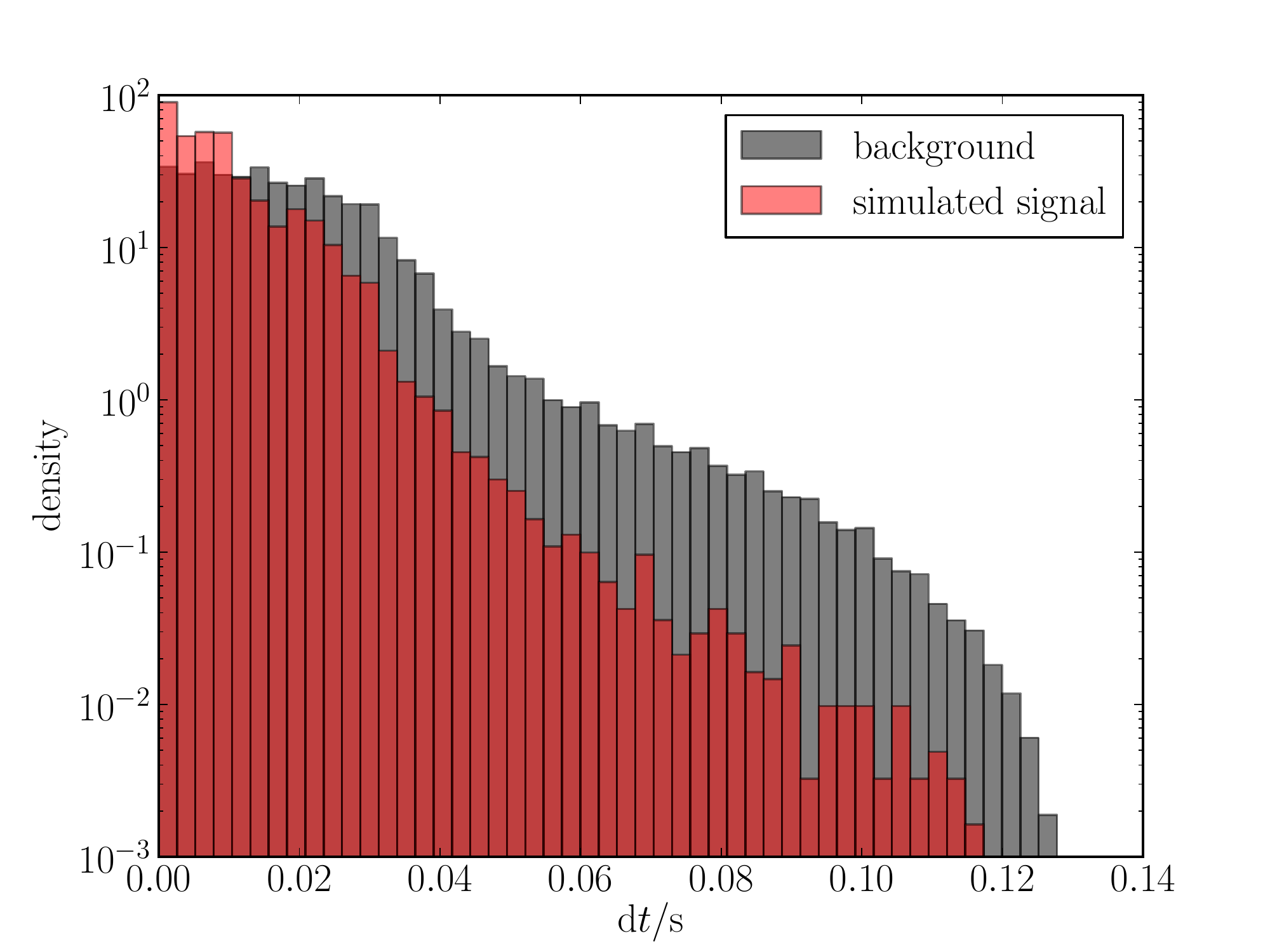}}
    \subfigure[]{\label{fig:bvol}\includegraphics[width=.45\textwidth]{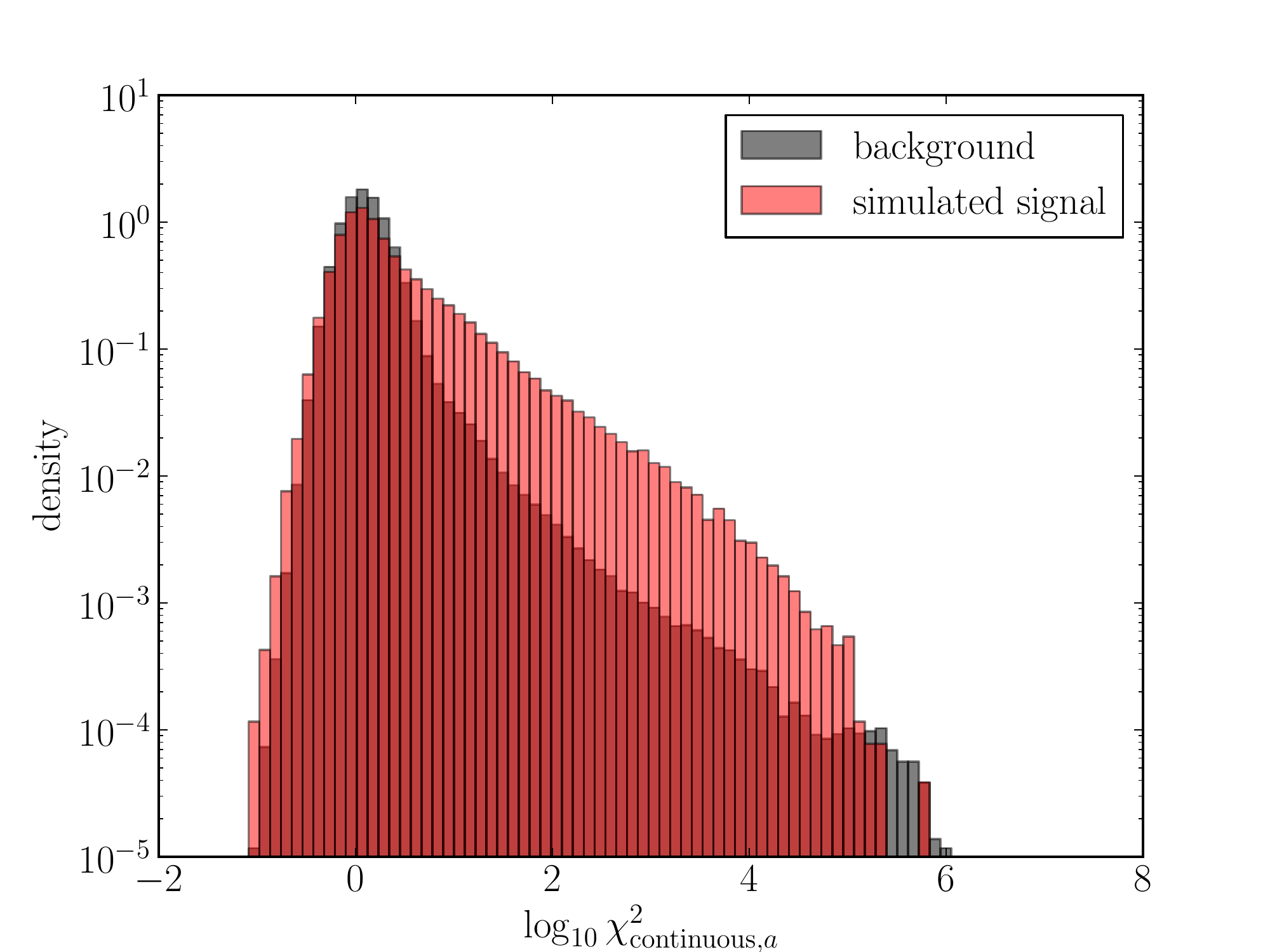}}
        \caption{Signal and background density distributions for a selection of feature vector statistics for the IMR search.}
    \label{fig:hmparams1}
  \end{figure*}

\begin{figure*}
    \centering
     \subfigure[]{\label{fig:bvol}\includegraphics[width=.45\textwidth]{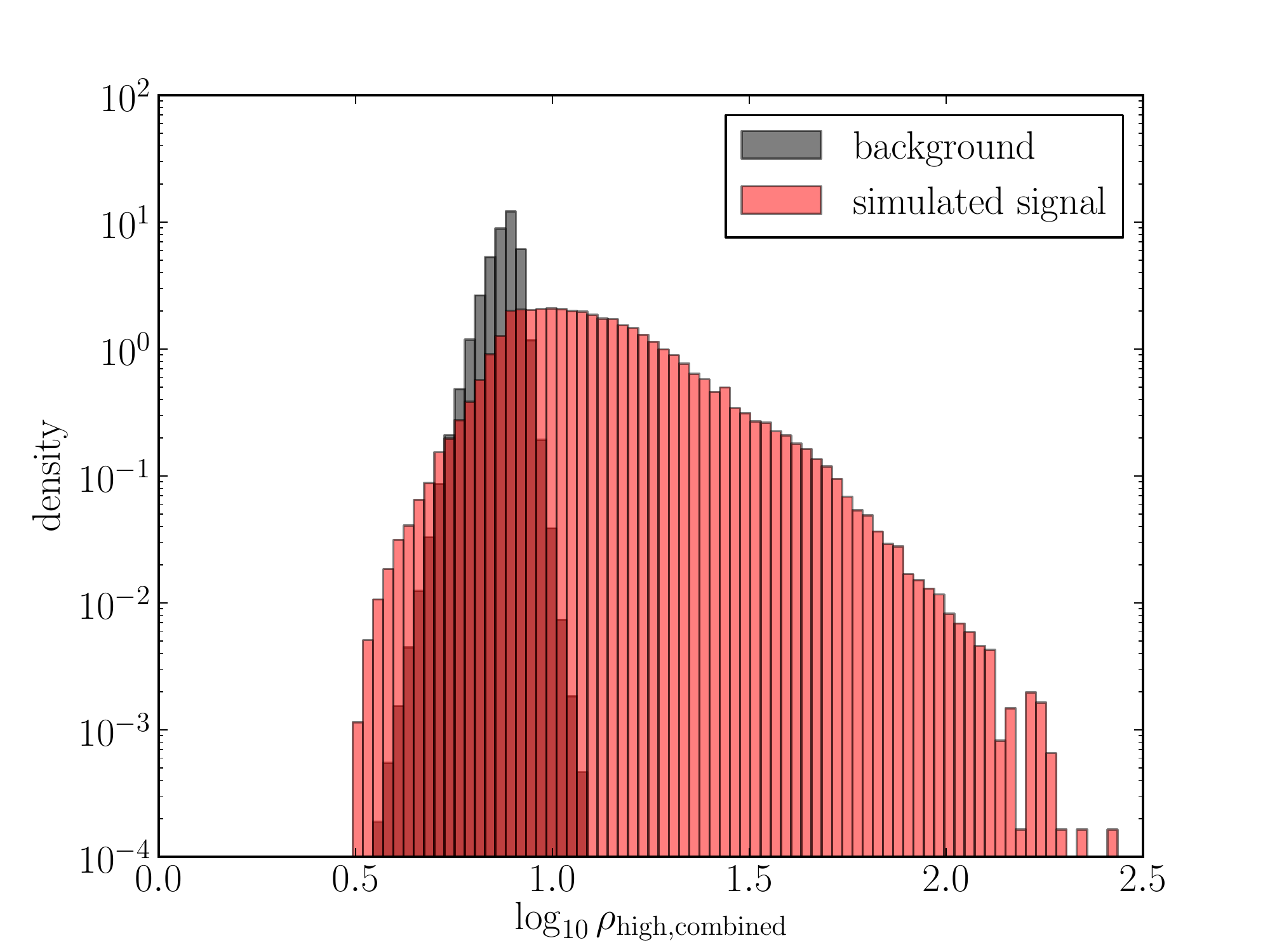}}
     \subfigure[]{\label{fig:avol}\includegraphics[width=.45\textwidth]{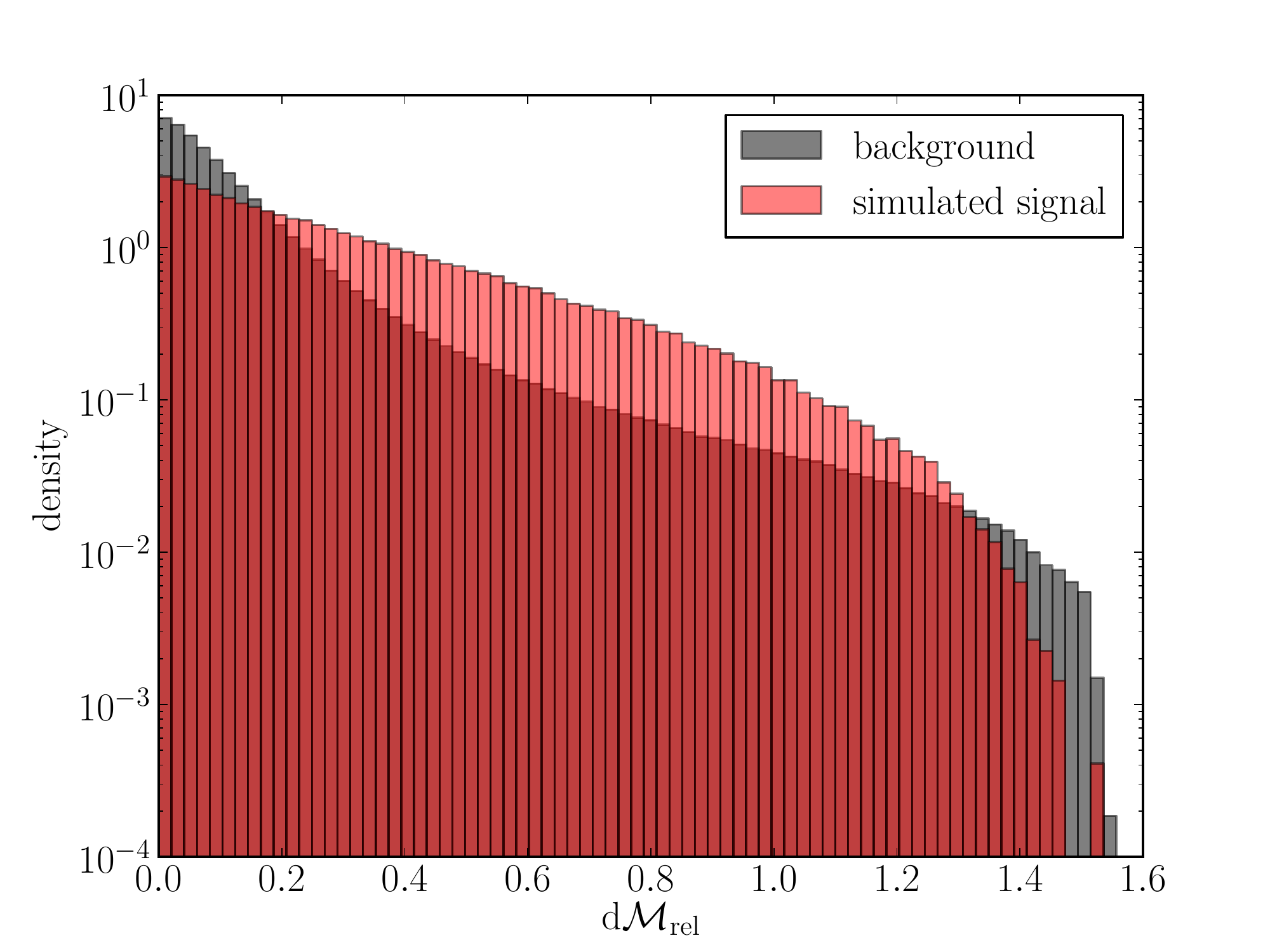}}
    \subfigure[]{\label{fig:bvol}\includegraphics[width=.45\textwidth]{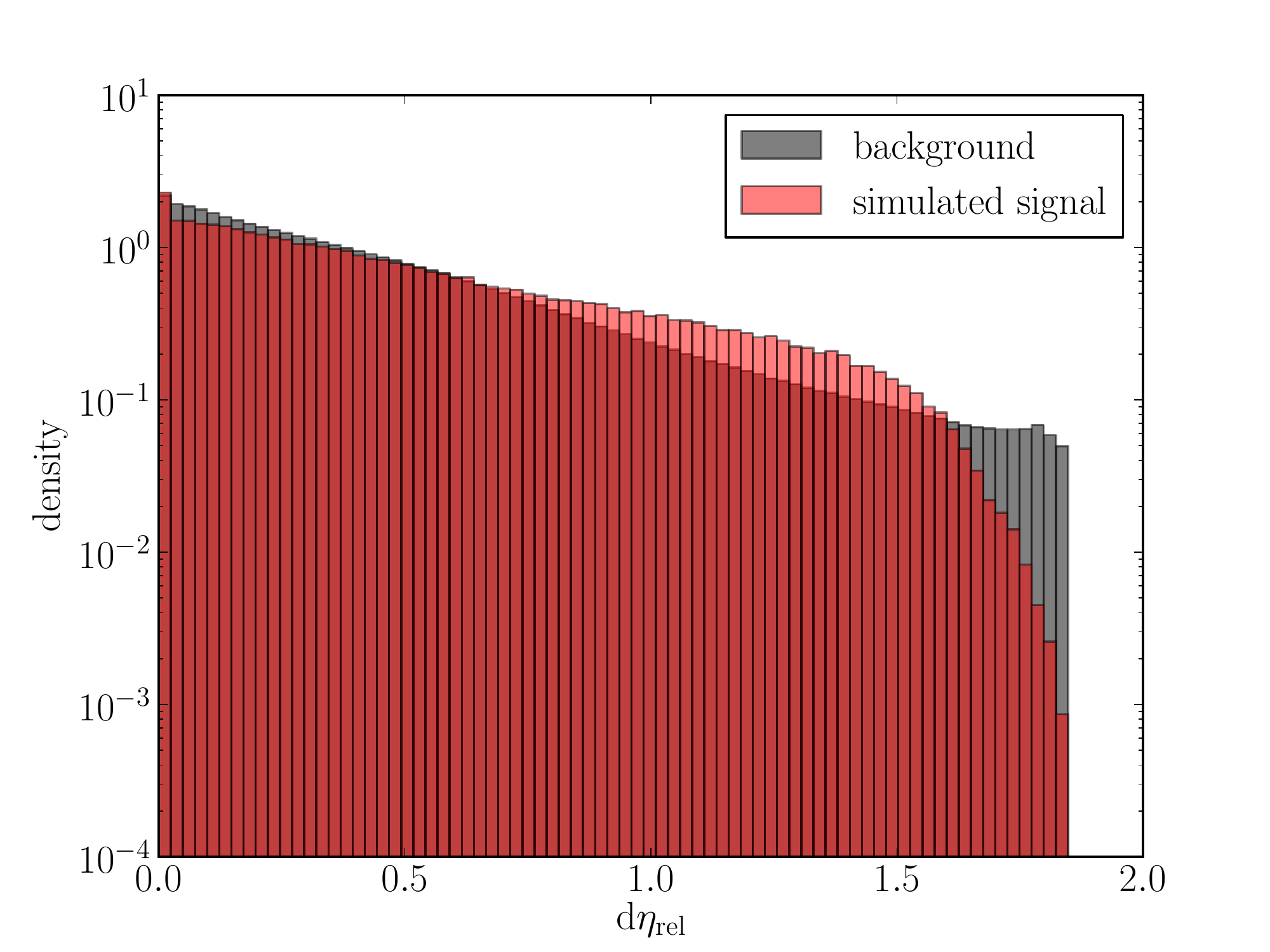}}
    \subfigure[]{\label{fig:bvol}\includegraphics[width=.45\textwidth]{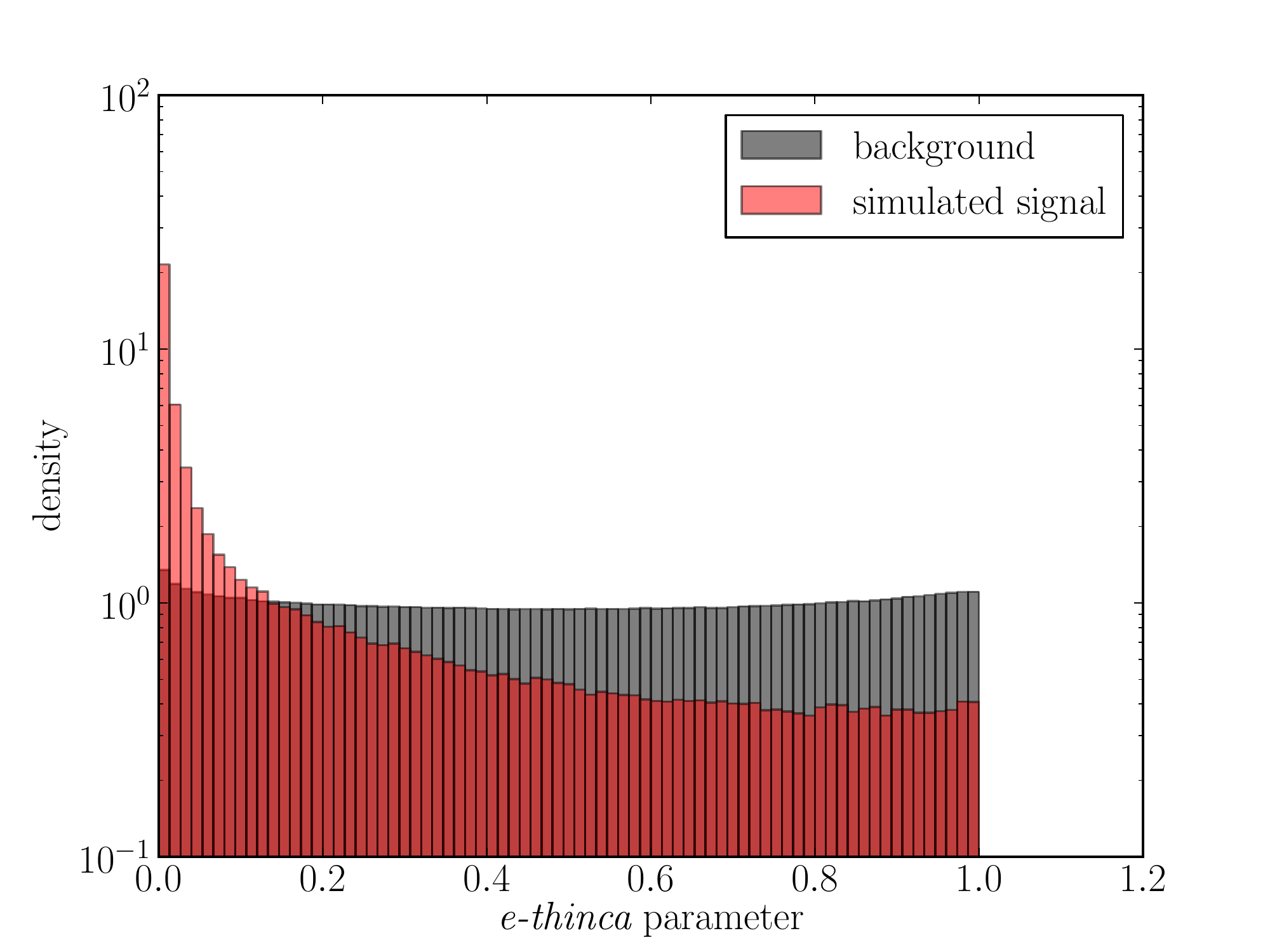}}
    \caption{Signal and background density distributions for a selection of feature vector statistics for the IMR search.}
    \label{fig:hmparams2}
  \end{figure*}

The full list of statistics used for the IMR search's feature vector are given in Table~\ref{tab:paramshm}.
In the following sections, we provide more detail on the definitions of each statistic.
Density distributions of these statistics for this search's simulated signal and background training sets are shown in Fig.~\ref{fig:hmparams1} and~\ref{fig:hmparams2}.

\subsection{Single trigger statistics}\label{sec:hmsngl_parameters}
Single trigger statistics are defined for each individual trigger that makes up each multi-detector coincidence.
For the IMR search, single trigger statistics added to the feature vector included the matched-filter SNR from each detector~\cite{findchirp}, the $\chi^2$ signal-consistency test for the matched-filter result in a number of frequency bins for each detector~\cite{2005allen}, the $r^2$ veto duration for which a weighted $\chi^2$ exceeds a pre-set threshold, and the $\chi_\mathrm{continuous}^2$ quantity for the residual of the SNR and autocorrelation time series in each detector.

More details on the matched-filter SNR for the IMR search templates are given in~\cite{2011s5highmass} and a definition is given in Table~\ref{tab:paramshm}.
\par

The $\chi^2$ signal-consistency test, only currently calculated for the low and high mass searches, tests how well the template waveform matches the data in various frequency bands.
The bins are constructed so that the matched template contributes an equal amount of SNR to each bin.
Then the following quantity is computed,
\begin{equation}
\chi^2=10\left[\sum\limits_{i=0}^{10} (\rho_{i}-\rho/10)^2\right]
\end{equation}
where $\rho_i$ is the SNR contribution from the $i$th bin.
\par

The $r^2$ veto duration measures the amount of time that the following quantity is above a threshold of 0.0002, within $6\,$s of the trigger,
\begin{equation}
r^2=\frac{\chi^2(t)}{p+\rho(t)^2}
\end{equation}
where $p=10$ bins are used.
This quantity is motivated by the fact that a signal is unlikely to exactly match a template so a non-centrality parameter is introduced to the distribution of the $\chi^2$ signal-consistency test.
Thus, rather than thresholding on $\chi^2$, we threshold on $r^2$.

The $\chi_\mathrm{continuous}^2$ calculation performs a sum of squares of the residual of the SNR time series and the autocorrelation time series of a single detector trigger.

\subsection{Composite statistics}\label{sec:hmcomp_parameters}
Composite statistics are defined by combining single trigger statistics in a meaningful way and are computed once for each coincidence.
Although the classifier can approximate such statistics in the multidimensional parameter space (e.g., if they are a combination of the $\rho$ and $\chi^2$), this ability is limited by the tree depth, the number of decision tree cuts before hitting the minimum leaf size.
Thus, if we have \emph{a priori} knowledge of a useful functional form for a ranking statistic, we should provide the classifier with this information.
By providing this information up front, a classifier can improve upon these good statistics rather than trying to construct them itself.

Some of these composite statistics have previously been used as ranking statistics when calculating combined FARs in searches.
For the IMR search, we include several previous ranking statistics in the feature vector.

The first of these is known as the effective SNR statistic and was used as a ranking statistic in~\cite{2011s5highmass},
\begin{equation}\label{eq:effsnr}
\rho_\mathrm{eff}=\frac{\rho}{\left[\frac{\chi^2}{10}\left(1+\rho^2/50\right)\right]^{1/4}}.
\end{equation}
The second is known as $\rho_\mathrm{high,combined}$, a $\chi^2$-weighted statistic described in detail in~\cite{2013s6highmass} for the IMR search.
Due to the different distributions of background triggers over SNR and $\chi^2$ for longer-duration versus shorter-duration templates, a different choice of ranking statistics was selected for each bin in~\cite{2013s6highmass}.
For long duration events, the following was used
\begin{equation}
   \hat{\rho} = \left\{
     \begin{array}{lr}
       \frac{\rho}{\left[ \left(1 + \left(\chi_r^2\right)^3 \right) \right]^{1/6}}& \mathrm{for}\,\,\chi_r^2>1\\
       \rho & \mathrm{for}\,\,\chi_r^2 \le1
     \end{array}
   \right.
\end{equation}
where $\chi_r^2\equiv \chi^2/(2p-2)$ for number of frequency intervals $p=10$.
For shorter duration events, Eq.~\ref{eq:effsnr} was used.
Thus, $\rho_\mathrm{high,combined}$ is a piecewise function of $\rho_\mathrm{eff}$ and $\hat{\rho}$ and is combined as a quadrature sum of single-detector statistics.
\par

Additionally, we calculate quantities that provide an indication of how close the pair of triggers from different detectors are in the metric space $\left(\mathcal{M},\eta,t\right)$ for the IMR search.
These include the difference in arrival time d$t$, the relative difference in chirp mass d$\mathcal{M}_\mathrm{rel}$, the relative difference in the symmetric mass ratio d$\eta_\mathrm{del}$, and a quantity known as the {\it e-thinca} test that combines these three by constructing error ellipsoids in time and mass space~\cite{ethinca}.
\par

\section{Ringdown-only feature vector}
\label{sec:appendix2}
\begin{figure*}
    \centering
     \subfigure[]{\label{fig:bvol}\includegraphics[width=.45\textwidth]{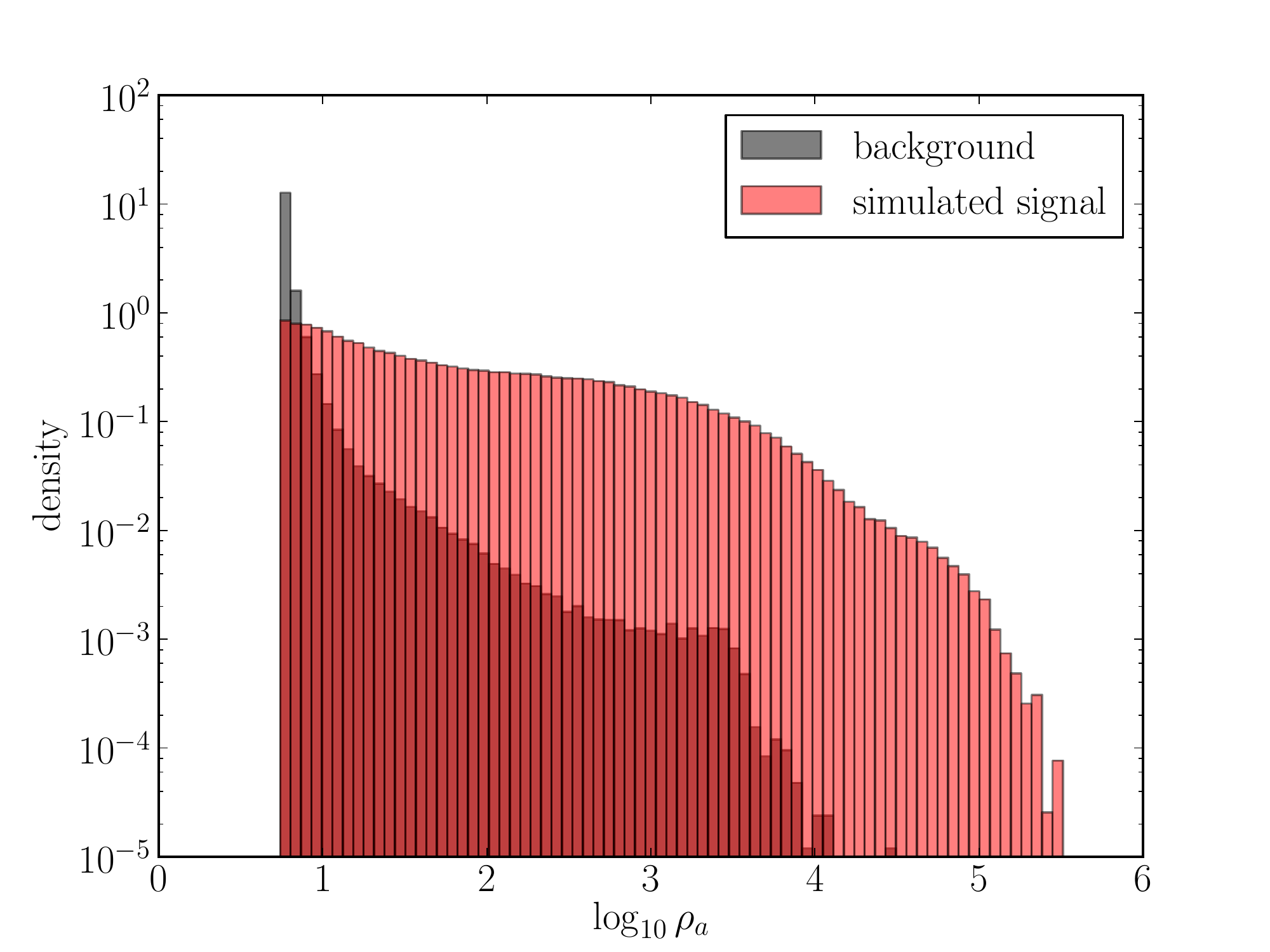}}
    \subfigure[]{\label{fig:avol}\includegraphics[width=.45\textwidth]{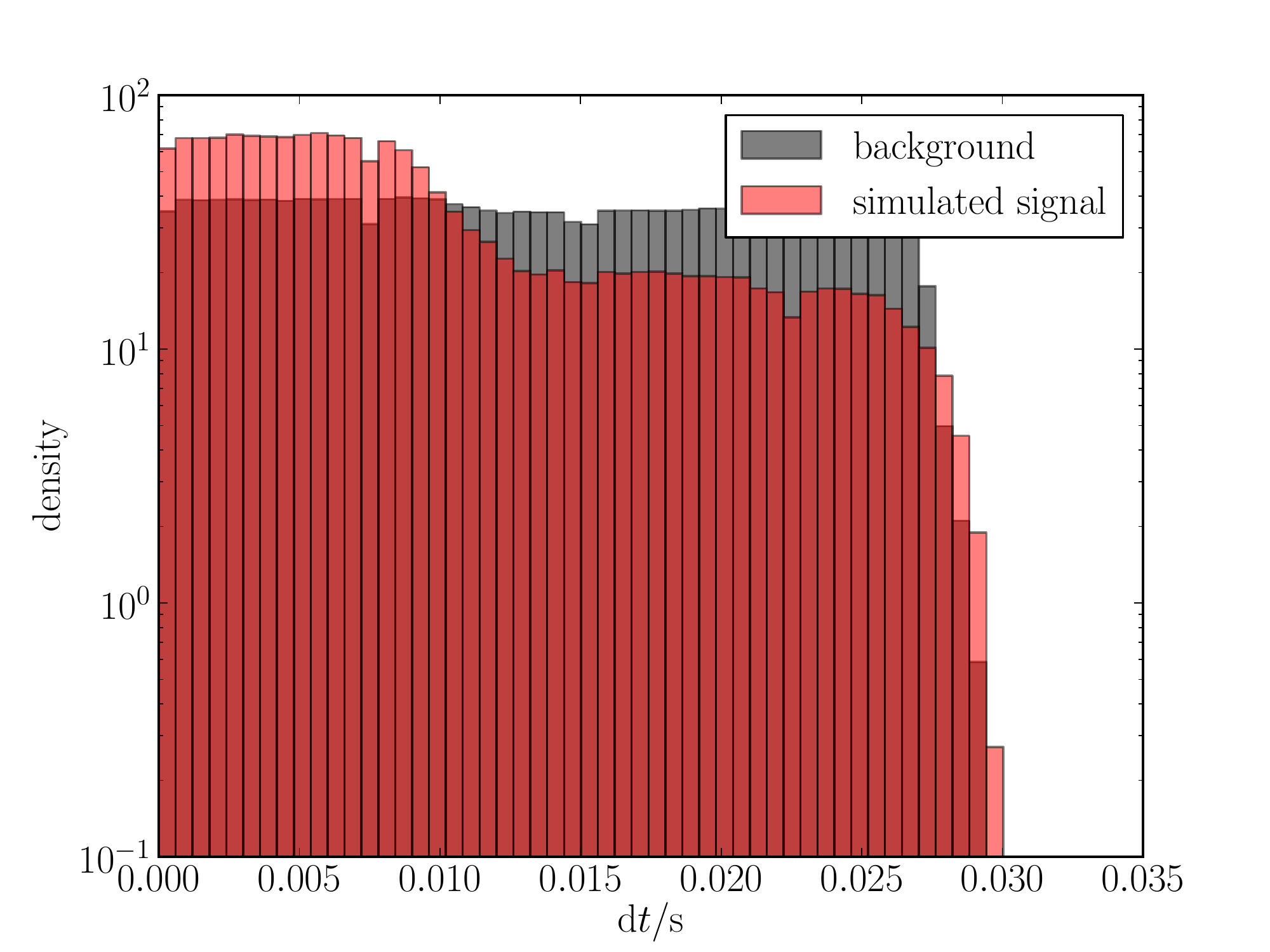}}
    \subfigure[]{\label{fig:bvol}\includegraphics[width=.45\textwidth]{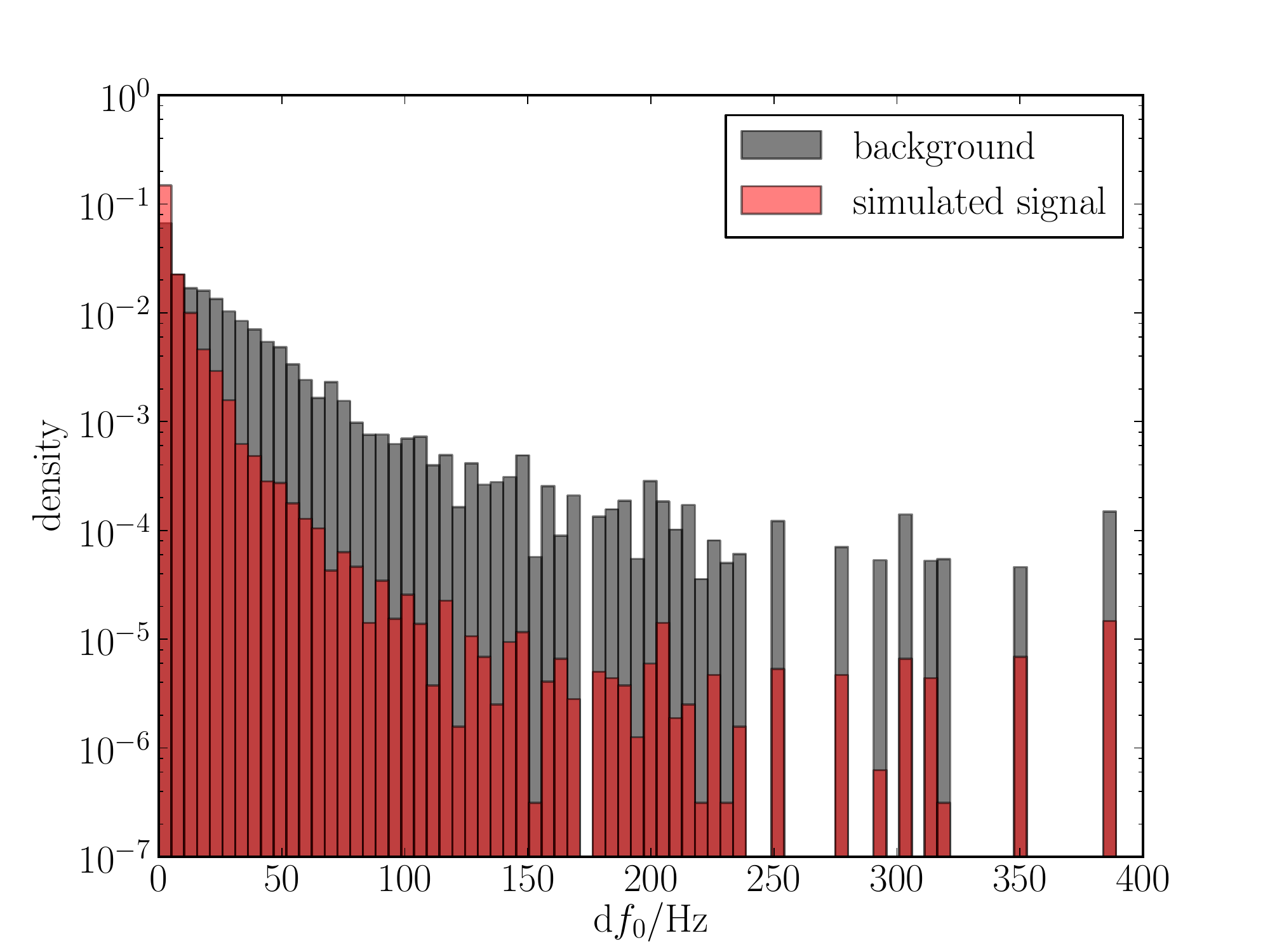}}
    \subfigure[]{\label{fig:bvol}\includegraphics[width=.45\textwidth]{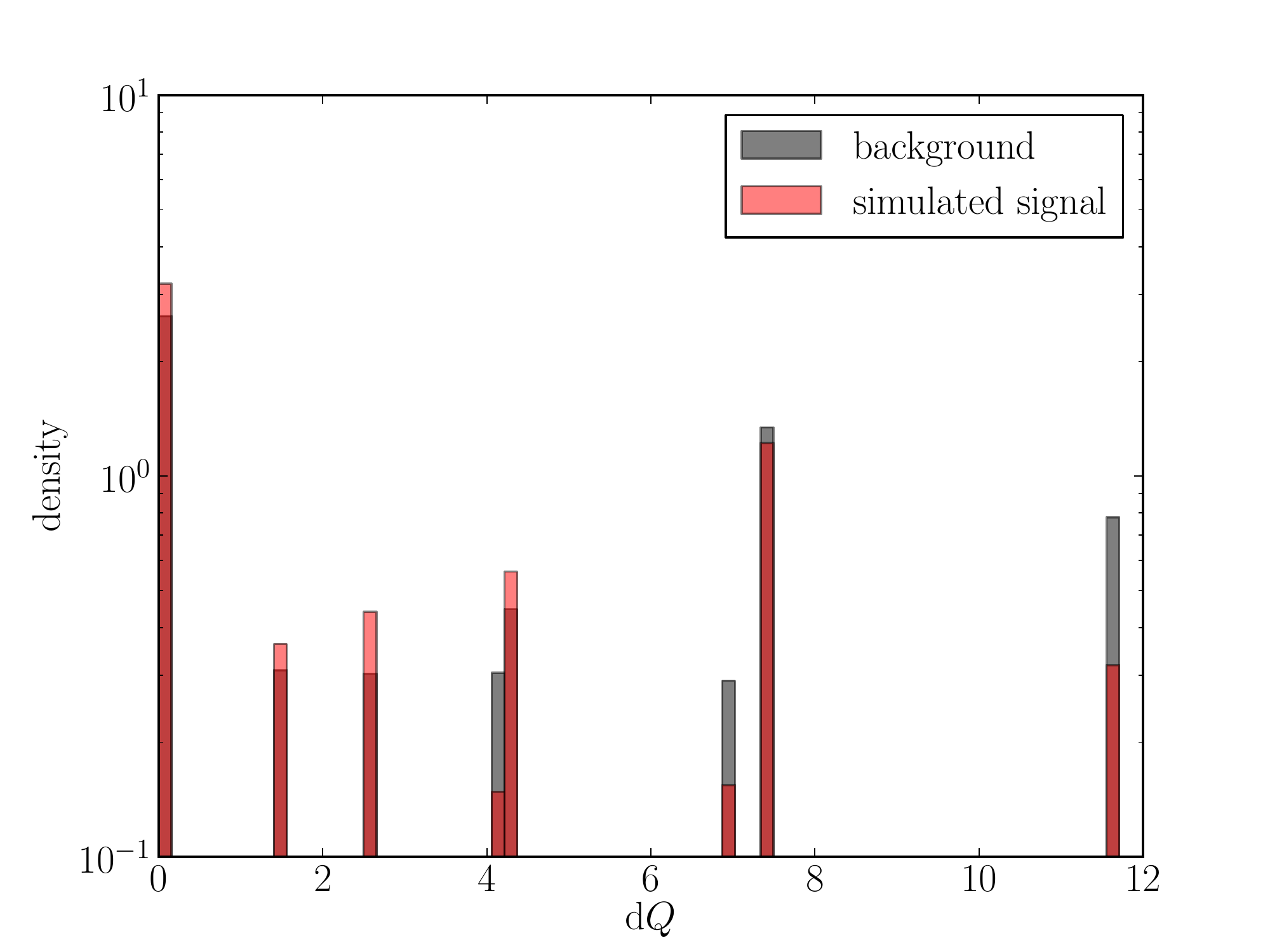}}
     \subfigure[]{\label{fig:bvol}\includegraphics[width=.45\textwidth]{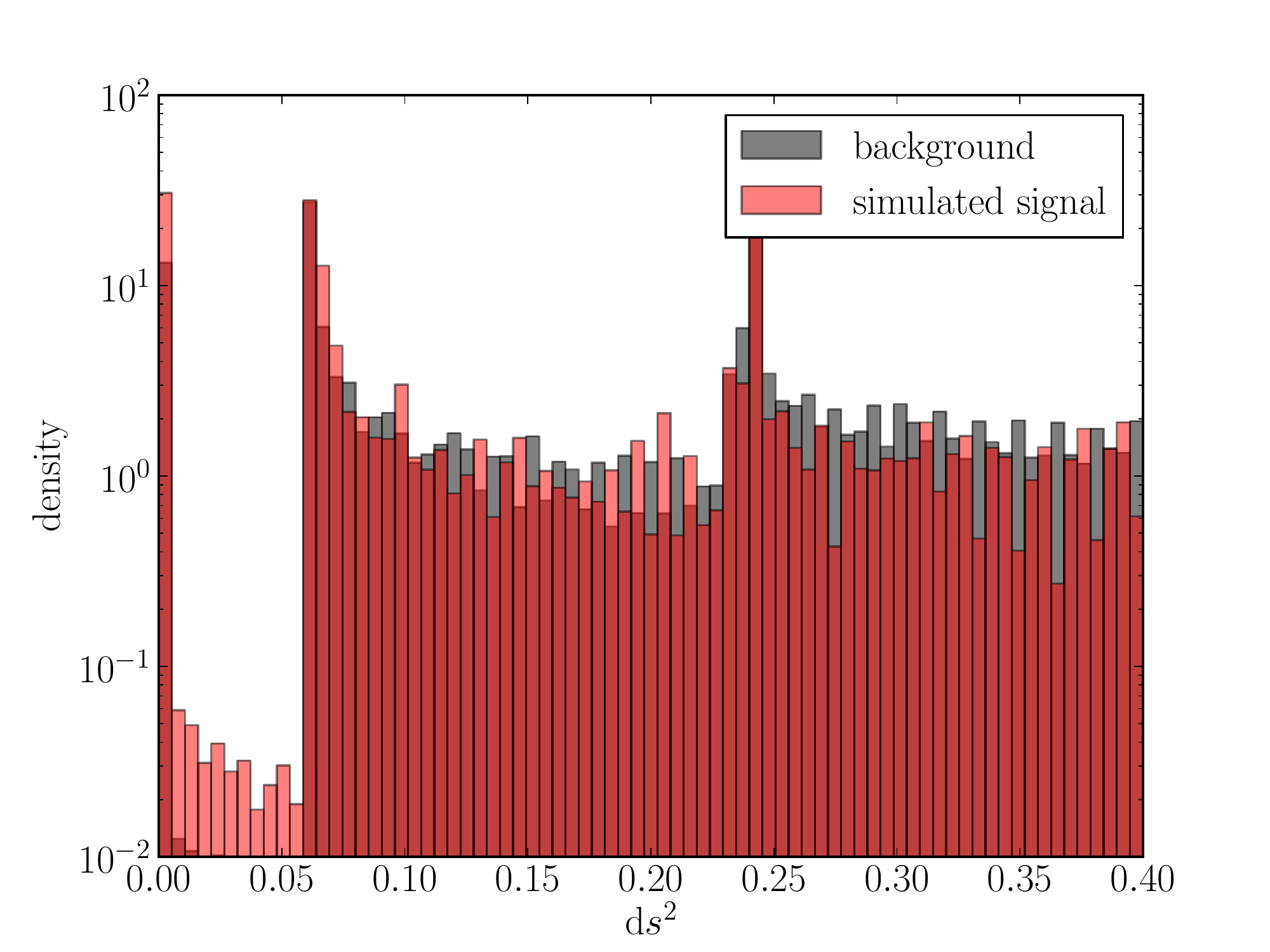}}
     \subfigure[]{\label{fig:bvol}\includegraphics[width=.45\textwidth]{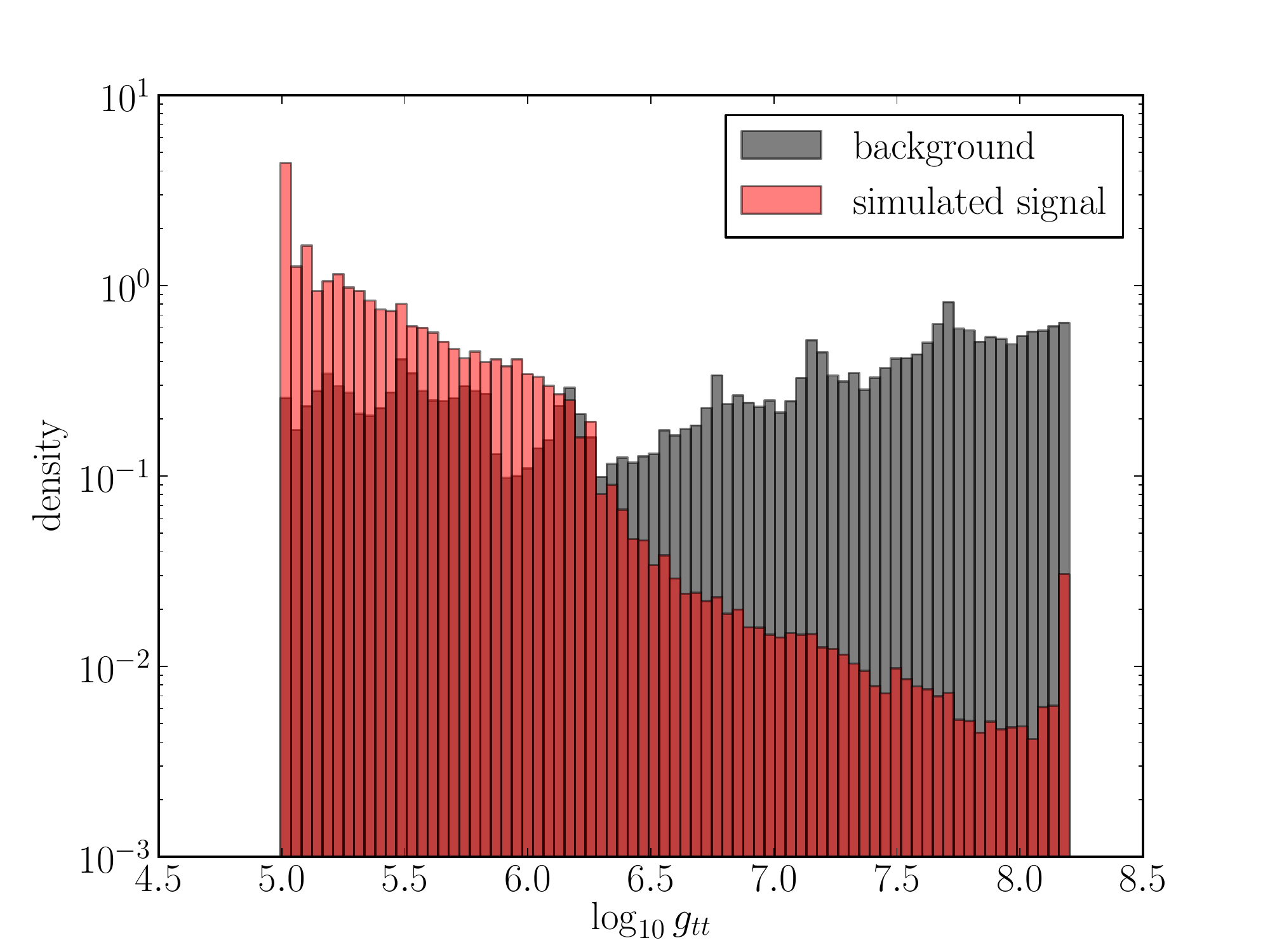}}
        \caption{Signal and background density distributions for a selection of feature vector statistics for the ringdown-only search.}
    \label{fig:rdparams1}
  \end{figure*}

\begin{figure*}
    \centering
     \subfigure[]{\label{fig:bvol}\includegraphics[width=.45\textwidth]{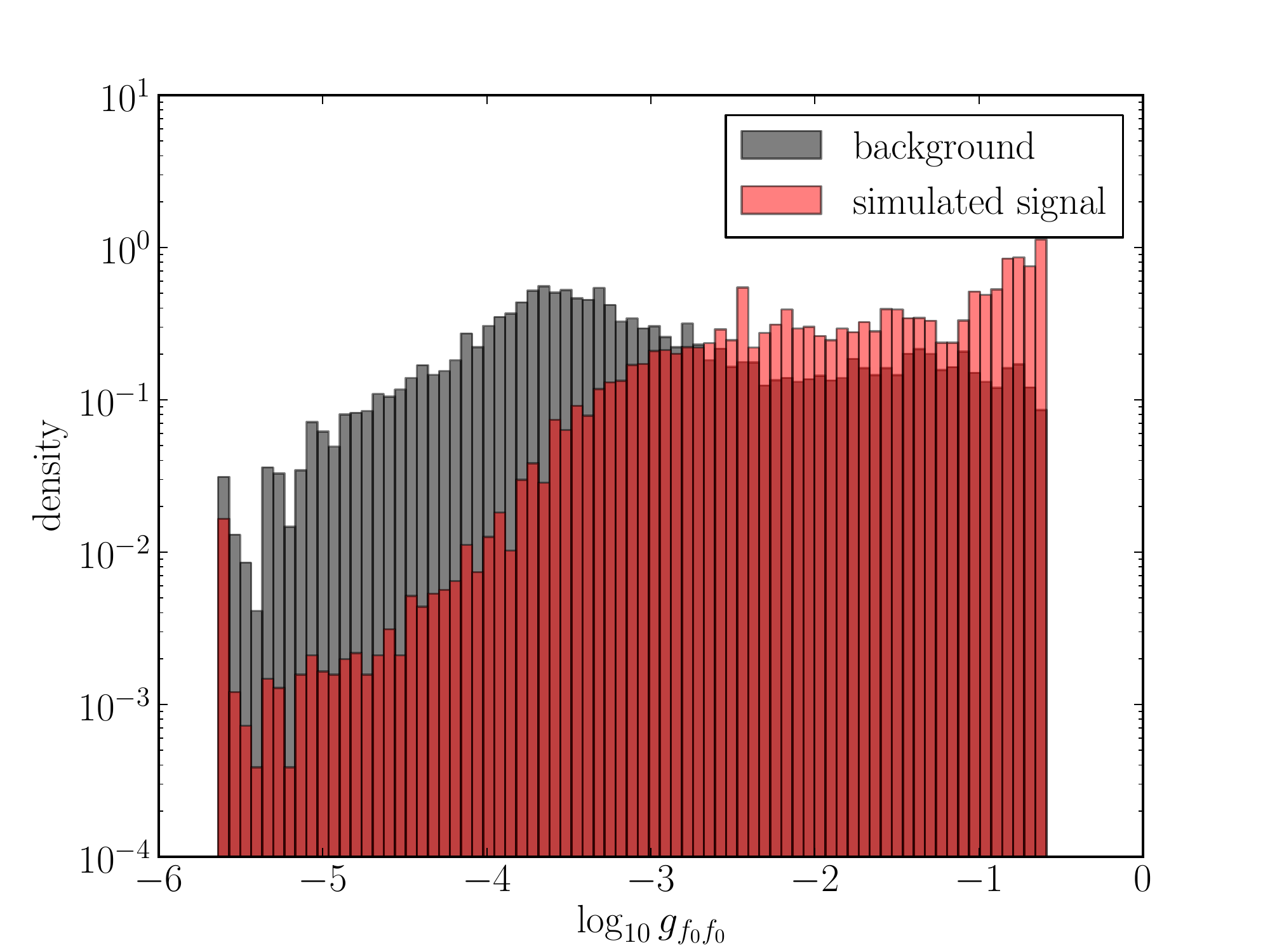}}
     \subfigure[]{\label{fig:avol}\includegraphics[width=.45\textwidth]{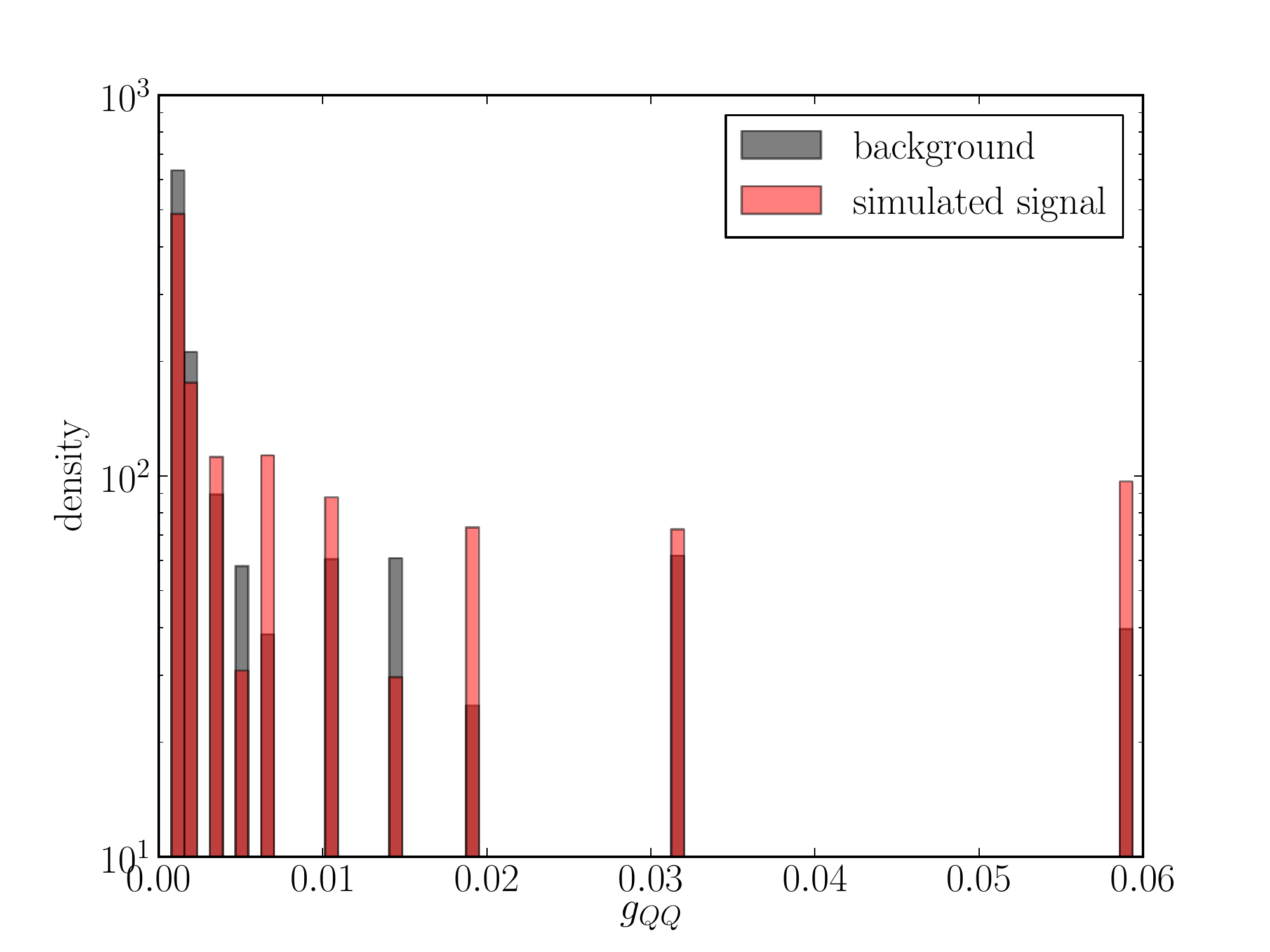}}
    \subfigure[]{\label{fig:bvol}\includegraphics[width=.45\textwidth]{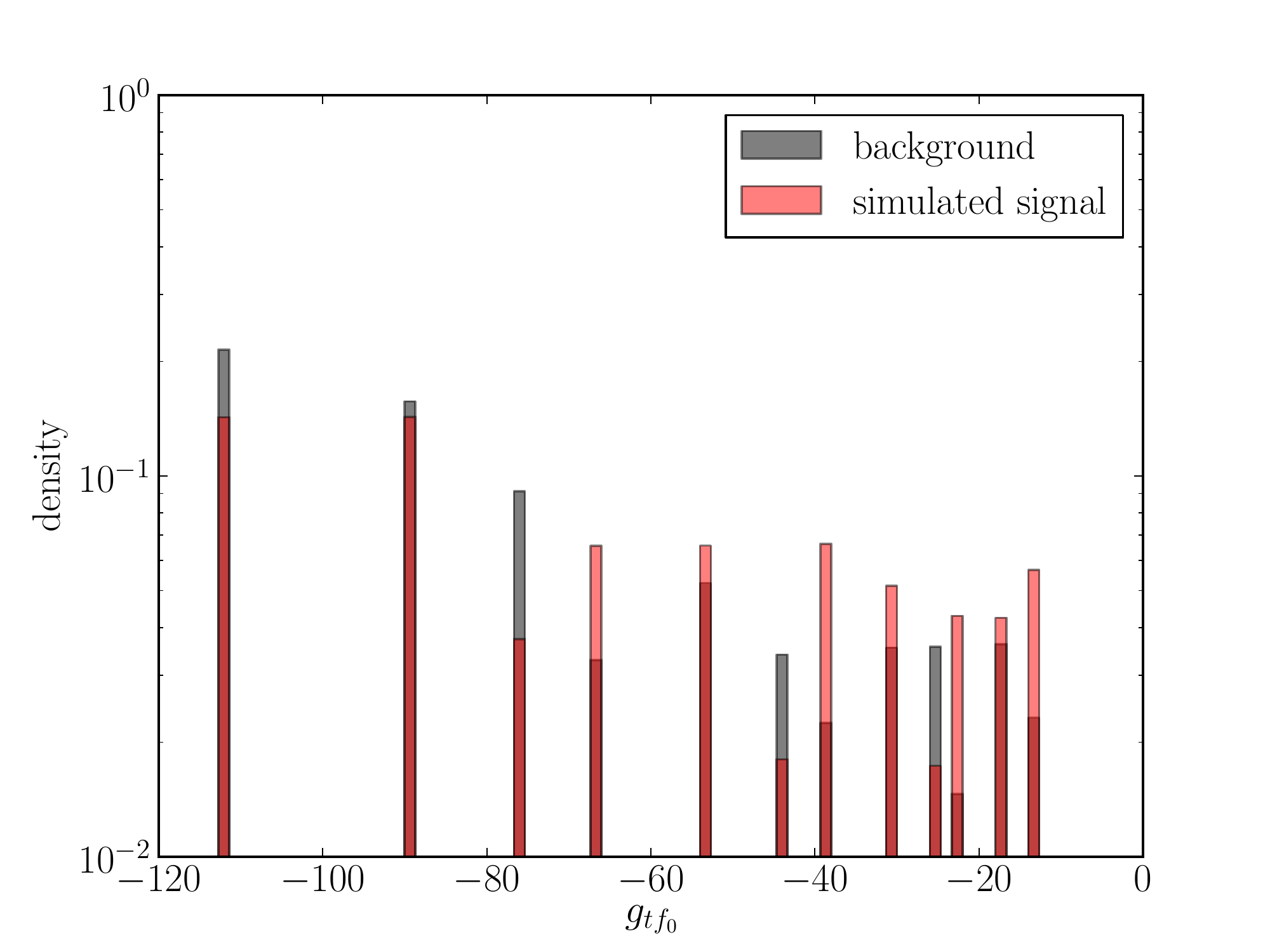}}
    \subfigure[]{\label{fig:bvol}\includegraphics[width=.45\textwidth]{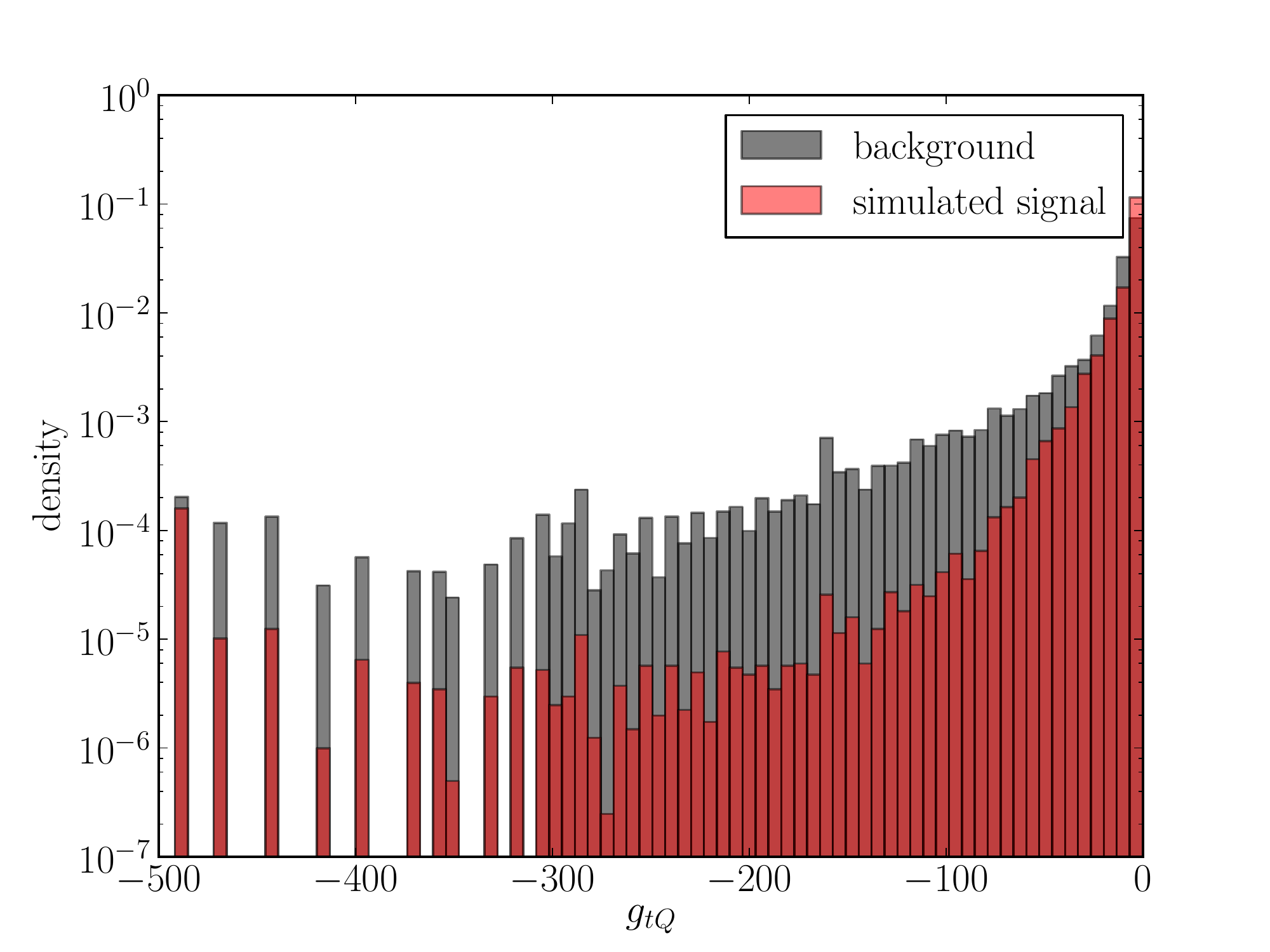}}
    \subfigure[]{\label{fig:bvol}\includegraphics[width=.45\textwidth]{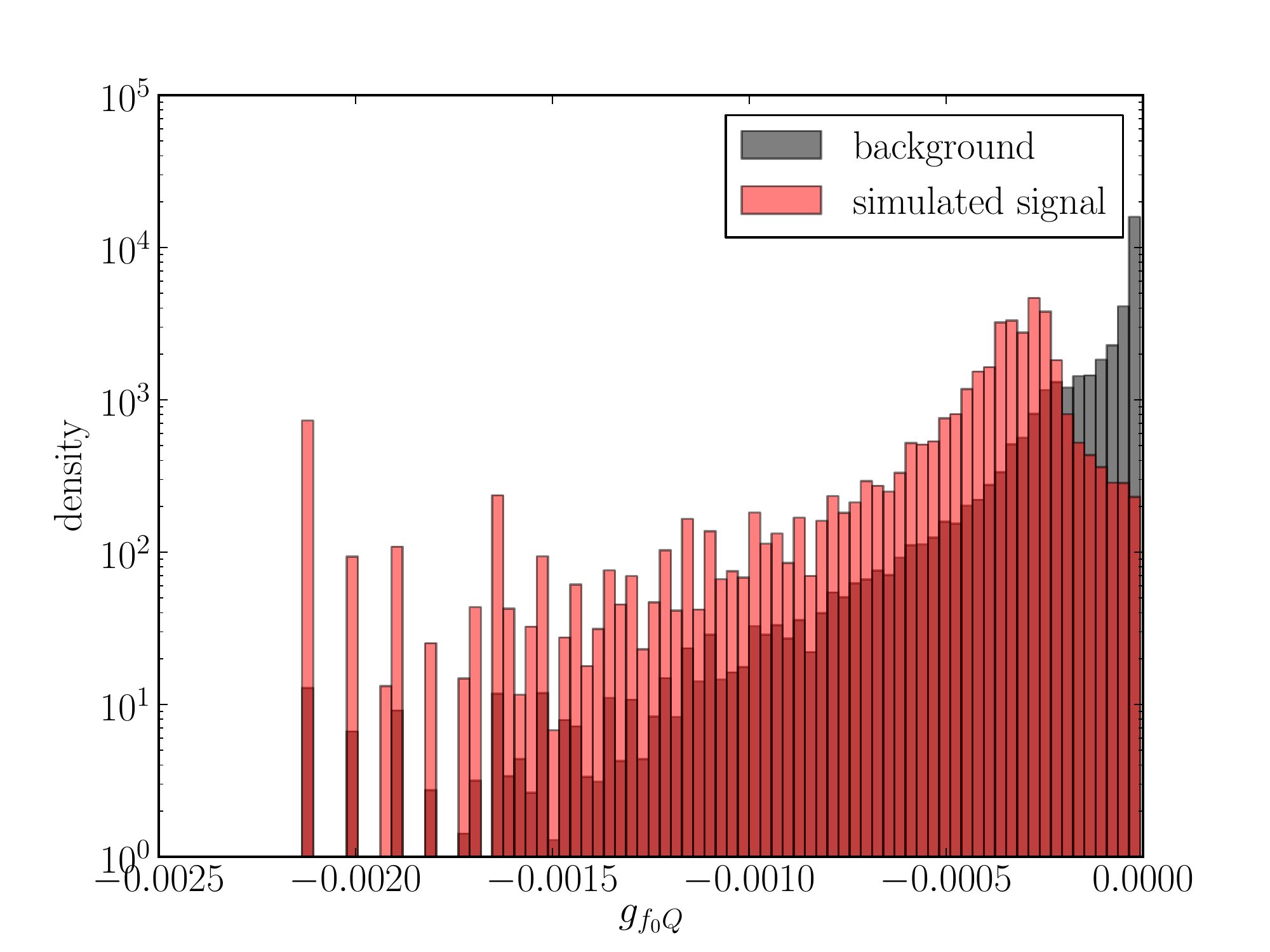}}
    \subfigure[]{\label{fig:avol}\includegraphics[width=.45\textwidth]{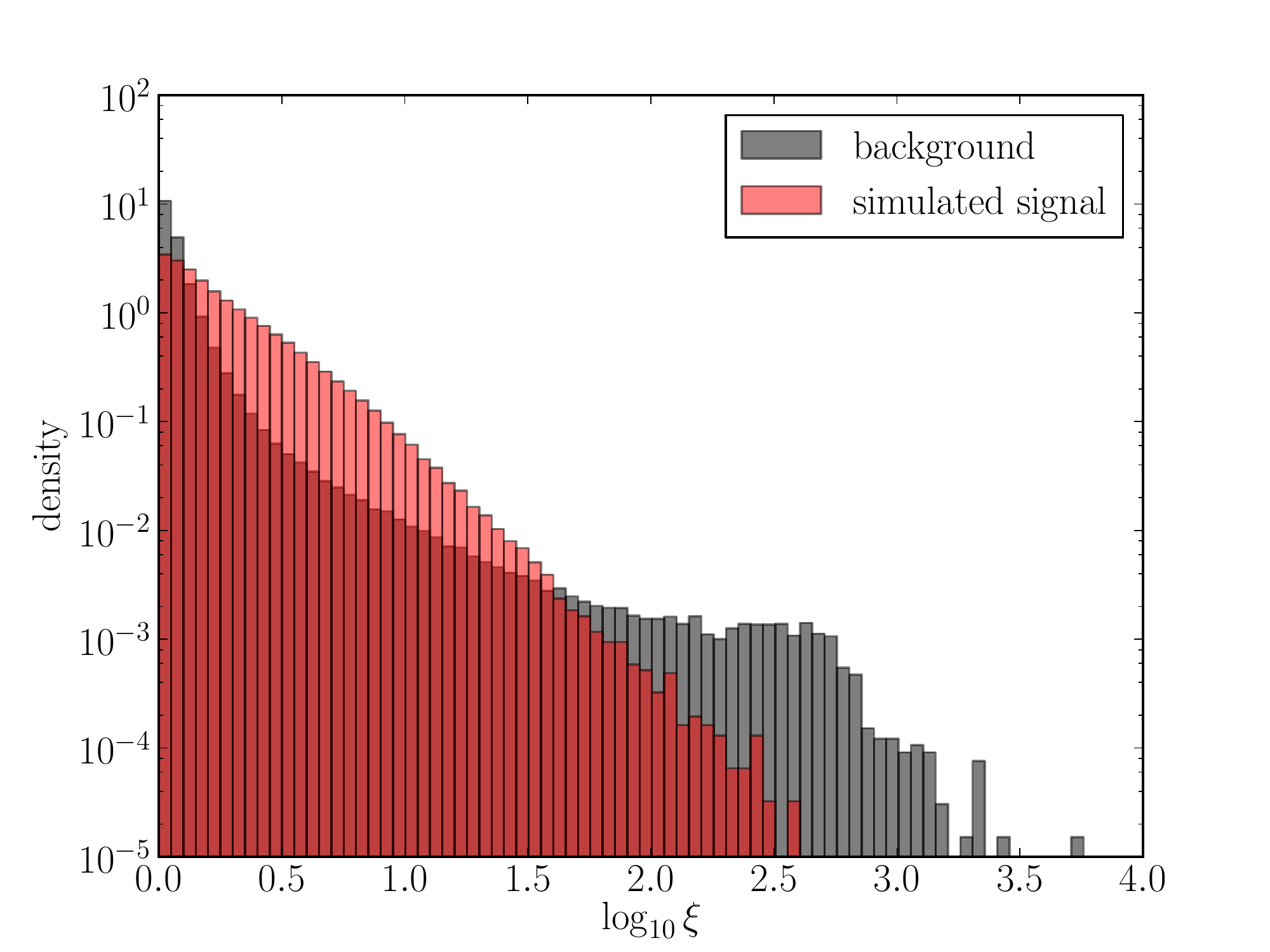}}
    \caption{Signal and background density distributions for a selection of feature vector statistics for the ringdown-only search.}
    \label{fig:rdparams2}
  \end{figure*}

\begin{figure*}
    \centering
    \subfigure[]{\label{fig:bvol}\includegraphics[width=.45\textwidth]{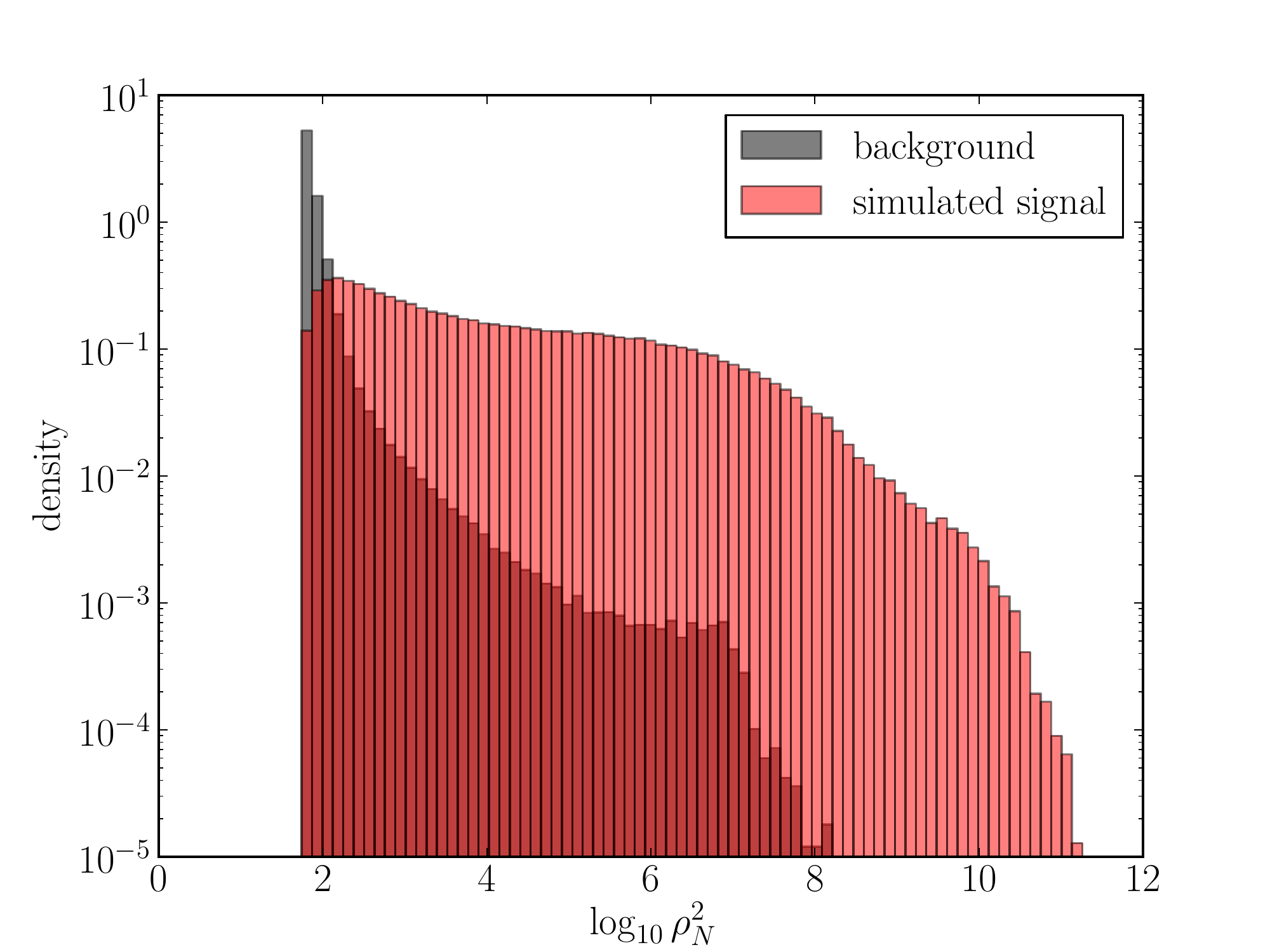}}
    \subfigure[]{\label{fig:bvol}\includegraphics[width=.45\textwidth]{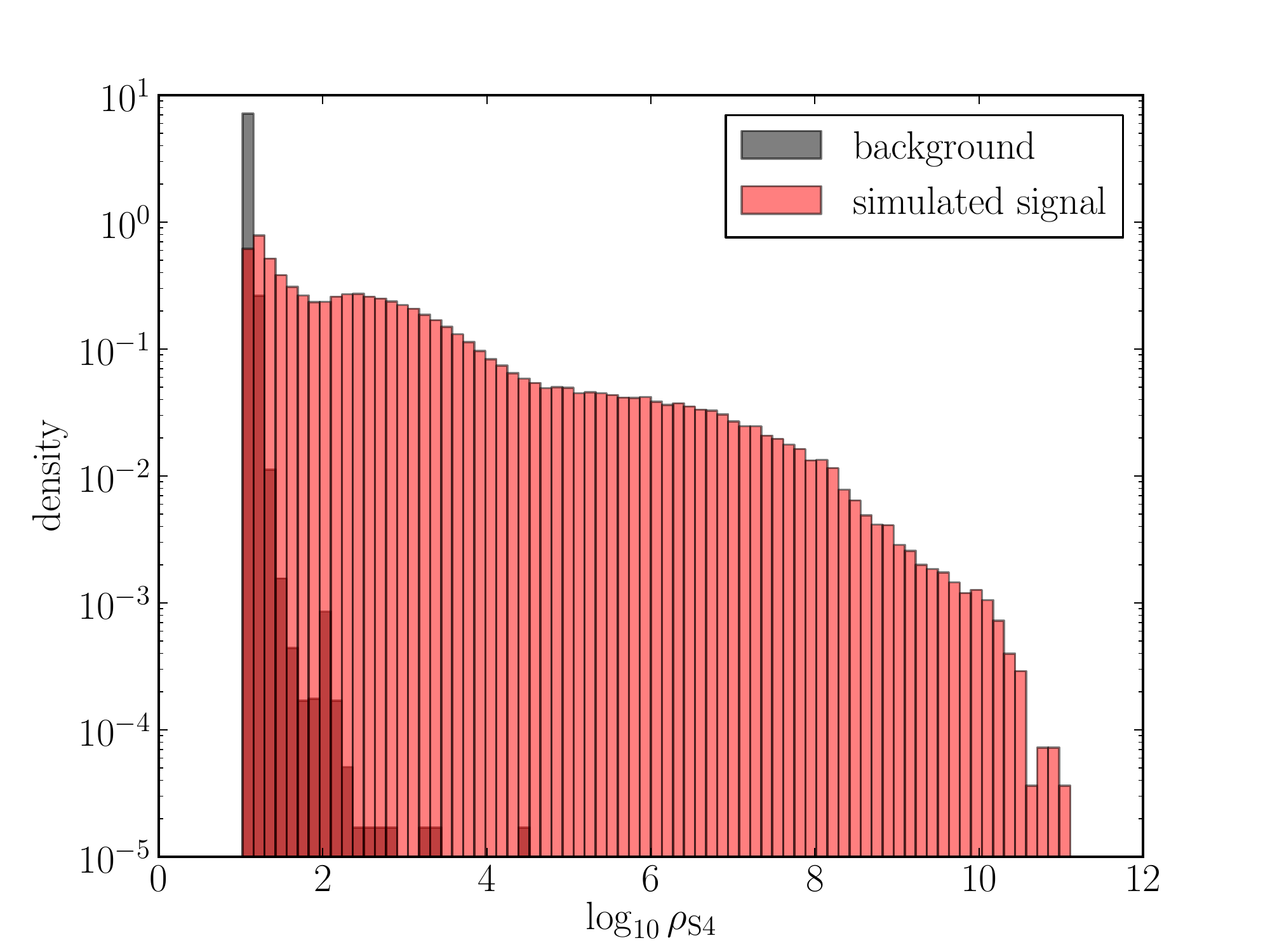}}
    \subfigure[]{\label{fig:bvol}\includegraphics[width=.45\textwidth]{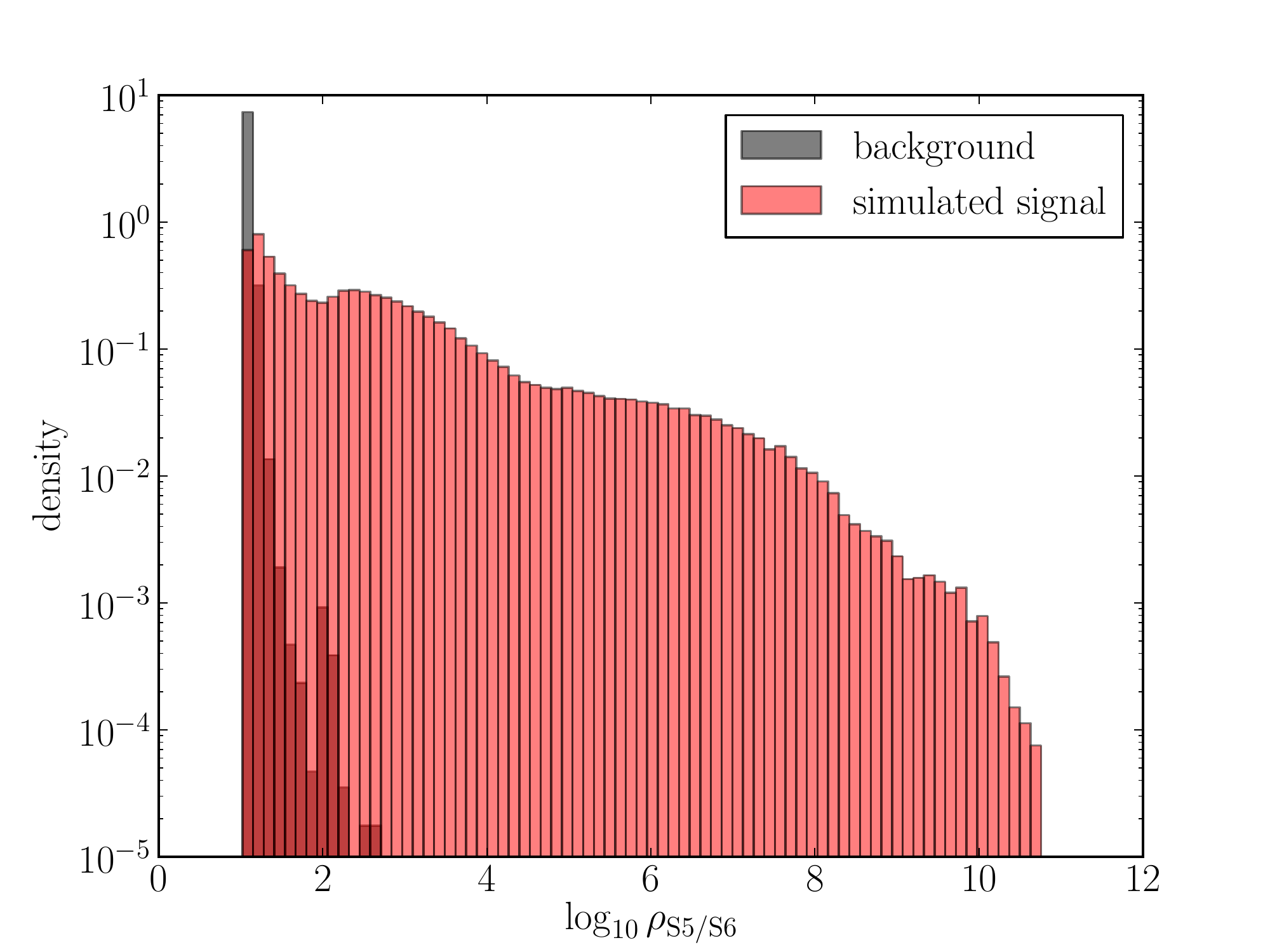}}
    \subfigure[]{\label{fig:bvol}\includegraphics[width=.45\textwidth]{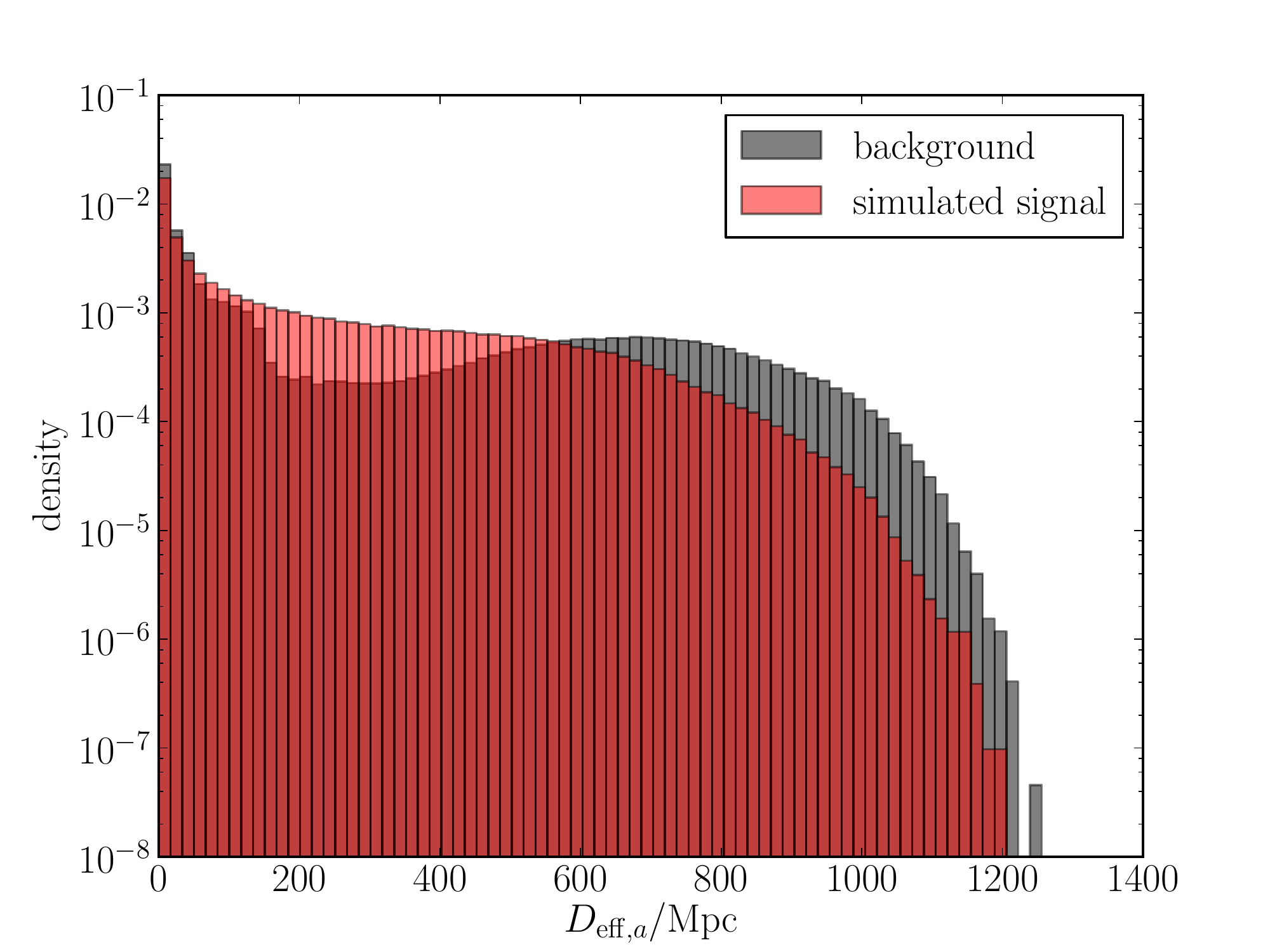}}
     \subfigure[]{\label{fig:avol}\includegraphics[width=.45\textwidth]{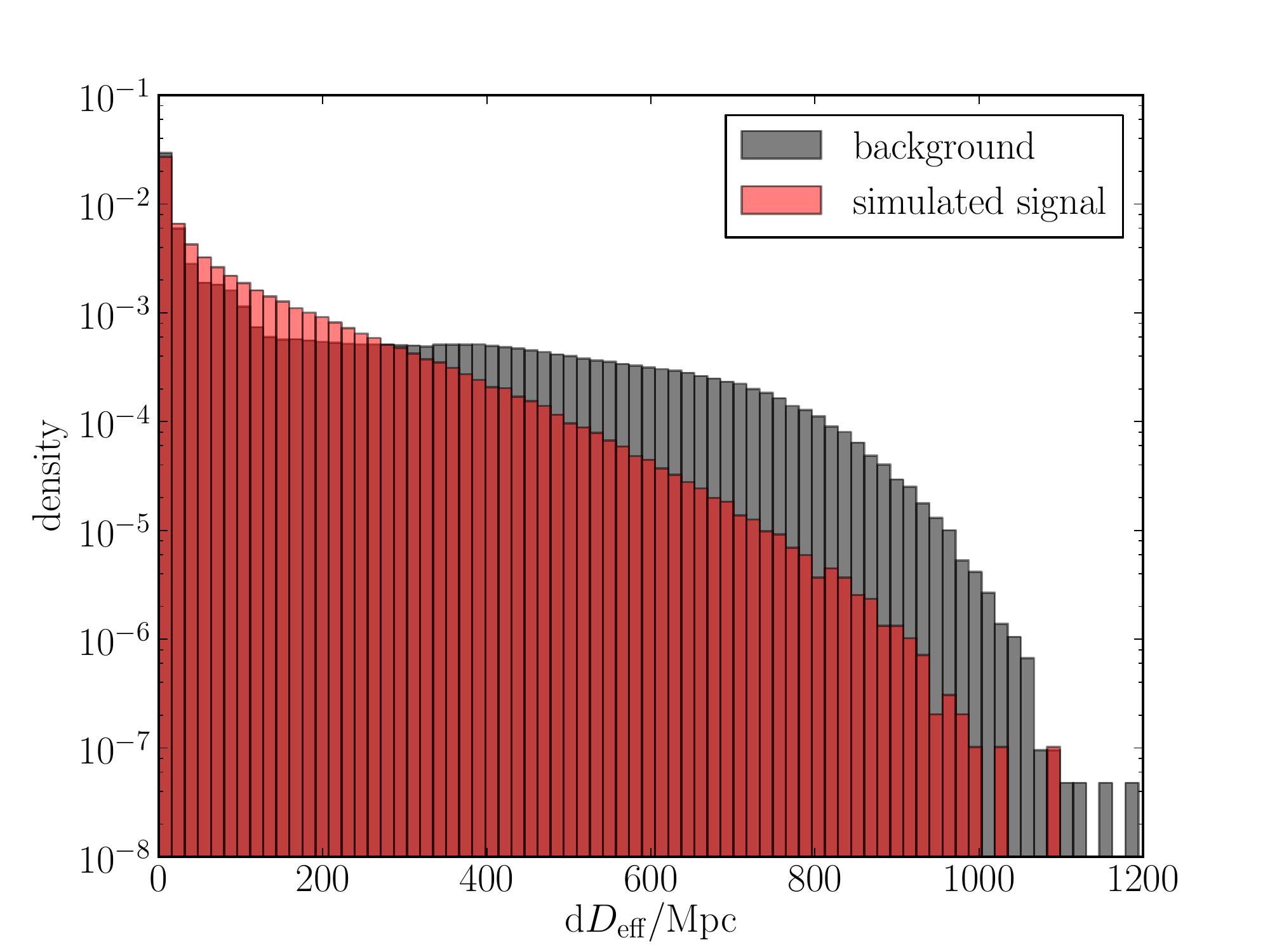}}
    \subfigure[]{\label{fig:bvol}\includegraphics[width=.45\textwidth]{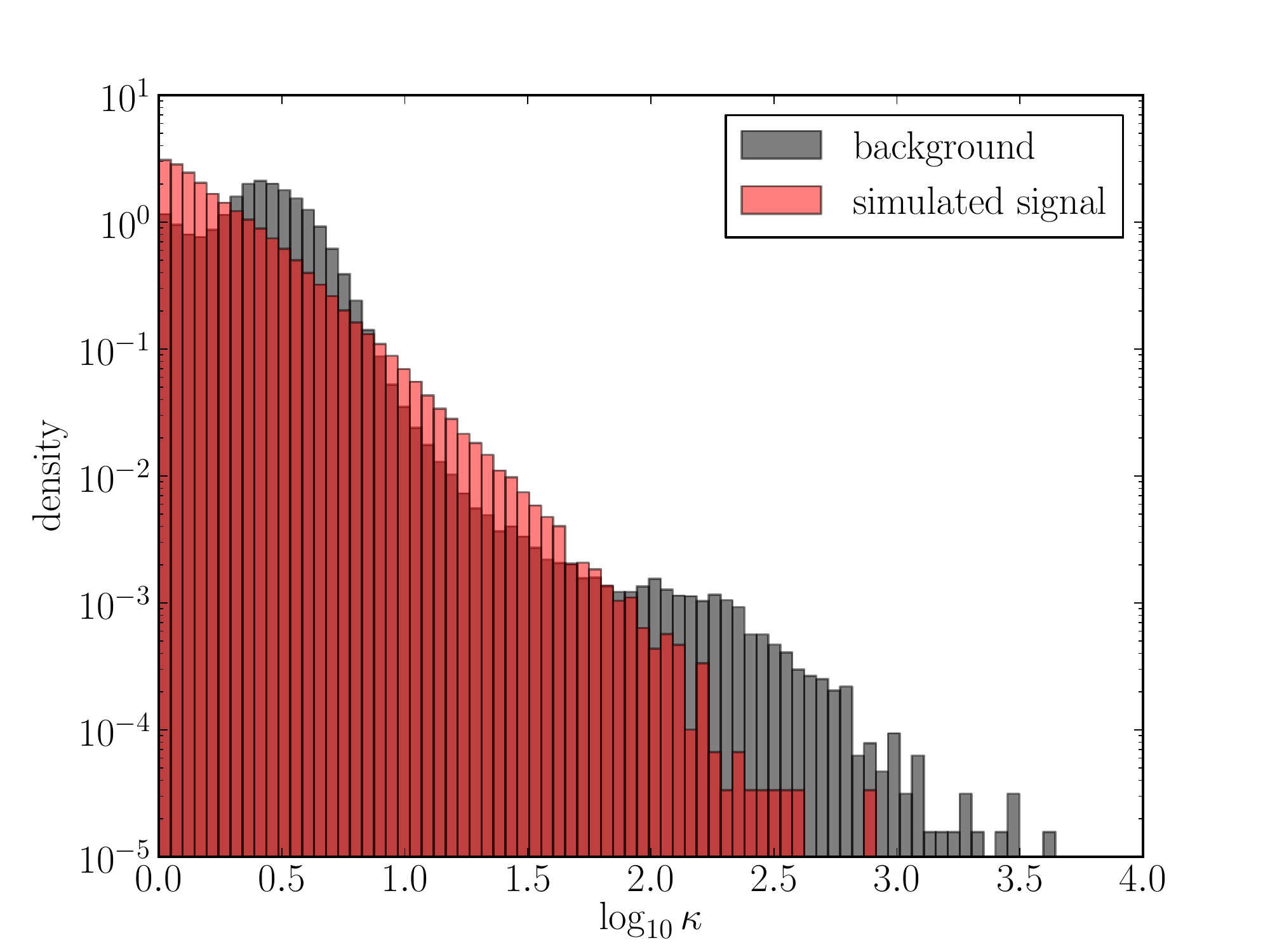}}
    \caption{Signal and background density distributions for a selection of feature vector statistics for the ringdown-only search.}
    \label{fig:rdparams3}
  \end{figure*}

\begin{figure*}
    \centering
    \subfigure[]{\label{fig:bvol}\includegraphics[width=.45\textwidth]{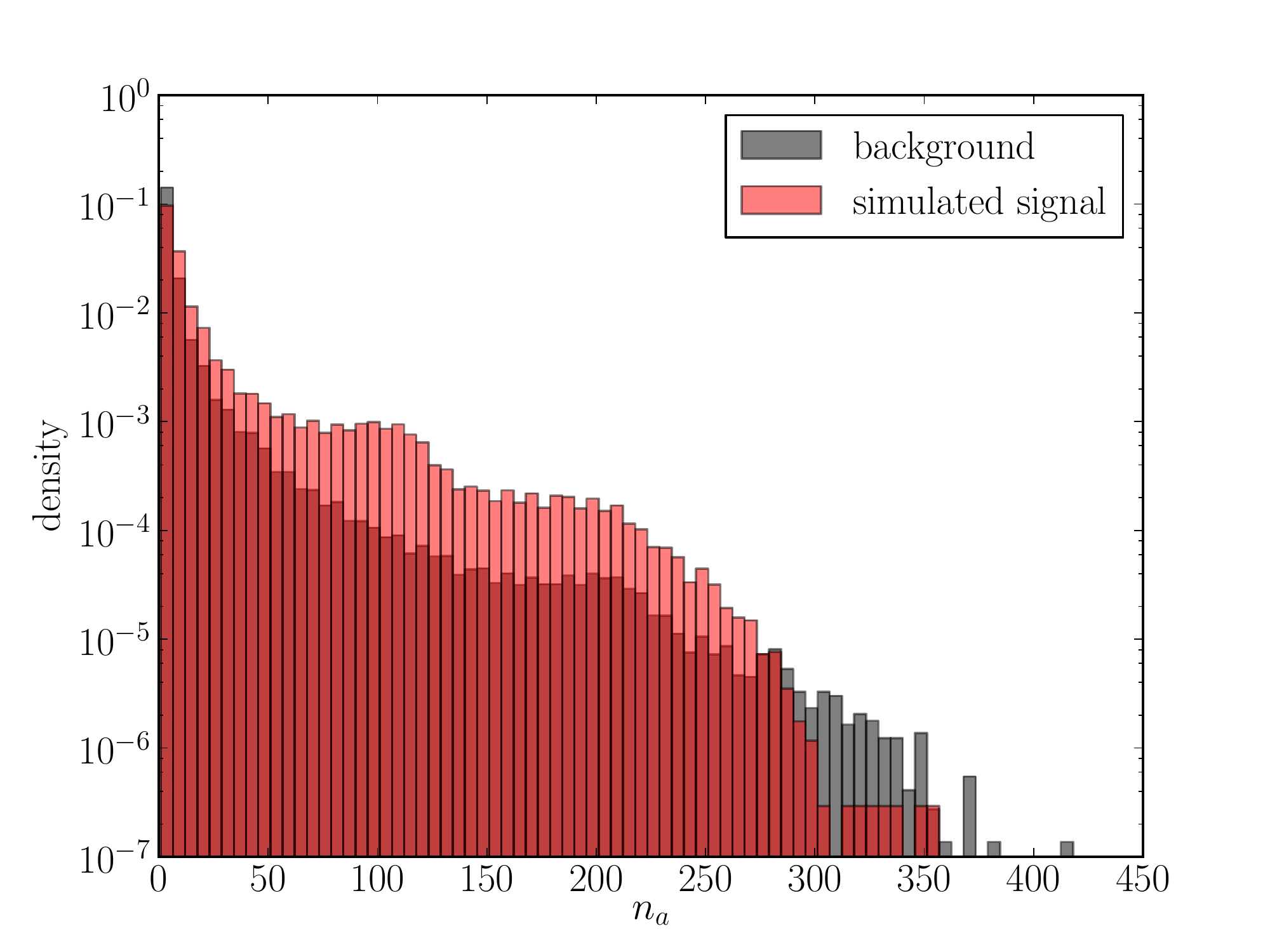}}
    \subfigure[]{\label{fig:bvol}\includegraphics[width=.45\textwidth]{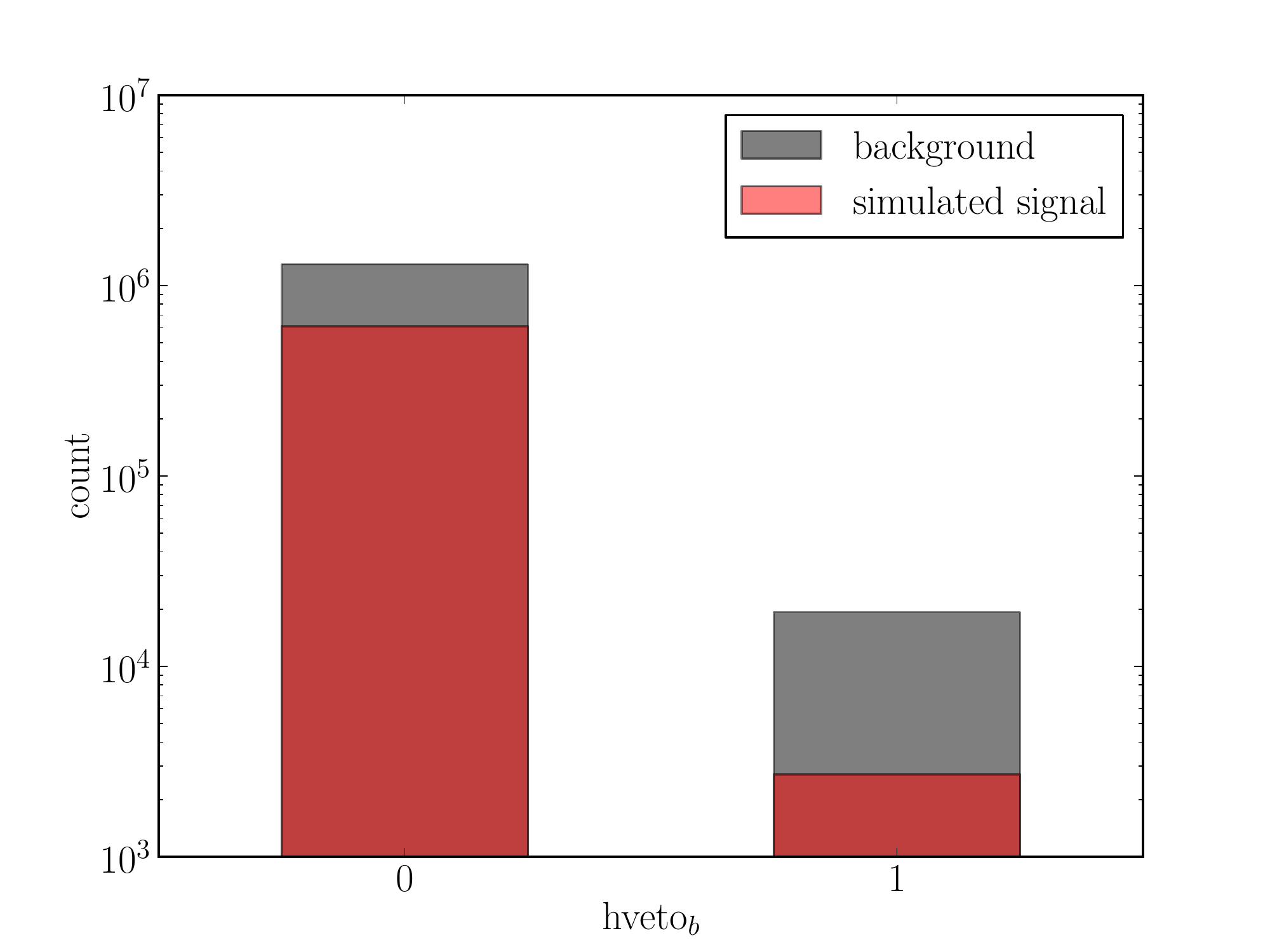}}
    \caption{Signal and background density distributions for a selection of feature vector statistics for the ringdown-only search.}
    \label{fig:rdparams4}
  \end{figure*}

The full list of statistics used for the ringdown-only search's feature vector are given in Table~\ref{tab:rdparams}.
In the following sections, we provide more detail on the definitions of each statistic.
Density distributions of these statistics for this search's simulated signal and background training sets are shown in Fig.~\ref{fig:rdparams1}-\ref{fig:rdparams4}.

\subsection{Single trigger statistics}\label{sec:rdsngl_parameters}
For the ringdown-only search, single trigger statistics added to the feature vector included the matched-filter SNR from each detector and the effective distance from each detector, $\mathrm{D}_{\mathrm{eff}}$.

More details on the matched-filter SNR, specifically for ringdown templates, are given in~\cite{2009s4ringdown, 2014rdsearch}.

The effective distance is equivalent to the distance $r$ to a source that is optimally oriented and located.
The theoretical formula for the effective distance is defined 
in terms of the $F_+$ and $F_\times$ detector antenna pattern functions and the inclination angle $\iota$,
\begin{equation}
D_\mathrm{eff}=\frac{r}{\sqrt{F_+^2\left( 1+ \cos^2\iota \right)/4 + F_\times^2\cos^2\iota}}.
\end{equation}
In practice, however, the effective distance is calculated from the power spectral density of the detector and the matched-filter SNR; see Table~\ref{tab:rdparams}.
\par

Although we did not include them here, additional single trigger statistics may be available to a search (e.g., coherent and null SNRs computed from coherent analyses~\cite{2011coherent, 2013talukder}).
\par

\subsection{Composite statistics}\label{sec:rdcomp_parameters}
Composite statistics included in the feature vector for the ringdown-only search include a combined network SNR, a detection statistic used in~\cite{2009s4ringdown}, and a ranking statistic detailed in~\cite{2012talukderthesis, 2012caudillthesis, 2013talukder}.

The combined network SNR for the $N$ detectors participating in the coincidence,
\begin{equation}\label{eq:ntwksnr}
\rho^2_N=\sum\limits_{i}^N\rho_i^2,
\end{equation}
where $\rho_i$ is the SNR in the $i$th detector, is the optimal ranking statistic for a signal with known parameters in Gaussian noise.

In the ringdown-only search in~\cite{2009s4ringdown}, due to the dearth of false alarms found in triple coincidence, a suitable statistic for ranking triple coincident events was found to be the the network SNR in Eq.~\ref{eq:ntwksnr} such that $\rho_\mathrm{S4,triple}=\rho^2_N$.
However, \cite{2009s4ringdown} found a high level of double coincident false alarms, often with very high SNR in only one detector.
While it is possible that a real gravitational-wave source could have an orientation that would produce an asymmetric SNR pair, the occurrence is relatively rare in comparison to the occurrence of this feature for false alarms.
The network SNR is clearly non-optimal in this case.
References~\cite{1999creightonchopL, 2008gogginthesis} found the optimal statistic in such a case to be a ``chopped-L" statistic,
\begin{equation}\label{eq:s4doub}
\rho_\mathrm{S4,double} = \min \left\{
  \begin{array}{c}
    \rho_{\mathrm{ifo1}}+\rho_{\mathrm{ifo2}} \\
    \alpha \rho_{\mathrm{ifo1}}+\beta \\
    \alpha \rho_{\mathrm{ifo2}}+\beta
  \end{array}
\right\}
\end{equation}
where the tunable parameters $\alpha$ and $\beta$ were set to 2 and 2.2, respectively, as described in~\cite{2009s4ringdown}.
We include this piece-wise detection statistic composed of $\rho_\mathrm{S4,triple}$ and $\rho_\mathrm{S4,double}$ in the feature vector.

For the most recent ringdown-only search~\cite{2014rdsearch}, due to a large increase in analysis time and lower SNR thresholds, a significant population of triple coincident false alarms were found.
Thus, an additional ``chopped-L"-like statistic was developed for triple coincidences,
\begin{equation}\label{eq:s5s6trip}
\rho_\mathrm{S5/S6,triple} = \min \left\{
  \begin{array}{c}
    \rho_N \\
    \rho_{\mathrm{ifo1}}+\rho_{\mathrm{ifo2}}+\gamma \\
    \rho_{\mathrm{ifo2}}+\rho_{\mathrm{ifo3}}+\gamma \\
    \rho_{\mathrm{ifo3}}+\rho_{\mathrm{ifo1}}+\gamma
  \end{array}
\right\}
\end{equation}
where the tunable parameter $\gamma$ was set to 0.75.
The development and tuning of this new statistic are described in detail in~\cite{2012talukderthesis, 2012caudillthesis, 2013talukder}.
Again, we include this piece-wise detection statistic composed of $\rho_\mathrm{S5/S6,triple}$ and $\rho_\mathrm{S4,double}$ in the feature vector.
\par

In addition to these three previous ranking statistics, we include the following simple composite statistics: the maximum of the ratios of the SNRs for triggers in each detector, the difference in recovered effective distances, and the maximum of the ratios of the recovered effective distances.
\par

Finally, we calculate quantities that provide an indication of how close the pair of triggers from different detectors are in the metric space $\left(f_0, Q, t\right)$ for the ringdown-only search.
These include the difference in arrival time d$t$, the template frequency difference d$f_0$, the template quality factor difference d$Q$, and the 3D-metric distance d$s^2$ between two triggers in $\left(f_0, Q, t\right)$ space~\cite{2014rdsearch,2003rdds2}.
Also included are the 3D-coincidence metric coefficients g$_{tt}$, g$_{f_0f_0}$, g$_{QQ}$, g$_{tf_0}$, g$_{tQ}$, and g$_{f_0Q}$ defined in Table~\ref{tab:rdparams}.
\par

\subsection{Other parameters}\label{sec:rdother_parameters}
Two additional parameters were added to the feature vector for the ringdown-only search in an effort to provide data quality information to the classifier.
\par

The first was a binary value used to indicate whether a trigger in a coincidence occurred during a time interval flagged for noise transients.
The flagged intervals were defined using the hierarchical method for vetoing noise transients known as {\it hveto} as described in~\cite{2011hveto}.
The LIGO and Virgo gravitational-wave detectors have hundreds of auxiliary channels monitoring local environment and detector subsystems.
The hveto algorithm identifies auxiliary channels that exhibit a significant correlation with transient noise present in the gravitational-wave channel and that have a negligible sensitivity to gravitational-waves.
If a trigger in the gravitational-wave channel is found to have a statistical relationship with auxiliary channel glitches, a flagged time interval is defined.
\par

The second additional parameter was a count of the number of single detector triggers clustered over a time interval of $0.5\,$ms using a SNR peak-finding algorithm described in~\cite{2008gogginthesis}.
The motivation behind this parameter comes from investigations that show that a glitch will be recovered with a different pattern of templates over time than a signal~\cite{2008chadthesis}.
Ideally, a $\chi^2$-based statistic could be computed.
However, in the absence of this test for the ringdown-only search, we simply provide a count of the number of templates in a small time window around each trigger giving a matched-filter SNR above the threshold.
\par

\end{document}